\documentclass[10pt,a4paper]{article}
\usepackage[latin1]{inputenc}
\usepackage{amsmath}
\usepackage{amsfonts}
\usepackage{amssymb}
\usepackage{graphics}
\usepackage{subfig}
\usepackage{psfrag}
\usepackage{epsfig}
\usepackage{pstricks}
\usepackage{float}
\usepackage{empheq}
\newcommand{\hiddensubsection}[1]{
    \stepcounter{subsection}
    \subsection*{\Alph{section}.\arabic{subsection}\hspace{1em}{#1}}
}
\newcommand{\m}[1]{\mathcal{#1}}
\newcommand{\ph}[1]{\phantom{#1}}
\newcommand{\w}{\wedge}
\newcommand{\ra}{\rightarrow}
\newcommand{\nn}{\nonumber}
\newcommand{\mb}[1]{\mathbb{#1}}

\newcommand{\os}{\overset{*}{=}}
\newlength\dlf  

\author{\\\\
Hans F. Westman$^1$\footnote{\texttt{hwestman74@gmail.com}}\; and Tom G. Zlosnik$^2$\footnote{\texttt{tom.zlosnik@gmail.com}}\\
{\small \it $(1)$ Instituto de F\'isica Fundamental, CSIC, Serrano 113-B, 28006 Madrid, Spain}\\
{\small \it $(2)$ Imperial College Theoretical Physics, Huxley Building, London, SW7 2AZ}
}
\date{\today}

\title{An introduction to the physics of Cartan gravity}
\begin{document}
\maketitle
\begin{abstract}

A distance can be measured by monitoring how much a wheel has rotated when rolled without slipping. This simple idea underlies the mathematics of Cartan geometry. The Cartan-geometric description of gravity consists of a $SO(1,4)$ gauge connection $A^{AB}(x)$ and a gravitational Higgs field $V^A(x)$ which breaks the gauge symmetry. The clear similarity with symmetry-broken Yang-Mills theory suggests strongly the existence of a new field $V^A$ in nature: the gravitational Higgs field. By treating $V^A$ as a genuine dynamical field we arrive at a natural generalization of General Relativity with a wealth of new phenomenology. Importantly, General Relativity is reproduced exactly in the limit that the $SO(1,4)$ norm $V^2(x)$ tends to a positive constant. We show that in regions wherein $V^{2}$ varies-but has a definite sign-the Cartan-geometric formulation is a particular version of a scalar-tensor theory (in the sense of gravity being described by a scalar field $\phi$, metric tensor $g_{\mu\nu}$, and possibly a torsion tensor ${\cal T_{\mu\nu}}^{\rho}$). A specific choice of action yields the Peebles-Ratra quintessence model whilst more general actions are shown to exhibit propagation of torsion. Regions where the sign of $V^2$ changes  correspond to a change in signature of the geometry. Specifically, a simple choice of action with FRW symmetry imposed yields, without any additional {\em ad hoc} assumptions, a classical analogue of the Hartle-Hawking no-boundary proposal with the big bang singularity replaced by signature change. Cosmological solutions from more general actions are described, none of which have a big bang singularity, with most solutions reproducing General Relativity, or its Euclidean version, for late cosmological times. Requiring that gravity couples to matter fields through the gauge prescription forces a fundamental change in the description of bosonic matter fields: the equations of motion of all matter fields become first-order partial differential equations with the scalar and Dirac actions taking on structurally similar first-order forms. All matter actions reduce to the standard ones in the limit $V^2\rightarrow const.$ We argue that Cartan geometry may function as a novel platform for inspiring and exploring modified theories of gravity with applications to dark energy, black holes, and early-universe cosmology. We end by listing a set of open problems.
\end{abstract}
\newpage
\tableofcontents
\section{Introduction}\label{intro}
Riemannian geometry forms the mathematical basis of Einstein's General Relativity. The metric representation of Riemannian geometry consists of the pair of variables $\{g_{\mu\nu},\Gamma^\rho_{\mu\nu}\}$. Whilst the metric tensor $g_{\mu\nu}$ encodes all information of distances between points on a manifold, the affine connection $\Gamma^\rho_{\mu\nu}$ encodes the information about parallel transport of tangent vectors $u^\mu$ as well as defining a covariant derivative $\nabla_\mu$ acting on tensors. Within Riemannian geometry the pair $\{g_{\mu\nu},\Gamma^\rho_{\mu\nu}\}$ must be {\em metric-compatible} and {\em torsion-free}:
\begin{itemize}
\item {\bf Metric compatibility:} $\nabla_\rho g_{\mu\nu}\equiv \partial_\rho g_{\mu\nu}-\Gamma^\sigma_{\rho\mu}g_{\sigma\nu}-\Gamma^\sigma_{\rho\nu}g_{\mu\sigma}=0$
\item {\bf Zero torsion:} $\Gamma^\rho_{\mu\nu}-\Gamma^\rho_{\nu\mu}=0$.
\end{itemize}
The affine connection can then be uniquely determined from the metric 
\begin{eqnarray}
\Gamma^\rho_{\mu\nu}=\frac{1}{2}g^{\rho\sigma}(\partial_\mu g_{\sigma\nu}+\partial_\nu g_{\mu\sigma}-\partial_\sigma g_{\mu\nu})
\end{eqnarray}
and it becomes natural to view the metric as the primary variable and the affine connection as a secondary, derived quantity. 

Despite its monumental success it has long been noted (see e.g. \cite{Weinberg:1996kr,Zee:2003mt}) that this description of the gravitational field is quite distinct from that of the force fields of the standard model, i.e. the electroweak and strong forces. The latter two are examples of standard Yang-Mills theories with the electroweak theory being an example of a symmetry-broken gauge theory. On the other hand, gravity in its traditional Riemannian formulation displays only a superficial similarity to a Yang-Mills field (see \cite{Randono:2010cq,Westman:2012zk} for  discussion of the differences). In \cite{Weinberg:100595} Weinberg writes:
\begin{quote}
{\em `\dots I believe that the geometrical approach has driven a wedge between General Relativity and the theory of elementary particles. As long as it could be hoped, as Einstein did hope, that matter would eventually be understood in geometrical terms, it made sense to give Riemannian geometry a primary role in describing the theory of gravitation. But now the passage of time has taught us not to expect that the strong, weak, and electromagnetic interactions can be understood in geometrical terms, and too great an emphasis on geometry can only obscure the deep connections between gravitation and the rest of physics.}'
\end{quote}
The aim of this article is to show that a lesser-known formulation of gravity, based on Cartan geometry -- whose mathematical ingredients are precisely those of a spontaneously-broken gauge theory -- can underpin a more general, alternative theory of gravity that reduces to General Relativity in a specific limit. We shall refer to that formulation as {\em Cartan gravity} although this name is also frequently used for the Einstein-Cartan formulation of General Relativity \cite{Trautman:2006fp}. Rather than driving a wedge between gravity and the other forces of the standard model, it describes gravity in the same language as the other forces, i.e as a Yang-Mills theory. The dynamical fields of Cartan gravity consist of a Yang-Mills gauge connection $A^A_{\ph AB}(x)={A_\mu}^A_{\ph AB}(x)dx^\mu$ and a symmetry-breaking field $V^A(x)$, where $A,B$ are $SO(1,4)$ gauge indices.\footnote{Cartan gravity can also be based on the anti-de Sitter $SO(2,3)$ or the Poincar\'e group $ISO(1,3)$. However, in this paper we will only consider the de Sitter gauge group $SO(1,4)$; as we shall see, this group is rather more naturally associated with a positive cosmological constant. In the case of $ISO(1,3)$, the field $V^{A}$ possesses no gauge-independent degrees of freedom and no new degree of freedom is introduced in the gravitational sector \cite{Leclerc:2005qc}.}
\begin{table}[H]
\begin{center}
\begin{tabular}{|l|l|l|}
\hline
& {\bf Electroweak theory} & {\bf Cartan gravity}\\
\hline Gauge connection: & ${W_\mu}^\alpha_{\ph\alpha\beta}(x)$ & ${A_\mu}^A_{\ph AB}(x)$\\
\hline Higgs field: & $\Phi^\alpha(x)$ & $V^A(x)$\\
\hline Symmetry group: & $U(1)_Y\times SU(2)_L$& $SO(1,4)$\\
\hline Stabilizer group: & $U(1)_{EM}$& $SO(1,3)$\\
\hline
\end{tabular}
\caption{\it \small The table displays a side-by-side comparison of the basic mathematical objects in electroweak theory and Cartan gravity. The remnant (stabilizer) symmetry group is defined by the subgroup that leaves the Higgs field invariant. In the case of electroweak theory the remnant symmetry group is the $U(1)_{EM}$ group of electromagnetism and in Cartan gravity it is $SO(1,3)$ local Lorentz invariance of the tangent spaces.}
\label{EWCGcomp}
\end{center}
\end{table}
When viewed alongside the electroweak theory it becomes undeniable that Cartan gravity is in its essence a symmetry-broken Yang-Mills theory. See Table \ref{EWCGcomp} for a side by side comparison between Cartan gravity and the electroweak theory.  However, there is a glaring discrepancy that would still drive a wedge between gravity and the other forces in nature. While the symmetry breaking Higgs field $\Phi$ of the electroweak theory is treated as a genuine dynamical field, with its quantum excitations corresponding to the recently detected Higgs particle, the gravitational Higgs field $V^A$ is commonly treated as a non-dynamical, absolute object \cite{Wise:2006sm}. Specifically, the norm $V^2$ is typically postulated to be a constant function on spacetime, i.e. $V^2(x)=const.$ Needless to say, this is not problematic from a mathematical standpoint. Nevertheless, the imposed constancy of $V^2$ contrasts sharply with the dynamics of the symmetry breaking Higgs field of the standard model.

To further drive home the analogy let us elaborate a bit more. Almost all components of the Higgs field $\Phi$ can be regarded as gauge degrees of freedom with the only gauge independent degree of freedom being its norm $|\Phi|^2$. This degree of freedom is untouched by $SU(2)_L\times U(1)_Y$ gauge transformations and is therefore a gauge invariant quantity. In standard presentations the unitary gauge $\Phi(x)\os (0,v(x))$ is often used to highlight the physical content. In complete analogy we see that almost all components of $V^A$ can be viewed as gauge degrees of freedom, with only the norm untouched by $SO(1,4)$ gauge transformations. To highlight the physical content it is convenient to work in the gauge $V^A\os \phi \delta^A_4$. \footnote{This gauge is attainable only in the Lorentzian case when $V^2>0$. As we shall see in this article, if $V^2<0$ the natural gauge choice is $V^A\os \phi \delta^A_0$ and we are dealing with a Euclidean geometry.}

Given these observations it becomes natural to propose the existence of a new field in nature, namely the gravitational Higgs field $V^{A}$ with its norm $V^2$ describing a new gauge independent physical degree of freedom subject to non-trivial equations of motion. We shall see in this paper that the limit $V^2\ra const.>0$ corresponds exactly to General Relativity in its Einstein-Cartan incarnation. Thus, Cartan gravity with $V^2=const.>0$ and Einstein-Cartan theory are two distinct mathematical formulations of the {\em same} physical theory. Therefore, we find it appropriate to reserve the term {\em Cartan gravity} to refer exclusively to the theory in which both $A^{AB}$ and $V^A$ are treated as genuine dynamical degrees of freedom. As we shall see Cartan gravity exhibits a rich phenomenology with interesting applications to cosmology.

Specifically, the norm $V^2$ can play a role as dark energy and indeed it corresponds exactly to the Peebles-Ratra slow rolling quintessence for a simple Cartan-geometric action. We shall exhibit simple actions that achieve the symmetry breaking $V^2\ra const.$ dynamically without any {\em ad hoc} restrictions imposed. Thus, we see that General Relativity can be seen as the symmetry-broken phase of Cartan gravity with $V^A$ fully dynamical. The quantum excitations of $V^2$ would then presumably correspond to a new type of Higgs boson whose imprints on the early universe should in principle be observable. We shall also see that a varying $V^2(x)$ corresponds to non-metricity $\nabla_\rho g_{\mu\nu}=\partial_\rho V^2 g_{\mu\nu}$ which can be exchanged for a scalar field $V^2(x)$ by metric redefinition $g_{\mu\nu}\ra \tilde g_{\mu\nu}=\frac{V_0^2}{V^2}g_{\mu\nu}$. The resulting theory is a form of scalar-tensor theory which may or may not have propagating torsion depending on what action principle we choose.

Apart from bringing more harmony by seemingly placing gravity as `just another gauge field' in nature, the mathematical machinery of Cartan geometry also has implications for the coupling of matter to gravity. Specifically, the coupling of a Yang-Mills field to a matter field, e.g. a Higgs field or a fermionic spinor field, follows the {\em gauge prescription}: i.e. the object we couple the gauge field to has a gauge index and the partial derivative is simply replaced by the gauge covariant derivative. As was detailed in \cite{Westman:2012zk}, if we require the coupling between gravity and matter fields to follow the same pattern we end up with a very different {\em first order} representation of matter fields which nonetheless reduces the the standard second order formulation in the General Relativistic limit $V^2\ra const.>0$.

The article is organized as follows:  In Section \ref{waywisermath} we develop the mathematical theory of \emph{idealized waywisers} which forms the mathematical basis of Cartan geometry. In order to facilitate visualization and build intuition, we first restrict attention to the case of two-dimensional manifolds embedded in a three-dimensional space. It is shown that all the basic mathematical objects of Riemannian geometry (i.e. $g_{\mu\nu}$, $\Gamma^\rho_{\mu\nu}$, $R_{\mu\nu\rho\xi}$, \dots) are recoverable from the mathematical objects that describe the idealized waywiser, the so-called {\em waywiser variables} $\{A^{AB},V^A\}$.  The notion of waywisers and the manner in which they probe geometry is immediately generalizable to manifolds of higher dimension. 
In Section \ref{higherd} we discuss the generalization of Cartan waywiser geometry to the physically important case of four dimensional spacetime manifolds and we clarify the relationship between the waywiser variables and the  aforementioned variables $e^{I}$ and $\omega^{IJ}$. In Section \ref{phenomenology} we explore the phenomenology of Cartan gravity: the emergence of Peebles-Ratra quintessence; the inevitability of dark energy having dynamics in Cartan gravity; the scope for the propagation of torsion; specific cosmological solutions involving a classical analogue of the Hartle-Hawking signature-change process; and the coupling of Cartan gravitational fields to the matter. Finally, in Section \ref{conclusions} we present our conclusions and suggest areas for further investigation. 

The language of differential forms is very helpful and simplifies the calculations and cleans up the notation immensely. However, the method of differential forms is not a standard tool for working physicists and cosmologists. In order to increase the accessibility we have therefore provided several appendices as to make the content of this paper as self-contained as possible.
\section{Introducing Cartan waywiser geometry}\label{waywisermath}
In this section we shall develop the mathematics of idealized waywisers. This conception of differential geometry treats both metric $g_{ab}(x)$ and affine connection $\Gamma^c_{ab}(x)$ as {\em derived} concepts constructed from the more basic waywiser variables $V^i(x)$ and ${A_a}^i_{\ph ij}(x)$ whose straightforward geometric interpretation to which we now turn. For a complementary and more mathematically sophisticated introduction to Cartan geometry see \cite{Wise:2006sm,SharpeCartan}.

\subsection{Idealized waywisers}
\label{whatarewaywisers}
\begin{figure}
    \centering
    \subfloat[Traditional waywiser]{{\includegraphics[width=7cm]{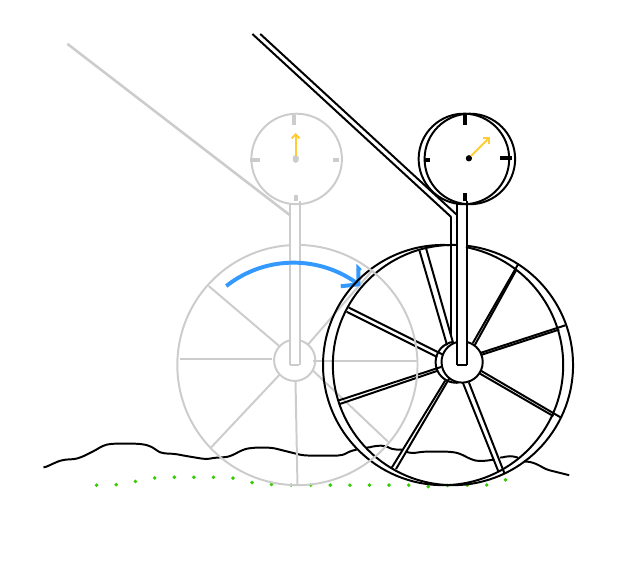} }}%
    \qquad
    \subfloat[Idealized waywiser]{{\includegraphics[width=7cm]{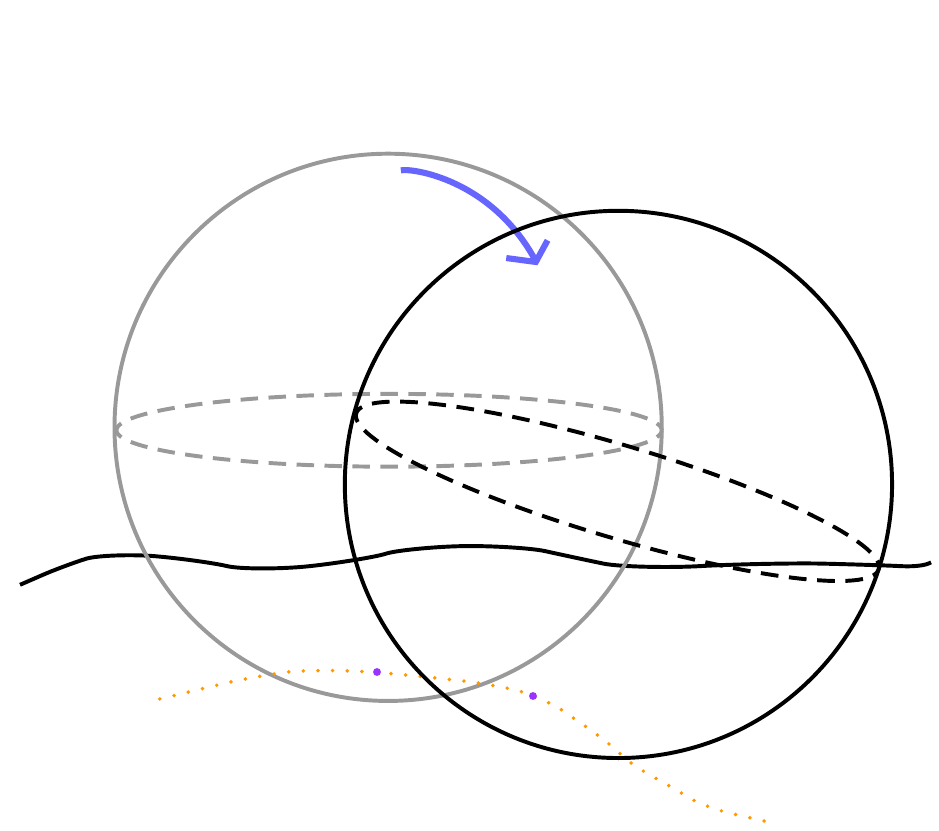} }}%
    \caption{The figure on the left is a traditional waywiser depicted rolling along on a two-dimensional surface. A mechanism converts the rolling of the wheel into a measure of distance traversed along the dotted path, as depicted by the changing orientation of the orange arrow. The picture on the right depicts a mathematical idealization and generalization. The wheel has been replaced by a symmetric space, a sphere, and the geometry of the manifold is now revealed by how this symmetric space has rotated when rolled (without slipping) along some path on a manifold.}%
    \label{fig:waywisers}%
\end{figure}

\begin{figure}[H]
\centering
\includegraphics[width=10cm]{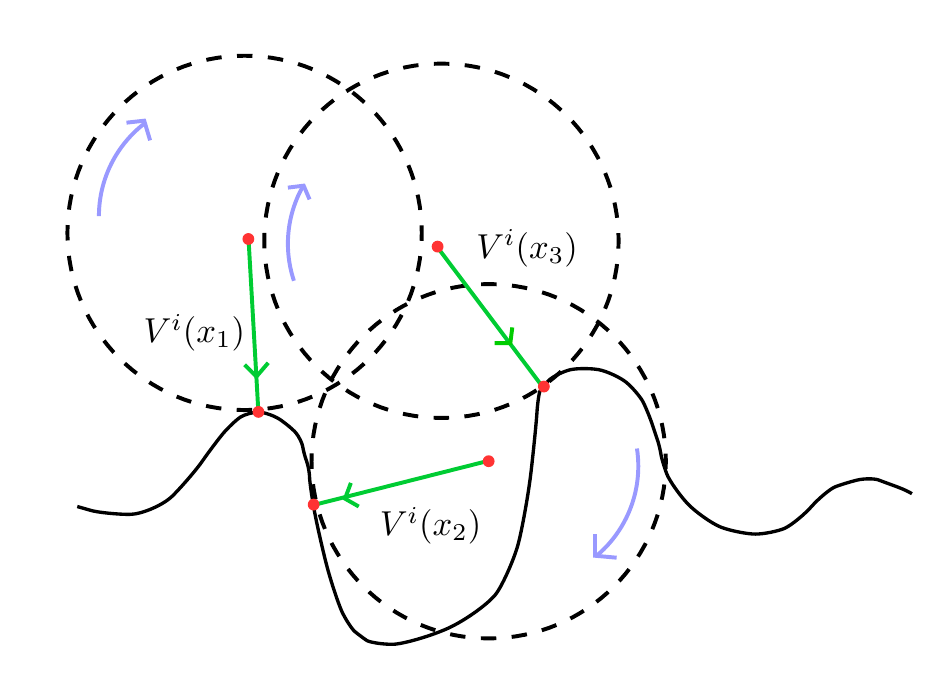} 
\caption {A figure demonstrating that the contact point between the ideal waywiser and manifold is the \emph{only} possible point of contact, and so the ideal waywiser at a given point is `invisible' to all other points. As such, the rolling of the ideal waywiser from $x_{1}\rightarrow x_{2} \rightarrow x_{3}$ is unhindered by features on the surface.}
\label{ghostwaywiser}
\end{figure}

Just as in the case of Riemannian geometry it is helpful for the sake of intuition to first invoke an embedding space. Consider then a two-dimensional surface/manifold $\m M\subset \mb R^3$ embedded in a three-dimensional Euclidean space $\mb R^3$ and some choice of coordinates $x^a$, $a=1,2$ that parametrized the surface. One may imagine `paths' $x^{a}(\lambda)$ on this surface. We define a waywiser as a device which one may attempt to `roll' along a path $x^{a}(\lambda)$ and in doing so yield information about the geometry of the manifold $\m M$. The amount of information that may be obtained will depend on the particular nature of the waywiser. The traditional waywiser depicted in Fig.  \ref{fig:waywisers} is suitable for measuring physical distances along paths $x^{a}(\lambda)$ on certain surfaces but is otherwise limited by the requirement that it may only roll along any path along the direction tangent to its wheel. A more versatile notion of a rolling object is the sphere $S^2$ of radius $\ell$. Clearly this is a more versatile object; for example, one may imagine a process of rolling such a sphere around a closed path ${\cal C}$. Upon returning the sphere may differ from its original, starting state by an arbitrary rotation, i.e. an $SO(3)$ transformation, which of course is a more general transformation than a traditional waywiser (i.e. a circle) is capable of whilst staying in contact with the surface. 
We shall be concerned with what we call an {\em idealized waywisers} with a symmetric space (with the same dimension as the manifold) as representing the `wheel'. These are `Platonic' creations of the mind with all irrelevant features, inherent in their material incarnations, have been removed. For example, no features in the embedded surface may obstruct or hinder the rolling of the idealized waywiser, see Fig. \ref{ghostwaywiser}.
\subsubsection{Mathematical representation of the contact point}
The first feature of an idealized waywiser is that it has a contact point between itself and the two-dimensional surface being probed. See the right figure in Fig. \ref{fig:waywisers} for an illustration. This is where the `rubber meets the road' as it were. Such a point of contact is itself a point on the sphere $S^2$. It is then convenient to represent the contact point by a {\em contact vector} $V^i$  satisfying $V^iV^j\delta_{ij}=\ell^2$ where $\delta_{ij}=diag(1,1,1)$. We visualize this vector as originating from the center of the sphere $S^2$ and ending at the point of contact where the sphere and manifold meet. The Latin index $i=1,2,3$ of the contact vector $V^i$ can conveniently be interpreted in many situations as referring to a three-dimensional Euclidean embedding space.

Picture now a sphere on top of {\em all the points} of the two-dimensional surface. For each coordinate $x^a$ we have a contact point   represented by $V^i(x)$. We note that the contact vector only depends on how the surface is embedded in the three-dimensional Euclidean space and is therefore the same regardless how the waywiser got there. In fact, using the three-dimensional Euclidean embedding space we see that the contact vector is always normal to the two-dimensional surface. Thus, it is then appropriate to introduce a {\em field} of contact vectors $V^i(x)$ for all the points on the surface. The contact vector $V^i(x)$ at some point $x^a$ we visualize as having its origin in the center of the sphere at the same point $x^a$.

To better convey the geometric picture we assume, at first, that $V^2(x)=\ell^2=const.$ The generalization to $V^2(x)\neq const.$ is then rather straightforward and, as we shall see, will be equivalent to introducing non-metricity $\nabla_c g_{ab}=\partial_c V^2 g_{ab}$.
\subsubsection{Rolling without slipping}
The second feature of the ideal waywiser is a prescription for how the sphere is rotated when rolled without slipping from one point to another along some path. Since it is a sphere $S^2$ the transformation group is $SO(3)$. Thus, the rolling of the waywiser corresponds to a succession of infinitesimal $SO(3)$ transformations. Mathematically these infinitesimal transformations can be specified by a $SO(3)$ connection $A_{a\ph ij}^{\ph ai}(x)$. The connection one-form $A_{a\ph ij}^{\ph ai}$, seen as a matrix $(A_a)^i_{\ph ij}$, is then a linear combination $(A_a)^i_{\ph ij}=A_a^{\alpha}(S_\alpha)^i_{\ph ij}$ of matrices $(S_\alpha)^i_{\ph ij}$ which satisfy the commutation relations $[S_\alpha,S_\beta]=2i\epsilon_{\alpha\beta}^{\phantom{\alpha\beta}\gamma}S_\gamma$ of the Lie-algebra $\mathfrak{so}(3)$. By feeding this connection an infinitesimal displacement $\delta x^a$ we obtain an infinitesimal rotation $\delta \Omega^i_j=\delta^i_j-\delta x^aA_{a\ph ij}^{\ph ai}$ \footnote{The minus sign in front of the connection is of course pure convention.}. This infinitesimal rotation characterizes mathematically the infinitesimal `response' of the idealized waywiser and how the point of contact consequently is altered, i.e. we have
\begin{align}\label{infroll}
V^i\ra\delta \Omega^i_jV^j=(\delta^i_j-\delta x^aA_{a\ph ij}^{\ph ai})V^j.
\end{align}

How can we check that the connection $A^i_{\ph ij}$ indeed corresponds to `rolling without slipping'? Well, without a metric $g_{ab}(x)$ already defined on the manifold $\m M$ this can in fact not be verified or checked. Instead, since no additional metric structure is present we are free to simply declare that the connection ${A_a}^i_{\ph ij}$ represents `rolling without slipping'. We shall see that a unique metric $g_{ab}(V,A)$ can be constructed from the waywiser variables $\{V^i,A^i_{\ph ij}\}$ so that the connection ${A_a}^i_{\ph ij}$ indeed corresponds to `rolling without slipping'.
\subsubsection{Physical content and the choice of representation}
On the group-theoretic side, we note that the representation of the contact point is nothing but the fundamental representation of $SO(3)$. A different way to represent the contact point would simply be by its spherical coordinate $(\theta(x),\varphi(x))$. However, the $SO(3)$ transformations then act non-linearly and inhomogeneously on the pair $(\theta,\varphi)$ and the clear link with the powerful mathematics of Lie group representations is lost. 

The particular choice of representation has implications for the physical content of the theory. It is therefore important to note that using the the fundamental representation introduces an additional degree of freedom, namely the norm  $V(x)=\sqrt{V^i(x)V^j(x)\delta_{ij}}$ which is invariant under $SO(3)$ transformations. In many presentations of Cartan geometry it is simply assumed that $|V(x)|^2=const.$ As we shall see, although this is perfectly fine from a mathematical point of view, this restriction on $V^i$ is from a physics perspective rather {\em ad hoc}. This becomes particularly clear when Cartan gravity is viewed alongside the electroweak theory. In the context of gravity we shall see that it is natural to let the contact vector to be a genuine dynamical degree of freedom subject to non-trivial equations of motion. As we shall see, although the scalar degree of freedom $V^2(x)$ may seem unwanted from a mathematical point of view it has the potential to play the role of a viable inflaton or quintessence candidate. In addition, we shall also see that a dynamical contact vector allows for exotic geometries with signature change. 
\subsection{Change in contact point and the metric tensor $g_{ab}(A,V)$}\label{contructmetricaffine}
In this section we are going to construct the metric tensor $g_{ab}(x)$ as a function of $V^i$ and ${A_a}^i_{\ph ij}(x)$. We assume in this section that $V^2=const.$ but will relax that condition in the next section.

\begin{figure}[H]
\centering
\includegraphics[width=14cm]{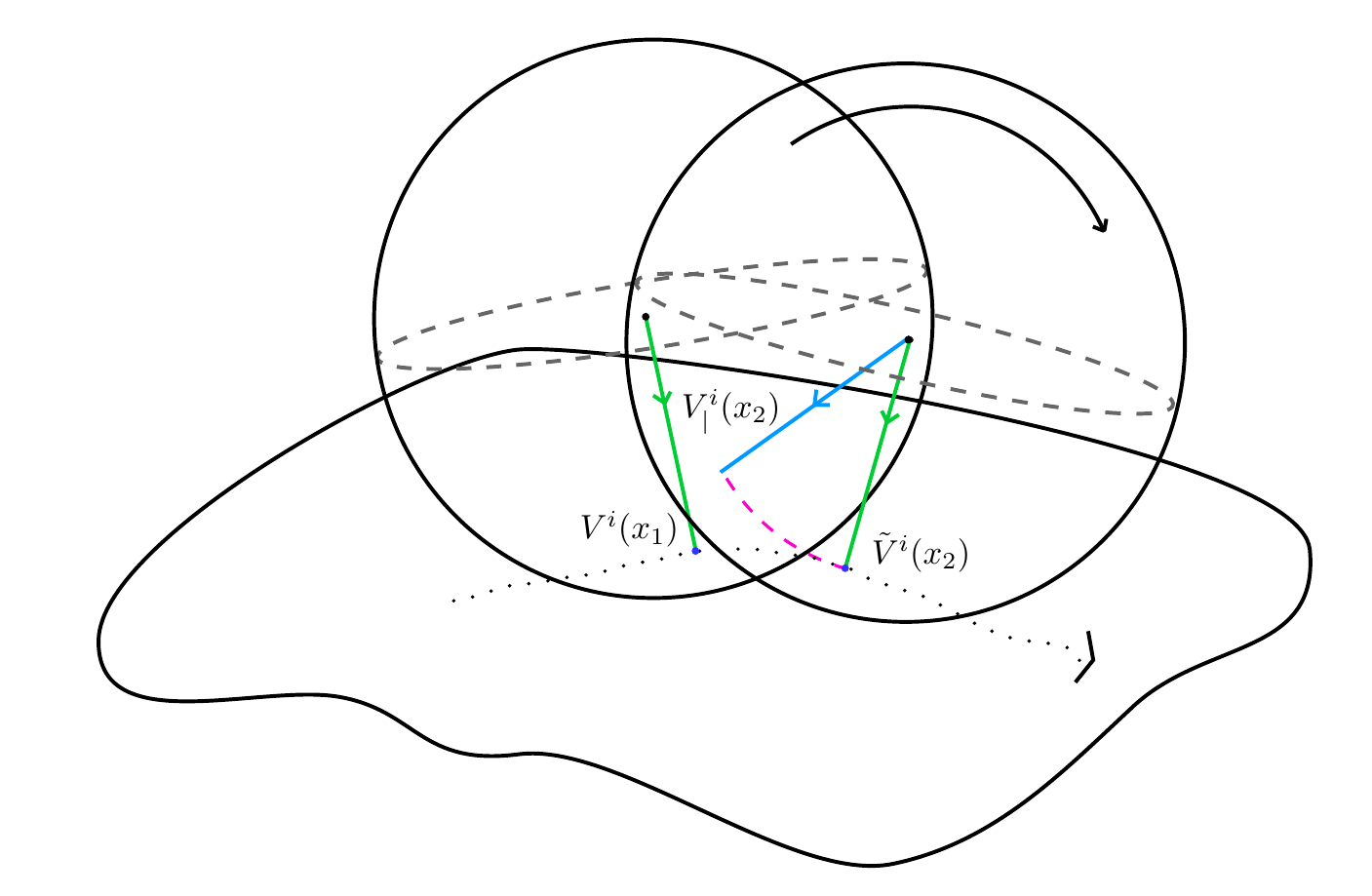} 
\caption{The figure illustrates how the `wheel' of the ideal waywiser is rotated when rolled on the surface from point $x_1$ to $x_2$.}
\label{idway}
\end{figure}

Let us now determine the distance between two neighbouring points $x_1^a$ and $x_2^a$ on the surface. In our mind's eye we now picture an idealized waywiser at $x_1$ (see Fig. \ref{idway} for a visualisation of this in an embedding picture where the manifold is regarded as a sub-manifold of $\mathbb{R}^{3}$). Before that ball is rolled we imagine a stick of length $\ell$ attached to the ball, with one end in the center of the ball and the other at the contact point $V^i(x_1)$. We denote this `stick-vector' $V^i_|$ which per definition coincides with the contact vector at $x_1$, i.e. $V^i_|(x_1)=V^i(x_1)$. Next we roll the ball in the direction $\delta x^a=x_2^a-x_1^a$ and put it to rest at $x_2^a$. Rolling the `stick-vector' is mathematically understood as a succession of infinitesimal $SO(3)$ transformations $\delta \Omega^i_j=\delta^i_j-\delta x^aA_{a\ph ij}^{\ph ai}$ acting on $V^i_|$. Thus, after an infinitesimal roll, we have according to equation \eqref{infroll}

\begin{eqnarray}
V^i_|(x_2)=\delta \Omega^i_jV^j_|(x_1)=(\delta^i_j-\delta x^aA_{a\ph ij}^{\ph ai})V^j_|(x_1)=V^i(x_1)-\delta x^aA_{a\ph ij}^{\ph ai}V^j(x_1)
\end{eqnarray}
where $A_{a\ph ij}^{\ph ai}$ is the $SO(3)$ connection dictating how much the ball has rotated when rolled without slipping.
Next, we can compare the rolled `stick-vector' $V^i_|(x_2)$ with the contact vector $V^i(x_2)$ at $x_2$ and compute the difference $\delta V^i\equiv V^i(x_2)-V^i_|(x_2)$:
\begin{eqnarray}
\delta V^i&\equiv& V^i(x_2)-V^i_|(x_2)=V^i(x_2)-(V^i(x_1)-\delta x^a A_{a\ph ij}^{\ph ai}V^j(x_1))=\delta x^{a}\partial_aV^i+\delta x^a A_{a\ph ij}^{\ph ai}V^j(x_1)\nonumber\\
&\equiv& \delta x^aD_a V^i
\end{eqnarray}
where we have introduced the gauge covariant derivative $D_a V^i\equiv\partial_aV^i+A_{a\ph ij}^{\ph ai}V^j$. The difference $\delta V^i$ represents the change in contact point. We note that because the contact vector satisfies $V^2=\ell^2$, we have $\delta_{ij}V^iDV^j=0$ and the object $\delta x^aD_aV^i$ therefore has no normal component and belongs to the tangent space of the surface at $x_1$. We now identify the distance $\delta s$ between the two points $x_1$ and $x_2$ as the Euclidean norm of the difference $\delta V^i$, or equivalently
\begin{eqnarray}
\delta s^2=\delta_{ij}\delta V^i \delta V^j=\delta x^a\delta x^b\delta_{ij}D_aV^i D_bV^j
\end{eqnarray}
The metric tensor $g_{ab}$, encoding all information about distances of the surface, can then be defined as
\begin{eqnarray}\label{metricdef}
g_{ab}=\delta_{ij}D_aV^i D_bV^j.
\end{eqnarray}
We always have gauge-freedom to select a gauge where $V^{i}= \ell \delta^{i}_{3}$. In this gauge $DV^{(3)}=0$ and $DV^{(1)}$ and $DV^{(2)}$ can be identified with the co-zweibein fields.

We can now clarify what it would mean to roll {\em with slipping}. Suppose the manifold was already equipped with a metric tensor $h_{ab}(x)$ in addition to $V^i$ and ${A_a}^i_{\ph ij}$. If we then find that $h_{ab}(x)\neq g_{ab}(V(x),A(x))$ then that indicates that the balls was slipping. Put differently, the condition of `rolling without slipping' translates mathematically into the requirement that the metric of the manifold is given by $g_{ab}(V,A)$ and not some other metric.

\subsection{Waywisers with variable size: $V^2=V^2(x)$}
Inspired by the structural similarities between Cartan geometry and a symmetry-broken Yang-Mills theory we shall later in this paper treat $V^i$ as a genuine dynamical field with no restrictions imposed apart from the equations of motion. We can then no longer impose  that $V^2(x)$ should be constant as a function on the manifold $\m M$. Thus we shall here generalize the derivation of the metric $g_{ab}$ and co-zweibein $e^i$ to the general case in which $V^2(x)$ is a non-trivial function of $x^a$. this only causes a minor difference in the derivation with the underlying geometric picture intact.

In fact, when we roll the contact vector from $x_1^a$ to $x_2^a$ we will find there a sphere with a different size. This means that the quantity $V^i(x_2)-V_|^i(x_2)$ will not to first order be a measure of the change in contact point. Nevertheless, $V^i(x_2)$ still points in the direction of the contact point. Thus, before subtracting we need to rescale $V^i(x_2)\rightarrow  \tilde V^i(x_2)$  as to have the same size as the ball we rolled there. The rescaled $V^i(x_2)$ to first order in $\delta x^a$ becomes
\begin{align}
\tilde V^i(x_2)=\sqrt{\frac{V_|^2(x_2)}{V^2(x_2)}}V^i(x_2)=V^i(x_1)-\frac{1}{2}\delta x^a\partial_a\log V^2 V^i(x_1).
\end{align}
The change of contact point is now calculated as
\begin{align}
\delta V^i=\tilde V^i-V_|^i=\delta x^aD_aV^i-\frac{1}{2}\delta x^a\partial_a\log V^2 V^i(x_1)
\end{align}
or more succinctly:
\begin{align}
\delta V^i=\delta x^a P^i_{\ph ij}D_aV^j.
\end{align}
where we have introduced the projector $P^{i}_{\ph{i}j}$:
\begin{equation}
P^i_{\ph ij}\equiv \delta^i_j-\frac{V^iV_j}{V^2}.
\end{equation}
The (squared) distance traveled becomes
\begin{align}
\delta V^2=\delta V^i\delta V^j\delta_{ij}=\delta x^a\delta x^b P^i_{\ph ik}D_aV^kP^j_{\ph jl}D_aV^l\delta_{ij}=\delta x^a\delta x^b P_{ij}D_aV^iD_bV^j.
\end{align}
from which we identify 
\begin{equation}
\boxed{g_{ab}\equiv P_{ij}D_aV^iD_bV^j}
\end{equation}
as the metric tensor and 

\begin{equation}
\boxed{e^{i}_{a} \equiv  P^{i}_{\ph{i}j}D_{a}V^{j}}
\end{equation}
as co-diad.

Consider now a sequence of neighbouring points $x_1^a,\dots,x_n^a$ along some trajectory $x^a(\lambda)$. The distance between $x^a_1$ and $x^a_{2}$ is then given by $\delta \ell^2(x_1)=g_{ab}(x_2^a-x_1^a)(x_2^a-x_1^a)$ which is determined by monitoring how much the ball at $x_n^a$ as rotated when rolled to $x_{n+1}^a$. To determine the distance between $x^a_2$ and $x^a_{3}$ we imagine discarding the ball we picked up at $x_1^a$ and instead make use of the one at $x_2^a$ which may have a different size. The ball at $x_2^a$ is then rolled to $x_3^a$ yielding the distance $\delta \ell^2(x_2)=g_{ab}(x_3^a-x_2^a)(x_3^a-x_2^a)$. And so on. Therefore, over a finite-length path $x^{a}(\lambda)$ the aggregate of infinitesimal rolls of the ball `against the surface' of the manifold yields a notion of the physical distance $l$ between two points $a$ and $b$ along the path:
\begin{eqnarray}
l =  \int_{a}^{b}\sqrt{ P_{ij}D_aV^iD_bV^j\frac{\delta x^{a}}{d\lambda}\frac{\delta x^{b}}{d\lambda}}d\lambda.
\end{eqnarray}
From this point on no restrictions on $V^i$ will be imposed and we shall allow for varying $V^2(x)$.
\subsection{Parallel transport and the affine connection: $\Gamma^a_{bc}(V,A)$}\label{contructaffine}
We have now understood how the metric tensor can be recovered from the waywiser variables $\{V^i,A^{ij}\}$ and that the metric directly corresponds to the change of contact point when the waywiser is rolled. However, the metric tensor cannot tell us how to parallel transport tangent vectors, $u^a$ say, along the surface. How vectors are parallel transported is something which is encoded in the affine connection $\Gamma^c_{ab}$; as follows, we illustrate how an affine connection is naturally recovered in Cartan geometry.
\subsubsection{The soldering of the tangent spaces and the soldering map}
In Cartan geometry we deal with two distinct manifolds: one is the manifold $\m M$ whose geometry we wish to characterize geometrically and the other one the symmetric space, i.e. the sphere $S^2$, we roll on top of the manifold $\m M$. Consider now some point $x\in \m M$ and the associated tangent space $T_x(\m M)$. On top of that point sits the sphere $S^2$ with an associated tangent space $T_V(S^2)$ at the contact point represented by $V^i$. The basic idea behind Cartan geometry is that these tangent spaces should be identified as one and the same. This is the {\em soldering} of the tangent spaces. 

Mathematically this means we need to introduce mapping $\mathfrak m:T_x(\m M)\rightarrow T_V(S^2)$ that associates a vector $u=u^a$ of $T_x(\m M)$ to a vector $u^i$ of $T_V(S^2)$. The latter vector must satisfy $u^iV_i=0$ in order to be orthogonal to the contact vector $V^i$ and so be a tangent vector. Regarding dimensionality of quantities, since we would like $u^{a}$ and $u^{i}$ to represent the same entity we must require them to have the same dimensions of length implying that the map $\mathfrak m$ itself is dimensionless.\footnote{We can attach different dimensions of length to $u^i$ and $u^a$ leading to a different soldering map which would have to compensate for that difference in dimension. However, this only complicates the derivation and is in fact not natural if we want to think of the objects $u^a$ and $u^i$ as representing the same vector.} As discussed in Appendix \ref{unitsnconventions}, $V^{i}$ has dimensions of length and the connection $A_{a}^{\ph{a}ij}$ as well as the manifold coordinates $x^a$ are dimensionless. If we take the tangent vectors $u^i$ and $u^a$ to be dimensionless we see that the map  $e^{i}_{a}=P^{i}_{\ph{i}j}D_{a}V^{j}$ has dimensions of length and is not appropriate as a soldering map since the dimensions in the expression $u^i=e^i_au^a$ does not add up. The appropriate object is instead the quantity $\hat{e}^{i}_{a}\equiv P^{i}_{\ph{i}j}D_{a}\hat{V}^{j}$  where $\hat{V}^{i} \equiv V^{i}/\sqrt{V^{2}}$. This object contains no information about the length of $V^{i}$ and is dimensionless. The map is now easily guessed: the map $\mathfrak m:T_x\rightarrow T_V(S^2)$ is simply given by $\hat{e}^{i}_{a}$ so that 
\begin{equation}
\boxed{u^i=\hat{e}^{i}_{a}u^a}.
\end{equation}
We see that $u^i$ indeed satisfies $u^iV_i=0$. This map is to be regarded as a {\em postulate} of Cartan geometry that needs no further justification.
\begin{figure}
\centering
\includegraphics[width=9cm]{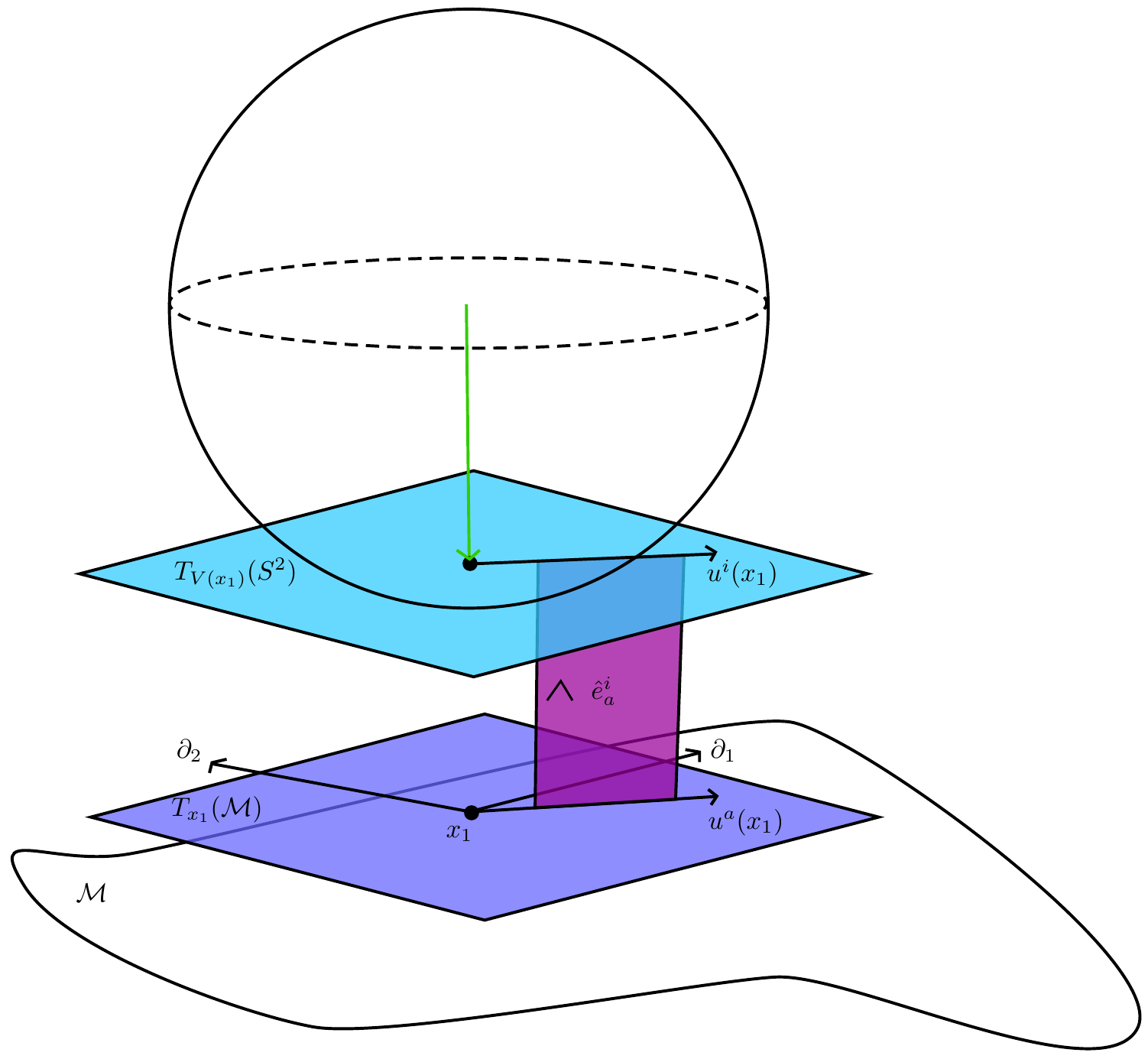} 
\caption{Illustration of the soldering-map $\hat{e}^{i}_{a}(x_{1})$ which maps vectors
$u^{a}(x_{1})$ living in the tangent-space $T_{x_{1}}({\cal M})$ to vectors $u^{i}(x_{1})$ living in the tangent-space
$T_{V(x_{1})}(S^{2})$.}
\end{figure} 

\subsubsection{Cartan-geometric parallel transport}
We can now ask how a tangent vector $u^a\in T_x(\m M)$ is parallel transported along a path on the manifold $\m M$. Normally, this is dictated by the affine connection  $\Gamma^c_{ab}$.  However, in Cartan geometry we only given the variables $V^i(x)$ and $A^{ij}(x)$. It thus behoves us to work out parallel transport from a Cartan-geometric perspective in terms of the sphere $S^2$ and its contact point. To do that we assume we have a tangent vector $u^i(x_1)\in T_V(S^2)$ at some point $x_1^a$. We wish to parallel transport this vector to the neighbouring point $x_2^a$ with $\delta x^a=x_2^a-x_1^a$ considered as an infinitesimal displacement vector. Since $u^i(x_1)$ is a tangent vector it satisfies $u^i(x_1)V_i(x_1)=0$.

\begin{figure}[H]
\centering
\includegraphics[width=15cm]{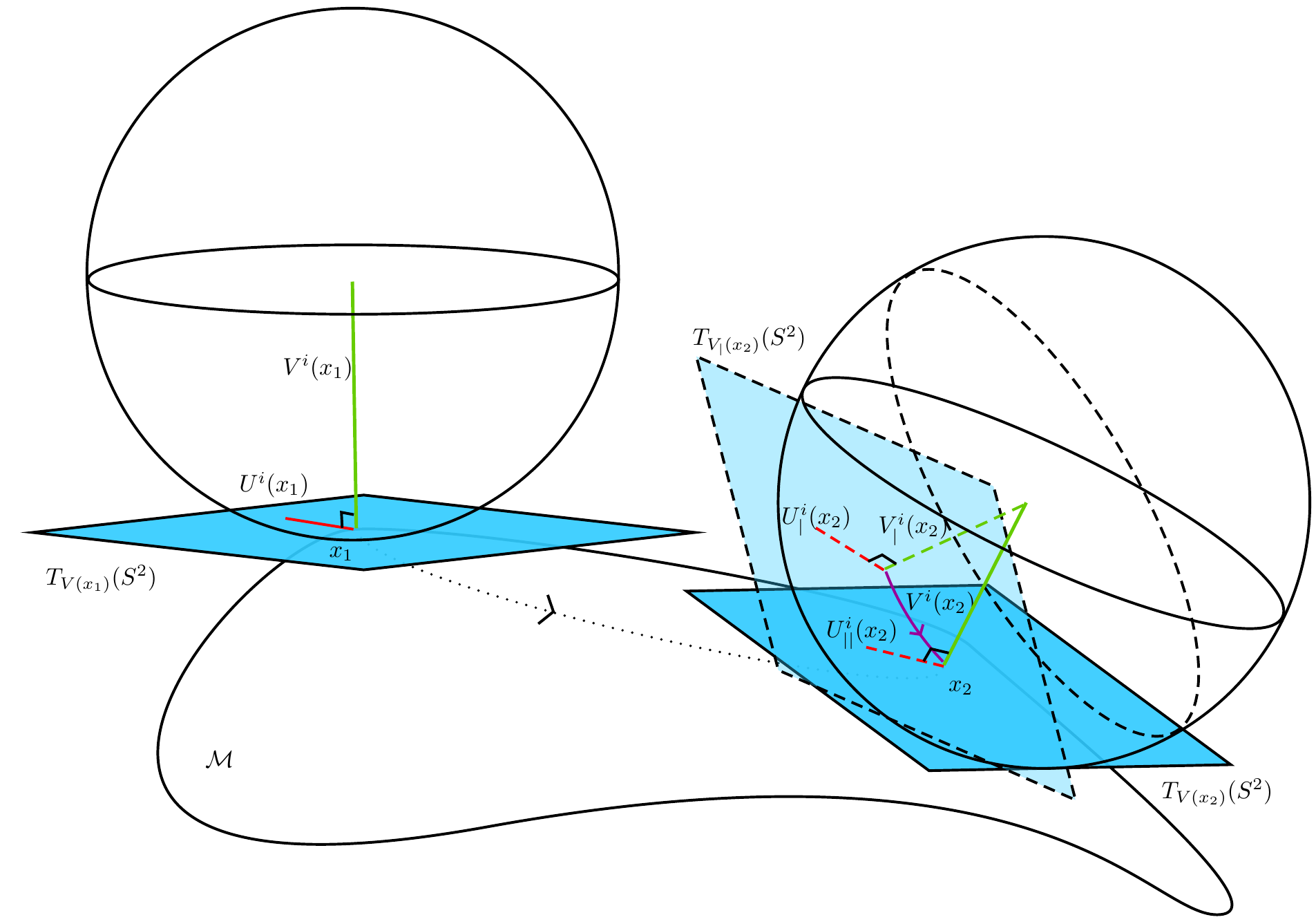} 
\caption{Illustration of how rolling the idealised waywiser with $A_{a\,j}^{\,\,i}$ from $x_{1}$ to $x_{2}$ generically `lifts' the tangent space $T_{V_{|}(x_{2}}(S^{2})$ away from $T_{V(x_{2})}(S^{2})$ and in doing so breaks the condition $U_{|}^{i}V_{i}=0$}
\label{roller}
\end{figure}

First we consider what happens if we roll the tangent vector to point $x_2$. This yields
\begin{align}
u^i_|(x_2)\equiv u^i(x_1)-\delta x^a {A_a}^i_{\ph ij}(x_1)u^j(x_1).
\end{align}
However, the object $u^i_|(x_2)$ is no longer a tangent vector since to first order in $\delta x^a$ we have
\begin{align}\label{normpart}
u^i_|(x_2)V_i(x_2)=(u^i(x_1)-\delta x^a {A_a}^i_{\ph ij}(x_1)u^j(x_1))(V_i(x_1)+\delta x^b \partial_b V_i(x_1))=\delta x^a D_aV^i(x_1)u_i(x_1)\neq0.
\end{align}
Thus, we see that the tangent vector property $u^iV_i=0$ is not preserved under an infinitesimal roll. In fact, the vector $u^i_|$ belongs, not to $T_{V(x_2)}(S^2)$ but to $T_{V_|(x_2)}(S^2)$, i.e. the tangent space at the point $V^i_|$ of the sphere (see Fig. \ref{roller}). In order to obtain a tangent vector at $x_2^a$ we have to slide the vector $u^i_|(x_2)$ down to the tangent space $T_{V(x_2)}(S^2)$. To first order in $\delta x^a$ this amounts to simply removing the part which is not normal to $V^i(x_2)$ calculated in \eqref{normpart}, i.e. we define
\begin{align}\label{CGPtrans}
u^i_\parallel (x_2)\equiv u^i_|(x_2)-\frac{V^i}{V^2}u^j_|(x_2)V_j(x_2)=u^i(x_1)-\delta x^a \left({A_a}^i_{\ph ij}(x_1)+\frac{V^i}{V^2}D_{a}V_{j}(x_1)\right)u^j(x_1)
\end{align}
A parallel transported vector should satisfy $u^i(x_2)=u^i_\parallel (x_2)$ and this leads us to the following definition of Cartan-geometric parallel transport of a vector $u^i(\lambda)$ along a path $x^a(\lambda)$:
\begin{equation}
\boxed{\frac{D_\parallel u^i}{D_\parallel\lambda}\equiv\frac{du^i}{d\lambda}+\frac{\delta x^a}{d\lambda}\omega_{a\ph{i}j}^{\ph{a}i}u^j=0}
\end{equation}
where (see also \cite{Shirafuji:1988ia})
\begin{equation}
\label{omegaeq}
\boxed{\omega_{a}^{\ph{a}ij} \equiv  A_{a}^{\ph{a}ij}+\frac{2}{V^{2}}V^{[i}D_{a}V^{j]}}.
\end{equation}
We are now ready to read off the affine structure $\Gamma^c_{ab}$ from the more basic Cartan-geometric variables $V^i$ and ${A_a}^i_{\ph ij}$.
This connection preserves the condition $u^{i}_{|| }V_{i}=0$ and preserves the norm $\delta_{ij}u^{i}_{||}u^{i}_{||}$. Thus it acts as an $SO(2)$ connection.

\subsubsection{Identifying the affine connection}\label{identconnection}
If we want to parallel transport the tangent vector $u^a(x_1)\in T_{x_1}(\m M)$ from $x_1$ to $x_2$ the standard expression would simply be
\begin{align}
u^c_\parallel(x_2)=u^c(x_1)-\delta x^a\Gamma^c_{ab}(x_1)u^b(x_1).
\end{align}
Given the soldering map $\hat e_a^i:T_x\rightarrow T_V(S^2)$ we now have a new way of obtaining $u^i_\parallel(x_2)$:
\begin{align}\label{soldpartrans}
u^i_\parallel(x_2)&=\hat{e}^{i}_{c} u^c_\parallel(x_2).
\end{align}
By requiring the two expressions \eqref{CGPtrans} and \eqref{soldpartrans} for $u^i_\parallel(x_2)$ to coincide we obtain the equation
\begin{align}
u^i_\parallel(x_2)&\equiv u^i(x_1)-\delta x^a \omega_{a\ph{i}j}^{\ph{a}i}u^j(x_1)\nn\\
&=\hat{e}^{i}_{c}(x_{2})\left( u^c(x_1)-\delta x^a\Gamma^c_{ab}(x_1)u^b(x_1)\right).
\end{align}
which then imposes the desired relationship between the Cartan-geometric variables $\{V^i,A^{ij}\}$ and the affine connection $\Gamma^c_{ab}$.
To first order in $\delta x^a$ we have:
\begin{align}
&u^i(x_1)-\delta x^a \omega_{a\ph{i}j}^{\ph{a}i} \hat{e}^{j}_{b}(x_{1}) u^b(x_1)\nn\\
&=\hat{e}^{i}_{c}(x_{2})u^c(x_1)-\hat{e}^{i}_{c}(x_1)\delta x^a\Gamma^c_{ab}(x_1)u^b(x_1)\nn\\
&=u^i(x_1)+\delta x^a\partial_a\left(\hat{e}^{i}_{c}(x_1)\right)u^c(x_1)-\hat{e}^{i}_{c}(x_1)\delta x^a\Gamma^c_{ab}(x_1)u^b(x_1)
\end{align}
This expression must hold for all, $x_1^a$, $u^b(x_1)$, and $\delta x^a$ which leads to the following identity
\begin{equation}
\partial_{a}\hat{e}^{i}_{b}-\hat{e}^{i}_{c}\Gamma^{c}_{ab}+\omega_{a\ph{i}j}^{\ph{a}i}\hat{e}^{j}_{b}= 0 
\end{equation}
Using the definition $e^{i}_{a}=  \sqrt{V^{2}} \hat{e}^{i}_{a}$ we may write that as:
\begin{equation}
\partial_{a}e^{i}_{b}-\frac{1}{2}\partial_{a}\log V^{2}e^{i}_{b} - e^{i}_{c}\Gamma^{c}_{ab}+\omega_{a\ph{i}j}^{\ph{a}i}e^{j}_{b}= 0 \label{soldere}
\end{equation}
If, furthermore, $e^{i}_{a}$ is `invertible' i.e. there exists a field $e^{a}_{i}$ such that $e^{i}_{a}e^{b}_{i}=\delta^{a}_{b}$ (and $e^{i}_{a}e^{a}_{j} = P^{i}_{\ph{i}j}$) then we may 
act on (\ref{soldere}) with $e_{i}^{d}$ to yield:
\begin{equation}
\label{gamma}
\boxed{\Gamma^{d}_{ab} = e_{i}^{d}\left(\partial_{a}e^{i}_{b} +\omega_{a\ph{i}j}^{\ph{a}i}e^{j}_{b}-\frac{1}{2}\partial_{a}\log V^{2} e^{i}_{b}\right)}
\end{equation}
This immediately implies that for $\Gamma^{c}_{ab}$ and $g_{ab} \equiv \delta_{ij}e^{i}_{a}e^{j}_{b}$ we have that:
\begin{eqnarray}
\label{metcomp}
\boxed{\nabla_{a}g_{bc} \equiv  \partial_{a}g_{bc} - \Gamma^{d}_{ac}g_{bd}-\Gamma^{d}_{ab}g_{dc}= \partial_{a}\log V^{2} g_{bc}}
\end{eqnarray}
where we have used the fact that $\omega_{aij}e^{j}_{(c}e^{i}_{b)}=0$. Thus, the covariant derivative $\nabla_{a}$ associated with $\Gamma^{c}_{ab}$ is not metric-compatible.

It is quite pleasing to see that both metric and affine connection, which play two distinct mathematical roles in Riemannian geometry, can be constructed from the more primary variables $\{V^i,A^{ij}\}$ which themselves admit a crisp geometric interpretation in terms of idealized waywisers. The force of (Riemannian) habit may make us uncomfortable with $V^i$ and $A^{ij}$ as the fundamental descriptors of geometry and we may also have an itch to translate back to the metric formulation with $g_{ab}$ and $\Gamma^c_{ab}$ to place ourselves on familiar mental ground. However, it should be clear from this point on that Cartan geometry provides an alternative description of geometry which is not only mathematically elegant but also rests on a strikingly simply underlying geometric picture in terms of idealized waywisers.

%
\subsection{Decomposition of the $SO(3)$ curvature}
Recalling equation (\ref{omegaeq}) we can readily work out the curvature two-form associated with $A^{ij}$. In the notation of differential forms, the curvature two-form $F^{ij}=\frac{1}{2}{F_{ab}}^{ij}dx^adx^b$ is defined as:
\begin{eqnarray}
F^{ij} &\equiv & dA^{ij}+A^{i}_{\ph{i}k}A^{kj}
\end{eqnarray}
Its spatial components are thus given explicitly by:
\begin{eqnarray}
{F_{ab}}^{ij}&=&  2 \left(\partial_{[a}A^{\ph{b]}ij}_{b]} + A_{[a\ph{i} k}^{\ph{[a}i}A_{b]}^{\ph{b]}kj} \right)
\end{eqnarray}
Using equation (\ref{omegaeq}) we have that:
\begin{eqnarray}
F^{ij} &=& R^{ij}-\frac{1}{V^{2}}e^{i}e^{j}  +\frac{2}{V^{2}}\left(T^{[i}V^{j]}+d\log V^{2} V^{[i}e^{j]}\right)-\frac{2}{V^{2}}D^{(\omega)}V^{[i}e^{j]}\\
 &=& R^{ij}-\frac{1}{V^{2}}e^{i}e^{j}  +\frac{2}{V^{2}}\left(T^{[i}V^{j]}-\frac{1}{2}d\log V^{2}e^{[i}V^{j]}\right)
\end{eqnarray}
where $R^{ij} \equiv d\omega^{ij}+\omega^{i}_{\ph{i}k}\omega^{kj}$, $T^{i}\equiv de^{i}+\omega^{i}_{\ph{i}j}e^{j}$ and we have used the fact that $D^{(\omega)}V^{i}= dV^{i} + \omega^{i}_{\ph{i}j}V^{j}  =  \frac{1}{2}d \mathrm{\log} V^{2}  V^{i}$. This decomposition of the curvature for $V^{2}(x)$ may also be found in \cite{Westman:2013mf,Jennen:2014mba}.
\subsection{Geometric interpretation of curvature}
Consider at some point $x^{a}$, an infinitesimal quadrangle with vertices at points $x_{0}^{a}=x^{a}$, $x_{1}^{a}=x^{a}+\delta x^{a}$, $x_{2}^{a}=x^{a}+\delta x^{a}+\delta y^{a}$, $x_{3}^{a}=x^{a}+\delta y^{a}$. First we consider transporting the contact vector $V^{i}(x^{a})$ around the path $x^{a}(\lambda):  x_{0}^{a}\rightarrow x_{1}^{a} \rightarrow x_{2}^{a}\rightarrow x_{3}^{a} \rightarrow (x_{4}^{a}=x_{0}^{a})$ using the $SO(3)$ transport equation i.e. 
\begin{eqnarray}
\frac{dx^{a}}{d\lambda}\left(\partial_{a}V_{|}^{i}+A^{\ph{a}i}_{a\ph{i}j} V_{|}^{j}\right)=0
\end{eqnarray}
where the initial condition is of course $V_{|}^{i}(x_{0}^{a})=V^{i}(x_{0}^{a})$.
Upon reaching $x_{4}^{a}=x_{0}^{a}$ again, the transported vector $V_{|}^{i}(p_{x}^{a})$ may generally differ from $V_{|}^{i}(x_{0}^{a})$. Retaining all terms up to first order in $\delta x^{a}$ and $\delta y^{a}$ yields the following result for 
$\Delta V^{a}\equiv V_{|}^{i}(x_{4}^{a}=x^{a})-V_{|}^{i}(x_{0}^{a})$: \footnote{We note that torsion can be simulated within condensed matter systems. The link \cite{Vaid:2013woa} between the theory of Cartan gravity based on the group $SO(5)$ and a recent gauge theory of  high $T_{C}$ superconductivity and anti-ferromagnetism based on the same group is suggestive of a possible broader mapping between condensed matter and gravitational phenomenology .}
\begin{eqnarray}
\Delta V^{i} &=&  {F_{ab}}^{ij}V^{j} \delta y^{a}\delta x^{b}\\
&=&\left({T_{ab}}^i-\partial_{[a}\mathrm{log}V^{2}e_{b]}^{i}\right)\delta y^{a}\delta x^{b}
\end{eqnarray}
where all quantities are evaluated at $x_{0}^{a}$. We can use the identity provided by equation (\ref{gamma}) to show the following:
\begin{equation}
\boxed{\Delta V^{i} =  2 e^{i}_{d}\Gamma^{d}_{[ab]}\delta y^{a}\delta x^{b}}
\end{equation}
Thus we see that $\Gamma^{a}_{[bc]}$, the antisymmetric part of the affine connection, measures the change that the contact vector experiences after being transported along an infinitesimal closed path using the connection $A^{i}_{\ph{i}j}$. Note that in Einstein-Cartan theory, $\Gamma^{a}_{[bc]}$ and $\frac{1}{2}e^{a}_{I}T_{bc}^{\ph{bc}I}$ are interchangeable; this is no longer the case for Cartan gravity due to the presence of $dV^{2}$. This may also be taken as an indication that ${\m  T_{ab}}^i={T_{ab}}^i-\partial_{[a}\mathrm{log}V^{2}e_{b]}^{i}$ is the more geometrically natural definition of torsion. 

Similarly,  we may consider parallel transporting a tangent vector around the same route. Thus we start with a vector $U_{||}^{i}$ defined to be a tangent vector living in $T_{V(x_{0}^{a})}(S^{2})$ and then transport it using the connection $\omega^{ij}$ (which, recall, preserves the condition $u_{||}^{i}V_{i}=0$ at all points along the path) i.e.
\begin{eqnarray}
\frac{dx^{a}}{d\lambda}\left(\partial_{a}u_{||}^{i}+\omega^{\ph{a}i}_{a\ph{i}j}u_{||}^{j}\right)=0.
\end{eqnarray}
The result for $\Delta u^{i}(x^{a})=u^{i}_{||}(x_{4}^{a})-u_{||}^{i}(x_{0}^{a})$ is:
\begin{equation}
\label{deltau}
\boxed{\Delta u^{i}={R_{ab}}^{i}_{\ph{i}j}u^{j}\delta y^{a}\delta x^{b}}
\end{equation}
Here we have an $SO(3)$-covariant expression of the familiar result that the change experienced by $SO(2)$ vectors when transported around an infinitesimal closed path using an $SO(2)$ connection ($\omega^{i}_{\ph{i}j}$)  is related to the $SO(2)$ curvature as given by (\ref{deltau}).
\begin{figure}[H]
\centering
\includegraphics[width=11cm]{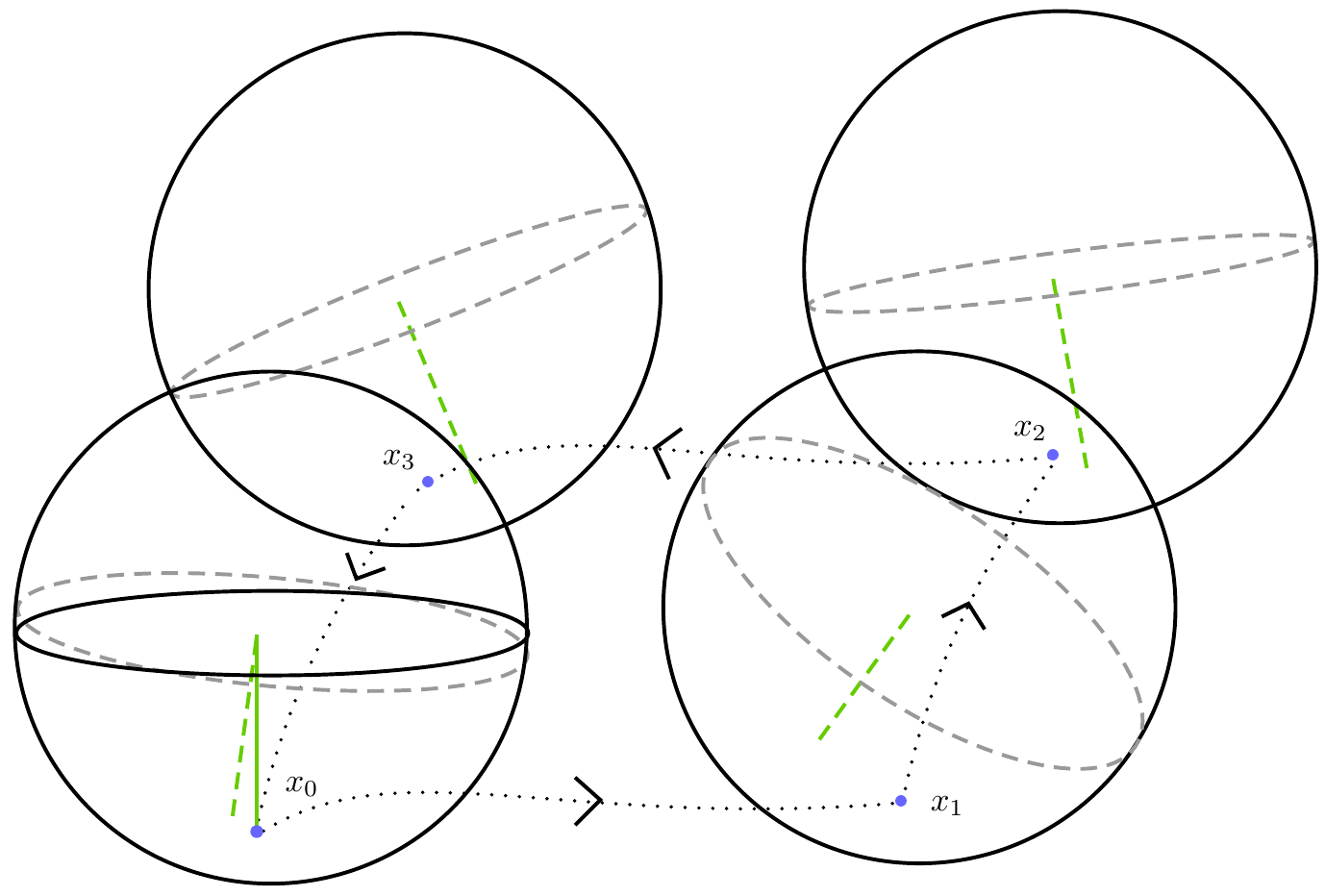} 
\caption{The effect of torsion as a failure of the $SO(3)$-transported $V_{|}^{i}$ to agree at the beginning and end of an infinitesimal path $x_{0}^{a}\rightarrow x_{1}^{a} \rightarrow x_{2}^{a} \rightarrow x_{3}^{a} \rightarrow x_{4}^{a}=x_{0}^{a}$. Thus, torsion represents the change in contact point when we roll a sphere around an infinitesimal closed loop. The solid-green line is $V_{|}^{i}(x_{0})$; at $x_{0}$ the dashed-line is $V_{|}^{i}(x_{4})$ and at the remaining locations the dashed-line represents $V_{|}^{i}$ at that point on the path.}
\end{figure} 

\subsection{Geometry of embedded surfaces and Cartan geometry}
Our development of Cartan geometry was guided by the picture of a sphere rolling on a manifold embedded in the real coordinate space $\mathbb R^3$. In this section we are going to make more precise the relationship with Cartan geometry and the geometry of embedded surfaces.
\subsubsection{Geometry of embedded surfaces}
First we start with a brief recapitulation of standard results concerning the geometry of submanifolds of $\mathbb R^{3}$. Let $X^i$ be the Cartesian coordinates of $\mathbb R^3$ and $X^i(x)$ be the parametrization of a 2D submanifold $\m M$ with $x^a$ its two coordinates. Then 

\begin{eqnarray}
\frac{\partial}{\partial x^{a}}=   \frac{\partial X^{i}}{\partial x^{a}}\frac{\partial}{\partial X^{i}}
\end{eqnarray}
represent the set of tangent vectors at the point $x^a$ on $\m M$ i.e. they comprise a set of two separate vectors, each with $\mathbb R^{3}$ components $\partial X^{i}/\partial x^{a}$. The normal can be constructed as follows: first we construct the density $U^i$:
\begin{align}
U^i \equiv\epsilon^i_{\ph ijk}\partial_aX^j\partial_bX^k\varepsilon^{ab}
\end{align}
which is merely orthogonal to the vectors $\partial/\partial x^{a}$ but not of unit length. The field of \emph{unit} normals $N^i(x)$ of $\m M$ is then defined by
\begin{align}
N^i\equiv \frac{U^i}{\sqrt{U^2}}
\end{align}
where $U^{2}\equiv \delta_{ij} U^{i}U^{j}$. The vector field $N^{i}$ can then be seen to satisfy $N_i\partial_{a}X^{i}=0$ and $N^2=1$. The induced metric $g_{ab}$ on the submanifold ${\cal M}$ is given by $\delta(\partial/\partial x^{a},\partial/\partial x^{b})=\delta_{ij}dX^{i}\otimes dX^{j}(\partial/\partial x^{a},\partial/\partial x^{b})$ i.e.
\begin{align}
g_{ab}=\partial_a X^i\partial_bX^j\delta_{ij}
\end{align}
where $\delta= \delta_{ij}dX^{i}\otimes dX^{j}$ is the metric on $\mathbb{R}^{3}$.

Next consider the parallel transport of vector field $u^i(x)=u^a(x) \partial_a X^i$ on $\Sigma$ required to satisfy $N_i u^i=0$, i.e. $u^i$ is everywhere a tangent vector of $\Sigma$. Since parallel transport in the Euclidean flat ambient space is trivial we may use this to induce a notion of parallel transport on $\Sigma$. 

Taking the difference of two neighboring points
\begin{align}
u^i(x+\delta x)-u^i(x)=\delta x^a\partial_a u^i
\end{align}
does not yield a vector that belongs to the tangent space of $\Sigma$ as we have $N_i \partial_a u^i\neq0$ in general. When we parallel transport a vector we must of course allow for the tangent vector to be tilted appropriately as to stay a tangent vector. Specifically, a vector is parallel transported only if the {\em only} change of it occurs in the normal direction.

To see this clearly we expand  the derivative of the tangent basis vector $\partial_aX^i$ as follows
\begin{align}\label{expand}
\partial_{ab}X^i=\Gamma_{ab}^{c}\partial_cX^i+K_{ab}N^i.
\end{align}
where $\Gamma^c_{ab}$ are called the Christoffel symbols and $K_{ab}$ the second fundamental form (the extrinsic curvature). Then we consider the partial derivative of a tangent vector field $u^i(x)$, i.e.
\begin{align}
\partial_au^i&=\partial_a (u^b\partial_b X^i)=\partial_a u^b\partial_b X^i+u^b(\Gamma_{ab}^{c}\partial_cX^i+K_{ab}N^i)\nn\\
&=(\partial_a u^c+u^b\Gamma_{ab}^{c})\partial_c X^i+K_{ab}u^bN^i.
\end{align}
Imposing that the only change $u^i$ can undergo under an infinitesimal parallel transport is in the normal direction implies that
\begin{align}
\nabla_au^c\equiv \partial_a u^c+\Gamma_{ab}^{c}u^b=0.
\end{align}
We can easily see that $\Gamma_{ab}^c$ is indeed the metric torsion free connection associated with $g_{ab}$ by the following manipulation
\begin{align}
\partial_cg_{ab}&=\partial_c(\delta_{ij}\partial_a X^i \partial_b X^j)=\delta_{ij}\partial_{ac}X^i \partial_bX^j+\delta_{ij}\partial_{bc}X^i\partial_aX^j\nn\\
&=\delta_{ij}(\Gamma_{ac}^d\partial_dX^i+Y_{ac}N^i)\partial_bX^j+\delta_{ij}(\Gamma_{bc}^d\partial_dX^i+Y_{bc}N^i)\partial_aX^j\nn\\
&=\delta_{ij}\Gamma_{ac}^d\partial_dX^i\partial_bX^j+\delta_{ij}\Gamma_{bc}^d\partial_dX^i \partial_aX^j=\Gamma_{ac}^dg_{db}+\Gamma_{bc}^dg_{ad}
\end{align}
which is nothing but the metricity condition
\begin{align}
\nabla_cg_{ab}=\partial_cg_{ab}-\Gamma_{ac}^dg_{db}-\Gamma_{bc}^dg_{ad}=0
\end{align}
\subsubsection{Constructing the Cartan variables}
We now seek to describe the geometry of ${\cal M}$ in terms of Cartan-geometric variables 
$\{V^{i}(x), A^{ij}(x)\}$. In turn, this will enable us, when possible, to map Cartan-geometric variables to the variables $X^{i}(x)$ from the embedding picture. The Cartan-geometric description is invariant under local $SO(3)$ transformations with the metric on ${\cal M}$  given by 
\begin{align}
g_{ab}=D_aV^iD_bV^j\delta_{ij}
\end{align}
where we have assumed to simplicity that $V^2=const$. We may now wonder what the relationship between the embedding $X^i(x)$ and the pair $\{V^i,A^{ij}\}$ is. Since we aim to describe the same geometry we impose the relation
\begin{align}
\partial_aX^i\partial_bX^j\delta_{ij}=g_{ab}=D_aV^iD_bV^j\delta_{ij}
\end{align}
which means that we must have
\begin{align}
\partial_aX^i=\Omega^i_{\ph ij}(x)\sigma^j_{\ph jk}D_aV^k
\end{align}
for some local rotation matrix $\Omega^i_{\ph ij}(x)$ and some discrete and constant transformation $\sigma^i_{\ph ij}$. Since it is constant we can write without loss of generality 
\begin{align}
\partial_aX^i=\Omega^i_{\ph ij}(x)D_aV^j
\end{align}
Furthermore, since the Cartan-geometric framework is fully locally $SO(3)$ covariant we may write this relation as
\begin{align}
\partial_aX^i=\bar D_a\bar V^i=\partial_a\bar V^i+{\bar A_a}^{ij}\bar V_j
\end{align}
where we have introduced 
\begin{align}
\bar V^i=\Omega^i_{\ph ij}V^j\qquad {\bar A_a}^{ij}=\Omega^i_{\ph ik}\Omega^j_{\ph jl}{A_a}^{kl}-\partial_a\Omega^i_{\ph ik}\Omega^{jk}
\end{align}
We now readily see that that we have
\begin{align}
\bar V_i \partial_aX^i=\bar V_i\bar D_a\bar V^i=\frac{1}{2}\partial_a\bar V^2=0
\end{align}
which means that the vectors $\bar V^i(x)$ are at each point normal to the embedded surface $\m M$. Thus, we know how to obtain the contact vector $V^i$ in a particular $SO(3)$ gauge where $\partial_{a}X^i=D_{a}V^i$. The explicit relation, which only holds in certain special $SO(3)$ gauges, is then given by
\begin{align}
V^i\overset{*}{=}-\ell N^i
\end{align}
where $\ell$ is as usual the radius of the sphere we are rolling. The minus sign comes from the fact that the normal points away from the manifold while the contact vector by definition points towards the point of contact on the manifold.

The relation $\partial_{a}X^i\overset{*}{=}D_{a}V^i$ yields the equations 
\begin{align}
\partial_a(X^i-V^i)={A_a}^{ij}V_j
\end{align}
from which some but not all the components of  $A^{ij}$ can be determined. In order to deduce the remaining ones we impose that the $SO(2)$ connection defined by \eqref{omegaeq} should yield the same parallel transport of tangent vectors as $\Gamma^c_{ab}$. Starting from the equation representing an infinitesimal parallel transport
\begin{align}
u_{||}^c(x_2)=u^c(x_1)-\delta x^a\Gamma_{ab}^c(x_1)u^b(x_1)
\end{align}
and then multiplying it with with $\partial_cX^i(x_2)$ yields
\begin{align}
u_{||}^i(x_2)=u^i(x_1)+\delta x^a \partial_{ab}X^i(x_1) u^b(x_1)-\delta x^a\Gamma_{ab}^c(x_1)u^b(x_1)\partial_{ab}X^i(x_1)
\end{align}
We then simplifying it using \eqref{expand} yielding
\begin{align}
u_{||}^i(x_2)=u^i(x_1)+\delta x^a K_{ab}(x_1)N^i(x_1) u^b(x_1)
\end{align}
On the other hand, $u^i$ can also be parallel transported with ${\omega_a}^i_{\ph ij}$, i.e.
\begin{align}
u_{||}^i(x_2)=u^i(x_1)-\delta x^a{\omega_a}^i_{\ph ij}(x_1)u^j(x_1)=u^i(x_1)-\delta x^a{\omega_a}^i_{\ph ij}(x_1)\partial_bX^ju^b(x_1)
\end{align}
which then yields the relation
\begin{align}
\boxed{K_{ab}N^i=-{\omega_a}^i_{\ph ij}\partial_bX^j}
\end{align}
from which we can determine ${\omega_a}^i_{\ph ij}$ using an inverse $(\partial X^{-1})^b_j$ satisfying 
\begin{align}
(\partial X^{-1})^b_j \partial_aX^j=\delta^b_a\qquad (\partial X^{-1})^b_j\partial_b X^i=P^i_{\ph ij}
\end{align}
where $P^i_{\ph ij}=\delta^i_j-N^iN^j$ is a projector.
\subsubsection{Remarks}
As we now have seen how one may construct the waywiser variables $\{V^i(x),A^{ij}(x)\}$ from the embedding $X^i(x)$ some remarks are in order.

\begin{itemize}
\item It is only in a particular $SO(3)$ gauge, the {\em embedding gauge}, wherein the contact vectors $V^i$ are normal to the embedded surface. Using the local $SO(3)$ gauge invariance of the Cartan geometric description we may nevertheless choose any other gauge which is convenient, e.g. the standard gauge in which $V^i=(0,0,\ell)$. This does not alter the geometry intrinsic to $\m M$. On the other hand, if we describe the geometry using the embedding variables $X^i(x)$ we loose the local $SO(3)$ gauge invariance. Specifically, the metric given by $g_{ab}=\partial_a X^i\partial_b X^j\delta_{ij}$ is not invariant under local $SO(3)$ transformations $X^i\rightarrow \Lambda^i_{\ph ij}(x)X^i$. Instead, the formalism is invariant under global $SO(3)$ transformations and translations $X^i\rightarrow X^i+\xi^i$.
\item Not all two-dimensional manifolds can be embedded into a three-dimensional Euclidean space. Thus, the embedding approach is more restrictive than the pure Cartan geometric description where the geometry is specified by postulating independently $V^i(x)$ and ${A_a}^{ij}(x)$.

\item As the ambient embedding space is torsion-free the {\em induced} affine structure $\Gamma_{ab}^c$, defined by \eqref{expand}, of any submanifold must also be torsion-free.  However, within Cartan geometry this condition is not a natural one as it imposes relationships on the variables $V^i$ and $A^{ij}$ which are naturally taken as independent ones. On the other hand, in the embedding approach both $V^i(\partial X)$ and $A^{ij}(\partial X,\partial^2 X)$ are variables built from the embedding variables $X^i$. There is however nothing that stops us from adding a contorsion field $C^{ij}$. Such a field is completely independent from the embedding variables $X^i(x)$ and will alter the notion of parallelism.

\item Thirdly, we may easily generalize the above discussion to waywisers of variable size, i.e. we have $V^2\neq const.$. However, it is clear that the quantity $V^2(x)$ is not determinable from $X^i$ signaling that it is not a geometric (Riemannian) quantity. This can be taken as an indication that one should impose $V^2=const.$ However, discussed below, the scalar $V^2$ naturally plays the role as dark energy in the form of quintessence. Thus, although the scalar $V^2$ may look awkward from a Riemannian point of view it is an object that naturally occurs within a Cartan-geometric description.

\item Finally, we note that in the embedding gauge in which $dX^i\overset{*}{=}DV^i=e^i$. This immediately shows that the frame field $e^i$ is integrable, i.e. $de^i=ddX^i=\frac{1}{2}\partial_{[ab]}X^idx^adx^b\equiv 0$. As far as we can tell this gauge exists whether we have torsion or not and may be useful. In the case of vanishing Cartan curvature $F^{ij}$ we are free to choose a gauge in which $A^{ij}\overset{*}{=}0$ and we see that $X^i$ and $V^i$ are equal up to a translation. This follows from the equation $0=\partial_a\ell^2=\partial_a(X^iX_i)=2X\partial_a X^i$ which says that $X^i$ is a normal. It would be interesting to explore if these scalars $X^i$ can be used to define a notion of local energy in Cartan gravity.
\end{itemize}

\subsection{Abstract Cartan waywiser geometries}
We can now forget about the embedding space which only served to facilitate visualization and helping intuition along. The situation is not different from Riemannian geometry where embedding spaces are invoked to facilitate visualization and does not indicate that the construction at the fundamental level invokes higher dimensions. The mathematical representation of an abstract Cartan waywiser geometry is simply the pair $\{V^i(x),A^{ij}(x)\}$ and no reference to an embedding space is required. From a mathematical point of view we see that we are dealing with a fiber-bundle structure where the base space is the manifold $\m M$ and the fiber is the group manifold $SO(3)$ with the vector $V^i\in \mathbb{R}^3$ belonging to the fundamental representation of the group.

The choice of representation is very important for determining the physical content of a theory. On that note we stress that the mathematical representation of both connection $A^i_{\ph ij}$ and contact point is the {\em fundamental representation} of the orthogonal group. It is the use of the fundamental representation that introduces the extra degree of freedom $V^2(x)$. Ironically, as we shall see, what could be considered from a mathematical point of view an unwanted scalar degree of freedom will, in a cosmological context, play the role of dark energy. Thus dark energy is in a Cartan-geometric description of gravity not an {\em ad hoc} degree of freedom that that needs to be added from the outside. Rather it is an integral part of the mathematical package of Cartan geometry. Put in a different way: dark energy can be understood  in Cartan gravity as an effect of a {\em gravitational Higgs field}.
\section{`Waywisers' for space-time theories}\label{higherd}
Now that we have gained some intuition about Cartan geometry and its geometric interpretation in terms of idealized waywisers, we turn to General Relativity. To accommodate spacetime geometries and relativistic theories we must adapt the above waywiser formalism accordingly. From a mathematical point of view the obvious change to make is to make use of symmetric spacetimes, rather than spaces, as idealized waywiser `wheels'. In the literature the symmetric spacetimes representing idealized relativistic waywiser wheels go by the name {\em model spaces} or {\em model spacetimes}. We shall from now on use those terms interchangeably.

In this article we will focus on the de Sitter spacetime as a model spacetime. We could also use an anti-de Sitter spacetimes or a flat Minkowski spacetime as model spacetime. The anti-de Sitter case is very similar to the de Sitter one but the choice of a flat model spacetime requires a slightly different mathematical representation \cite{Gronwald:1995em} of the contact point and we will not discuss that option in this paper \cite{Gronwald:1995em,Wise:2006sm}.
\subsection{De Sitter spacetime as model spacetime}
\begin{figure}
\centering
\includegraphics[width=9cm]{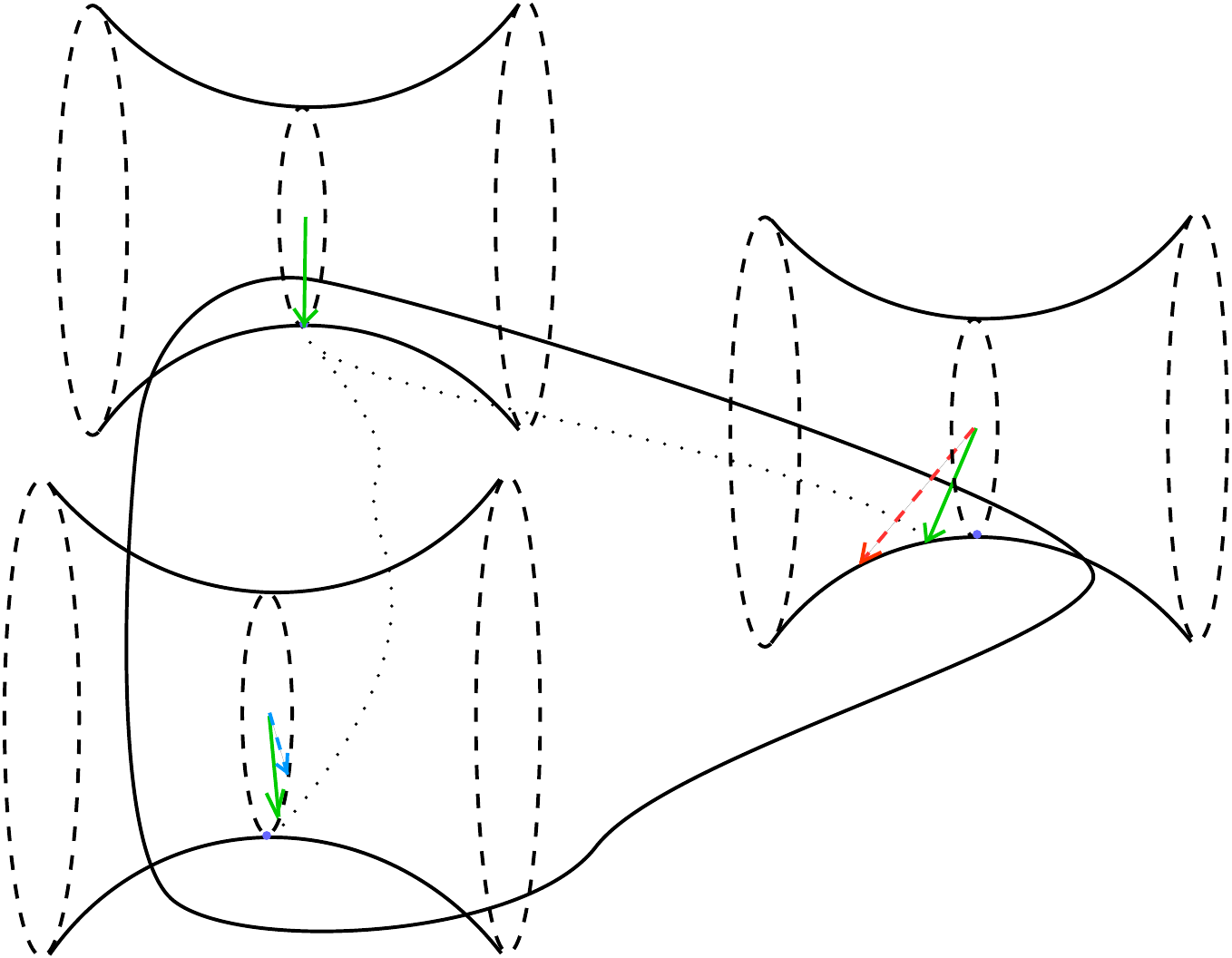} 
\caption{Visual depiction of the embedding picture of rolling for the case where the
embedding space is $\mathbb{R}^{(1,2)}$ and $g_{ab}$ has signature $(-,+)$. Solid arrows
represent $V^{i}$ in a gauge where it is normal to the submanifold ${\cal M}$ at each point whereas
dotted arrows represent the parallelly transported $V_{|}^{i}$ along the paths denoted
in the figure. As a guide to intuition, spacelike (according to $g_{ab}$) displacements
along ${\cal M}$ involving rolling of the de Sitter space (corresponding to spacelike displacements on the de Sitter space) whereas timelike (according to $g_{ab}$) displacements along ${\cal M}$ involve a `hyper-rolling' (corresponding to timelike displacements on the de Sitter space). }
\label{hyperroll}
\end{figure}

As a first mathematical realization of the idealized `relativistic wheel', i.e. model spacetime, we consider the de Sitter spacetime  which may be defined as a hypersurface in a five-dimensional Minkowski spacetime satisfying:
\begin{eqnarray}
-t^2+x^2+y^2+z^2+w^2=\ell^2
\end{eqnarray}
where $\ell$ is a real constant. The symmetry group of isometries on this surface defines the group $SO(1,4)$, i.e. all transformations that leave the metric $\eta_{AB}=\mathrm{diag}(-1,1,1,1,1)$ invariant where $A=0,\dots ,4$. A point on the surface may be represented by a spacelike contact vector $V^A$ which breaks the $SO(1,4)$ symmetry. The `rolling without slipping' is specified by a $SO(1,4)$ connection $A^{AB}=A_\mu^{\ph \mu AB}dx^\mu$ ($\mu=0,\dots 3$) and the geometry of the manifold $\m M$ is now completely characterized by the pair $\{V^A(x),A^{AB}(x)\}$. See Figure \ref{hyperroll} for an illustration of what is rolled and how in the embedding picture for the lower dimensional case of the group $SO(1,2)$ where a two-dimensional de Sitter space is rolled on a two dimensional submanifold $\m M$ of $\mathbb{R}^{1,2}$.

The indefinite character of the metric $\eta_{AB}$ makes the situation more complex than in the Euclidean positive definite case with $\delta_{ij}$. In this section we have assumed until now that the contact vector $V^A$ is spacelike, i.e. $V^2>0$. But we could, of course, also consider timelike $V^2<0$ or null $V^2=0$ contact vectors. If we allow for a possible dependence of $V^2(x)=\eta_{AB}V^{A}(x)V^{B}(x)$ upon spatial coordinate $x^\mu$ one may conceive of a norm $V^{2}$ which may not only vary in magnitude but also sign over the space-time manifold. Thus, at some points we may find that $V^{2}<0$ and that it can be regarded as a point on one of the sheets of a higher-dimensional hyperboloid of two-sheets defined by
\begin{eqnarray}
-t^{2}+x^{2}+y^{2}+z^{2}+w^{2} = -\ell^{2}.
\end{eqnarray}
The subgroup of transformations that leave the components of the timelike contact vector ($V^{2}<0$) invariant is the orthogonal group $SO(4)$. Finally one may imagine regions where the contact vector is null ($V^{2}=0$) and non-vanishing; in which case the subgroup of transformations that leave the vector invariant is the Poincar\'e group $ISO(3)$. The three groups $SO(1,3)$, $SO(4)$, and $ISO(1,3)$ are sometimes referred to as the stabilizer group corresponding the particular norms of $V^{A}$. 

As we shall see, the norm $V^{2}$ dictates the signature of the metric tensor. If the dynamics of a gravitational theory based on the Cartan-geometric variables  $\{V^A(x),A^{AB}(x)\}$ force the sign of $V^{2}$ to vary over the space-time manifold then signature change of the metric is inevitable. In fact, in Section \ref{cosmo} we shall showcase both analytical and numerical solutions to the Cartan-geometric equations of motion exhibiting signature change.
\subsection{Relation to standard notation}\label{standnotation}
Though the results of Section \ref{waywisermath} were derived for the case of a two-dimensional manifold and `rolling-group' $SO(3)$, they are immediately extendable to the physically relevant case of a four-dimensional manifold and the groups $SO(1,4)$  over any region where $V^{2}\neq 0$ (so that the generalization of the projector $P^{i}_{\ph{i}j}$ is well defined). We will use $A,B,C,...$ to denote $SO(1,4)$ indices; $I,J,K,..$ to denote indices in representations of the subgroup of transformations that leave $V^{A}$ invariant; and Greek letters will be used to denote space-time manifold co-ordinate indices. Hence we can immediately define the projector $P^{A}_{\ph{A}B}$, co-tetrad $e^{A}_{\mu}$, metric $g_{\mu\nu}$, $V^{A}$ invariant subgroup spin-connection $\omega_{\mu}^{\ph{\mu}AB}$, and affine-connection $\Gamma_{\alpha\beta}^{\gamma}$:
\begin{empheq}[box=\fbox]{align}
P^{A}_{\ph{A}B}&=\delta^{A}_{\ph{A}B}-\frac{V^{A}V_{B}}{V^{2}}\nn\\
e^{A}_{\mu}&=P^{A}_{\ph{A}B}D_{\mu}V^{B}\nn\\
g_{\mu\nu}&=P_{AB}D_{\mu}V^{A}D_{\nu}V^{B}\\
\omega_{\mu}^{\ph{\mu}AB}&=A_{\mu}^{\ph{a}AB}+\frac{2}{V^{2}}V^{[A}e^{B]}_{\mu}\nn\\
\Gamma^{\gamma}_{\alpha\beta} &=e^{\gamma}_{A}\left(\partial_{\alpha}e^{A}_{\beta}+\omega_{\alpha\ph{A}B}^{\ph{\alpha}A}e^{B}_{\beta}-\frac{1}{2}\partial_{\alpha}\mathrm{log}V^{2} e^{A}_{\beta}\right)\nn 
\end{empheq}
In the case $V^2>0$ we can adopt the gauge $V^A\os\phi\delta^A_4$ in which the co-tetrad $e^A$ and spin-connection take on the form
\begin{align}
e^A\os (e^I,0)\qquad  \omega_{\mu}^{\ph{\mu}AB}\os \left(\begin{array}{cc}\omega^{IJ}&0\\0&0                                                                                          \end{array}\right)
\end{align}
The metric $g_{\mu\nu}=e^A_\mu e_{\nu A}$ has then a Lorentzian signature $(-,+,+,+)$ and the connection $\omega^{IJ}$ behave as the standard $SO(1,3)$ spin connection. However, if $V^2<0$ the gauge $V^A\os\phi\delta^A_4$ is not attainable. Instead we can adopt the gauge $V^A\os\phi\delta^A_0$ in which case we have ($\bar I,\bar J=1,\dots,4$)
\begin{align}
e^A\os (0,e^{\bar I})\qquad  \omega_{\mu}^{\ph{\mu}AB}\os \left(\begin{array}{cc}0&0\\0&\Omega^{\bar I\bar J}                                                                                          \end{array}\right)
\end{align}
The metric $g_{\mu\nu}=e^A_\mu e_{\nu A}$ in this case has then a Euclidean signature $(+,+,+,+)$ and $\Omega^{\bar I\bar J}$ behaves as an $SO(4)$ connection. Finally, the null case $V^2=0$ defines a degenerate and non-invertible metric.

As in the case of $SO(3)$, the $SO(1,4)$ curvature two-form $F^{AB}$ can be decomposed as follows:
\begin{eqnarray}
F^{AB} &=& R^{AB}-\frac{1}{V^{2}}e^{A}e^{B}+\frac{2}{V^{2}}\left(T^{[A}V^{B]}-\frac{1}{2}d\log V^{2} e^{[A}V^{B]}\right)
\end{eqnarray}
where $R^{AB}\equiv d\omega^{AB}+\omega^{A}_{\ph{A}B}\omega^{BC}$ and $T^{A}\equiv d e^{A}+\omega^{A}_{\ph{A}B} e^{B}$.

We note that while the definition of the co-tetrad $e^A$ includes a gauge covariant exterior derivative, this is not the case for the spin-connection $\omega^{AB}$. This signals a significant mathematical difference between the two objects. In particular, while the spin connection $\omega^{AB}$ transforms inhomogeneously under a $SO(1,4)$ gauge transformation, the same is not true for the co-tetrad $e^A$. For this reason the co-tetrad $e^A$ cannot be thought of as a gauge connection in this context.\footnote{We contrast our approach to Poincar\'e gauge theory \cite{Hehl:1994ue} in which the co-tetrad is commonly conceptualized as a gauge connection with respect to local (or `soft') translations.} Specifically, the co-tetrad should not be thought of as a gauge connection related to the `translational' symmetry of the de Sitter model spacetimes.\footnote{A more accurate term is {\em transvections} \cite{Randono:2010cq}.} Rather, the co-tetrad is best understood as the quantifying the change of contact point when the idealized waywiser wheel is rolled; something which is not a gauge quantity. We also note that the `internal' translations, i.e. $SO(1,4)$ transformations that change $V^A$, are both conceptually and mathematically distinct from diffeomorphisms which can be thought of as `external' translations on the manifold. By considering specific examples it becomes clear that the action of an internal translation cannot in general be viewed as, or equated with, the action of a diffeomorphism on $\m M$. Nevertheless we shall see in Section \ref{SRlimit} how global Poincar\'e symmetry and the co-tetrad becomes intimately linked together in the special relativistic limit.

\subsection{Making contact with scalar-tensor theories}
A wide class of models called {\em scalar-tensor theories} are frequently used in cosmology to model dark energy, dark matter, and inflation. It is therefore of interest to note that relativistic Cartan geometry with a fully dynamical gravitational Higgs field $V^A$ is nothing but a scalar tensor theory in regions of spacetime where $V^2(x)$ has a definite sign.

As we noted in Section \ref{identconnection} affine connection $\Gamma^\rho_{\mu\nu}(A,V)$ was not metric with respect to $g_{\mu\nu}(A,V)$, i.e. we have
\begin{align}
\nabla_{\rho}g_{\mu\nu}=\partial_\rho\log V^2 g_{\mu\nu}.
\end{align}
However, we can instead introduce the new metric $\tilde g_{\mu\nu}=\frac{V_0^2}{V^2}g_{\mu\nu}$ where $V_0$ is an arbitrary constant of dimension length. The new metric then satisfies
\begin{align}\label{rescale}
\nabla_{\rho}\tilde g_{\mu\nu}=0
\end{align}
which then renders the affine connection $\Gamma^\rho_{\mu\nu}$ a metric one with respect to $\tilde g_{\mu\nu}$. At this point we recognize that all degrees of freedom are encoded in the metric compatible pair $\{\tilde g_{\mu\nu},\Gamma^\rho_{\mu\nu}\}$ and the scalar function $V^2(x)$. Thus, we are dealing with a scalar-tensor theory. This opens up the possibility of applications towards dark energy, dark matter, and inflation. In fact, we shall see in detail below how the extensively studied Peebles-Ratra slow rolling quintessence model of dark energy comes out from a very simple and polynomial Cartan-geometric action for the variable $V^A$ and $A^{AB}$.

Although the affine connection $\Gamma^\rho_{\mu\nu}$ is a metric compatible with respect to $\tilde g_{\mu\nu}$ it by no means implies that the affine connection is torsion-free. In fact, it is not since the field $V^A$ sources torsion. Depending on which action principle we choose this can happen in two distinct ways.
\begin{enumerate}
 \item Torsion is completely determined by the form of $V^A$: we are dealing with a standard scalar-tensor theory completely characterized by the metric tensor $g_{\mu\nu}$ and the scalar field $V^2(x)$.
 \item Torsion is a genuine dynamical degree of freedom, i.e. we are dealing with a generalized scalar-tensor theory with propagating torsion. In this case the theory is characterized completely by the metric $g_{\mu\nu}$, the scalar field $V^2(x)$, and a torsion tensor ${\m T_{\mu\nu}}^\rho$.
\end{enumerate}
As will be seen below, the most well-known action for Cartan gravity, the MacDowell-Mansouri action, yields a scalar tensor type theory with propagating torsion.

The above rescaling can only be done in an open set $\m U\subset\m M$ where $V^2$ has a definite sign. Indeed, if $V^2$ changes sign in $\m U$ then the conformal transformation \eqref{rescale} is singular. The possibility that $V^2$ can change sign immediately implies a change of signature, e.g. from Lorentzian $(-,+,+,+)$ to Euclidean $(+,+,+,+)$. Geometries with signature change have been studied in detail and it is quite surprising that a smooth process of signature change comes out naturally from Cartan gravity with dynamical gravitational Higgs field $V^A$. Thus, this theory of gravity is far more exotic than a scalar-tensor theory.  

We also note that the non-metricity comes in the form of an integrable Weyl field $c=\frac{1}{2}d\log V^2$, i.e. a one-form  introduced by Weyl to study conformal theories (see e.g. \cite{Blagojevic:2002du,westman2009first}). The integrability here refers to the fact $c$ is exact and thus a closed one-form $dc=0$. Indeed it may be shown that Cartan geometry with dynamical $V^a$ is nothing but a limit of a generalization of Cartan geometry that may be named {\em conformal Cartan geometry} which operates with the conformal group $C(1,3)\simeq SO(2,4)\simeq SU(2,2)$ and a pair of symmetry breaking fields $\{\m O^a,\infty^a\}$ which geometrically represents a `contact point' as well as conformal infinity (or alternatively a field $W^{ab}$ in the adjoint representation of $SO(2,4)$). This observation leads to a novel way of implementing scale invariance in physical theories  \cite{WestmanZlosnik2014}. \footnote{See also \cite{Gryb:2012qt} for a study of conformal structure in $2+1$ gravity.}
\subsection{Poincar\'e invariance and the special relativistic limit}\label{SRlimit}
In order to be empirically viable, any theory must contain special relativity as limiting case. To see how that comes about in Cartan gravity let us write the Cartan curvature two-form $F^{AB}$ in the gauge $V^A=\phi\delta^A_4$ (which is attainable only if $V^2>0$): 
\begin{eqnarray}
F^{IJ}&\overset{*}{=}& R^{IJ}-\frac{1}{\phi^{2}}e^{I}e^{J}\qquad F^{I4}\phi=de^I+\omega^I_{\ph IJ}e^J-d\log \phi e^I.
\end{eqnarray}
Suppose we study an open region $\m U$ in spacetime of the typical length scale $\ell$. If we adapt our units so that $\ell=1$ and consider smaller and smaller regions, then $\phi^2\ra\infty$ in these units. Similarly, in the limit of smaller and smaller $\m U$ we find that the {\em components} ${F_{\mu\nu}}^{AB}$ in a coordinate system $x^\mu$ adopted to the increasingly smaller size of the region $\m U$, tends to zero ${F_{\mu\nu}}^{AB}\ra0$.\footnote{The two-form $F^{AB}=\frac{1}{2}{F_{\mu\nu}}^{AB}dx^\mu dx^\nu$ is on the other hand of course completely coordinate independent. Thus, the limit $F^{AB}\ra 0$ cannot be interpreted in the same way as above, i.e. as the restriction to smaller and smaller regions $\m U$ of spacetime. Instead, that refers to changing the curvature form and thus changing the physical situation and not the size of the region in question. We also note that what is being measured is invariably the components  of a tensor in some physical coordinate system.} Thus, we consider the limit
\begin{align}\label{limit}
\phi\ra const.\ra\infty \qquad F^{I4}\phi\ra0 \qquad F^{IJ}\ra0
\end{align}
which then yields
\begin{eqnarray}
R^{IJ}\ra0\qquad de^I+\omega^I_{\ph IJ}e^J \ra 0
\end{eqnarray}
The vanishing of the Riemannian curvature two-form implies the existence of a special $SO(1,3)$ gauge in which $\omega^{IJ}\os0$. This in turn implies that the co-tetrad $e^I$ in that particular gauge is a closed one-form, i.e.
\begin{align}
de^I\os0.
\end{align}
Since $e^I$ is closed we can locally find four scalar fields $q^I(x)$ such that $e^I\os dq^I$. The choice of these four scalar fields $q^I$ is unique up to a transformation of the type $q^I\rightarrow q^I+a^I$ where $a^I$ are constants. In addition, the gauge condition $\omega^{IJ}\overset{*}{=}0$ is left invariant under a {\em global} Lorentz transformation. Since the formalism is also invariant under local Lorentz transformations we see that the scalars $q^I$ are unique up to a global Poincar\'e transformation, i.e.
\begin{align}
q^I(x)\rightarrow \Lambda^I_{\ph IJ}q^J(x)+a^I \label{poincaretrans}
\end{align}
It is now clear that it is the scalar fields $q^I(x)$ that play the role of the Cartesian coordinates in special relativity. Note also that in contrast to the manifold coordinates $x^\mu$, the Cartesian coordinates $q^I(x)$ have dimensions of length in agreement with their operational significance as length/time measurements in special relativity. 

To make this point clearer let us consider the concrete case of a massless Klein-Gordon field $\Phi(x)$ coupled to gravity. The standard  action is given by
\begin{align}
S_{KG}=\int d^4xe\eta^{IJ}e_I^\mu e^\nu_I\partial_\mu\Phi(x)\partial_\nu\Phi(x).
\end{align}
which in the special relativistic limit takes the form
\begin{align}\label{parKG}
S_{KG}=\int d^4x \left|\frac{\partial q}{\partial x}\right|\eta^{IJ}\frac{\partial x^\mu}{\partial q^I}\frac{\partial x^\nu}{\partial q^J}\partial_\mu\Phi\partial_\nu\Phi.
\end{align}
However, this is just the standard Klein-Gordon action in flat spacetime but written in a general coordinate system $x^\mu$. If we adapt our coordinates so that $(q^0,q^1,q^2,q^3)=(x^0,x^1,x^2,x^3)$ we find that the action reduces precisely to that of the Klein-Gordon action in Minkowksi spacetime.
\begin{align}
S_{KG}=\int d^4q \eta^{IJ}\partial_I\Phi(x(q))\partial_J\Phi(x(q)).
\end{align}
To sum up the essentials: first we find that in the limit of zero torsion and Riemannian curvature we find that the co-tetrad becomes integrable $e^I=dq^I$. This allows for the introduction of a coordinate system $q^I$ which is unique up to a Poincar\'e transformation $q^I(x)\rightarrow \Lambda^I_{\ph IJ}q^J(x)+a^I$. These four scalars $q^I(x)$, which exist {\em only} in the special relativistic limit, are then identified with the standard coordinates of special relativity, not to be conflated with the $x^\mu$'s. Thus, in the special relativistic limit there emerges a natural coordinate system $q^I$ and a formalism which is invariant, not under general coordinate transformations (or in an active view point: diffeomorphisms), but only under the Poincar\'e group. Varying the action \eqref{parKG} with respect to $q^I$ yields an equation which in the coordinate system $(q^0,q^1,q^2,q^3)=(x^0,x^1,x^2,x^3)$ is nothing but the conservation $dT^I=0$ of the energy-momentum three-form $T^I$. Furthermore, the symmetry of the action under global transformations of the form (\ref{poincaretrans}) yield conserved Noether charges corresponding to field energy, momentum, angular momentum, and three-charges due to invariance under Lorentz boosts.

The above shows in detail how the diffeomorphism group is broken down to the Poincar\'e group. This symmetry-breaking process $diff\ra ISO(1,3)$ is not aided by a Higgs or Stueckelberg fields but instead happens in the limit where the one-forms $e^I=dq^I$ become integrable and the theory exhibits global Poincar\'e invariance. $q^I(x)\ra \Lambda^I_{\ph IJ}q^J(x)+a^I$.

\section{Action principles for gravity}
\label{actions}
The Einstein-Hilbert action ${\cal S}_{EH}=\int \sqrt{-g} g^{\mu\nu}R_{\mu\nu}d^4x$ is a rather complicated action. It is manifestly non-polynomial in its basic dynamical variable $g_{\mu\nu}$ (since it involves the square root $\sqrt{-g}$ of the metric determinant $g=det\ g_{\mu\nu}$) as well as the inverse metric $g^{\mu\nu}$ which required to exist. The action is further complicated by the fact that it contains second order partial derivatives with respect to the metric tensor. This makes it necessary to add, in the case of non-compact spaces, a compensating boundary term (the Gibbons-Hawking term) in order to ensure that the Einstein-Hilbert action is indeed extremized whenever the field equations are satisfied \cite{Wald:1984rg}.

On the other hand, we shall see that the natural actions for gravity using the waywiser variables $\{V^A,A^{AB}\}$, are {\em polynomial} in the basic waywiser variables, and are, from a mathematical point of view, rather elegant. No restrictions on the variables are needed, e.g. requiring that the metric $g_{\mu\nu}(V,A)$ be invertible. 

Since an action is per definition an integration over a four-form, the construction of the simplest actions possible in Cartan waywiser geometry is just an exercise in `wedging' together the various forms we can construct from the waywiser variables.\footnote{Non-polynomial actions for General Relativity based on gauge connections can of course be considered \cite{Krasnov:2011pp,Krasnov:2012pd} but we shall restrict attention to polynomial actions.} Building an action is very much like playing with Lego \cite{Lego}: You only have but a few basic pieces (the forms) and the only task is to find out how to fit the pieces together to create four-forms with no rolling indices left un-contracted.

We note that where are two approaches with distinct physical content and phenomenology. These are:
\begin{itemize}
\item {\bf Non-dynamical:} $V^A$ is regarded as a non-dynamical {\em \`a priori} postulated variable, also called an absolute object \cite{AndersonRelativity,WestmanSonego2007b}. We simply pick some contact field $V^A(x)$ subject to the only constraint $\eta_{AB}V^{A}V^{B}= \ell^2$.  No equations of motion given for the contact vector $V^A$.
\item {\bf Dynamical:} $V^A$ is regarded as a genuine dynamical variable on par with $A^{AB}$ which have its own equations of motion and should be varied with respect to in an action principle. This yields a natural generalization of Cartan gravity which yields a wealth of new phenomenology. In this formulation gravity resembles a Yang-Mills theory with dynamical symmetry breaking.
\end{itemize}
In the following we shall pursue only the latter view. The following sections will make heavy use of the variational calculus of forms. For an exposition of all necessary ideas and techniques of the variational calculus of forms we point to the Appendices.
\subsection{The polynomial family of actions}
Let us then contemplate what kind of Lagrangian polynomial four-forms $\mathcal{L}$ may be constructed. To do that we should first list the basic building blocks we have at our disposal.
\begin{itemize}
\item the waywiser variables $\{V^A,A^{AB}\}$ from which the gauge covariant objects $F^{AB}$ and the one-form $DV^{A}$ can be constructed
\item the `internal' Minkowski metric $\eta_{AB}$ and Levi-Civita symbol $\epsilon_{ABCDE}$ associated with the orthogonal group $SO(1,4)$.
\end{itemize}
The most general polynomial gravitational action that can be constructed is
\begin{equation}\label{genaction}
\boxed{S[A^{AB},V^{A}]=\int\left( a_{ABCD} F^{AB} F^{CD} + b_{ABCD} DV^A DV^B F^{CD}+c_{ABCD} DV^A DV^B DV^C DV^D\right)}
\end{equation}
where 
\begin{eqnarray}
a_{ABCD} &=& a_{1}\epsilon_{ABCDE}V^{E}+a_{2} V_{A}V_{C}\eta_{BD} +a_{3} \eta_{AC}\eta_{BD} \\
b_{ABCD} &=& b_{1}  \epsilon_{ABCDE}V^{E}+b_{2} V_{A}V_{C}\eta_{BD}+b_{3}\eta_{AC}\eta_{BD}\\
c_{ABCD} &=&  c_{1}\epsilon_{ABCDE}V^{E}
\end{eqnarray}
In general the quantities $a_{i},b_{i},c_{i}$ may depend on the scalar $V^{2}=V_{E}V^{E}$. 
Though this action may look unfamiliar, we can see that it takes on a rather more familiar form in regions where $V^{2} \neq 0$. Specifically we will now look at the case where the group is $SO(1,4)$ and $V^{2}>0$ i.e. the sub-group that leaves $V^{A}$ invariant is then $SO(1,3)$. For ease of comparison to other models of gravity, we furthermore gauge fix to a gauge where $V^{A} = \phi(x^{\mu})\delta^{A}_{\ph{A}4}$. Hence, indices of quantities with vanishing projection along $V^{A}$ (e.g. $\omega^{AB}$ and $e^{A}$) can simply be written with $SO(1,3)$ indices $I,J,K,\dots$ and we have the following decomposition of the $SO(1,4)$ curvature $F^{AB}$:

\begin{eqnarray}
F^{IJ} &=&  R^{IJ}-\frac{1}{\phi^{2}}e^{I}e^{J} \\
F^{I4} &=& \frac{1}{\phi}\left(T^{I}-\frac{1}{2}d\log \phi^{2} e^{I}\right)
\end{eqnarray}
The action (\ref{genaction}) then takes the following form:

\begin{eqnarray}\label{4action}
S[e^{I},\omega^{IJ},\phi] &=&  \int  \frac{1}{32\pi G(\phi)}\left(\epsilon_{IJKL}\left(e^{I}e^{J}R^{KL}-\frac{\Lambda(\phi)}{6}e^{I}e^{J}e^{K}e^{L}\right)-\frac{2}{\gamma(\phi)}e_{I}e_{J}R^{IJ}\right) \nonumber \\
   && + \bigg({\cal C}_{1}(\phi)\epsilon_{IJKL}R^{IJ}R^{KL}+{\cal C}_{2}(\phi)R_{IJ}R^{IJ}+{\cal C}_{3}(\phi)(T^{I}T_{I}-e_{I}e_{J}R^{IJ})\bigg) \label{act1}
\end{eqnarray}
where 
\begin{eqnarray}\label{funcs}
16\pi G(\phi) &=&  \frac{\phi}{2\left(-2a_{1}+b_1\phi^{2}\right)},  \quad
\Lambda(\phi) =  6\frac{\left(a_{1}-b_{1}\phi^{2}+c_{1}\phi^{4}\right)}{\phi^{2}\left(2a_{1}-b_{1}\phi^{2}\right)} , \nonumber\\
\gamma(\phi) &=& 2\frac{\left(2a_{1}-b_{1}\phi^{2}\right)}{(a_{2}+b_{3})\phi},\quad 
{\cal C}_{1}(\phi) =  a_{1}\phi ,  \quad
{\cal C}_{2} (\phi)=  a_{3},  \nonumber\\
{\cal C}_{3}(\phi) &=& \frac{2a_{3}}{\phi^{2}}+\int^{\phi} \left(\frac{2a_{3}}{\phi^{4}}+\frac{a_{2}}{\phi^{2}}+\frac{b_{2}}{2}+\frac{b_{3}}{\phi^{2}}\left(1-\frac{\phi^{2}}{b_{3}}\frac{\partial b_{3}}{\partial \phi^{2}}\right)\right) d\phi^{2}+\left(a_{2}+b_{3}\right)
\end{eqnarray}
and where $T^{I} \equiv de^{I}+\omega^{I}_{\ph{I}J}e^{J}$ is the torsion. The perhaps surprising presence of the integral in ${\cal C}_{3}(\phi)$ is due to the following:  terms in the $abc$ action may originally be of the form $f(\phi)d\phi^{2}e_{I}T^{I}$; if we can write such terms as $dg(\phi)$ then we may rewrite this term as a boundary term plus the a `${\cal C}_{3}$' term via:  $dg(\phi) e_{I}T^{I}= d(g(\phi)e_{I}T^{I})- g(\phi)\left(T_{I}T^{I}-e_{I}e_{J}R^{IJ}\right)$. Therefore given $f(\phi)$ we may find $g(\phi)$ via the equation  $f(\phi)d\phi^{2} =  dg(\phi)$, the solving of which yields the above integration.

\subsection{General Relativistic limit}
Any new proposed theory must contain the older verified ones as limiting cases. Thus we must ask if General Relativity can be found in some limit of this new theory. The answer is simple: the limit $V^2=\phi^2\ra const.>0$ corresponds exactly to General Relativity in its Einstein-Cartan incarnation. Indeed, when $V^2=\phi^2=const.$ the three last terms in \eqref{4action} are topological terms and thus do not contribute the the equations of motion. The three first terms we recognize as the Einstein-Hilbert, cosmological constant, and Holst terms. The Holst term $e_Ie_JR^{IJ}$ modulates the amount of torsion which is induced by fields coupled to the spin-connection (for instance the spin density of fermionic fields) and is not a topological term.

In addition to demonstrating that General Relativity is contained as a limiting theory we must also show that there is a dynamical mechanism that drives $V^2\ra const.$ In fact, below we show that this is a typical behaviour in a cosmological setting for the action with $a_1$ and $b_2$ non-zero. A study of the General Relativistic limit for the most general action is still an open problem which should be addressed. 

Finally we note that the constancy of $\phi$ can be  can be achieved by simply adding a Lagrange multiplier to the action (\ref{genaction}) \cite{Stelle:1979va,Pagels:1983pq,Leclerc:2005qc,Randono:2010ym}:
\begin{eqnarray}
{\cal S}_{\lambda}[\lambda,V^{A}] =  \int \lambda \left(V^{2} - \ell^{2}\right)
\end{eqnarray}
Requiring that the action is stationary with respect to small variations of the Lagrange multiplier four-form $\lambda$ then produces the required fixed norm constraint. But this procedure is artificial since rather than enforcing equations of motion of dynamical variables, the equations of motion for $V^{E}$ simply amount to a definition of $\lambda$.

\subsection{Relation to other proposed modifications}
Note that $\phi$ appears only algebraically in \eqref{4action}, but in fact this is merely a relic of the first-order formalism. Sub-cases of (\ref{4action}) correspond to scalar-tensor theories when converted into second-order language (see, for instance, \cite{Mercuri:2009zi}). This ``algebraic relic'' is analogous to the fact that $e^{I}$ appears only algebraically in the Palatini action of Einstein-Cartan gravity but the metric formed using $e^{I}$ appears in the Einstein-Hilbert action via its first and second derivatives. The reason for this is that the equations of motion stemming from the Palatini action constrain  $\omega^{IJ}$ to be equal to derivatives of $e^{I}$. Upon inclusion of a $\phi$ dependence on ${\cal C}_{3}$ it can be shown that $\omega^{IJ}$ will additionally depend upon derivatives of $\phi$.
However, if it is ${\cal C}_{1}$ and/or ${\cal C}_{2}$ which contain a dependence on $\phi$, it may be shown that one can no longer solve algebraically for all $\omega^{IJ}$: parts exist that obey their own differential equation of motion. In these theories then, parts of the spin-connection (specifically parts of the `contorsion form')  propagate and represent new degrees of freedom in the gravitational sector.

It is worth noting that the various terms in the action \eqref{4action} have already separately been explored in the literature:
\begin{itemize}
\item If it is only $\gamma$ that depends on $\phi$, then we recover the dynamical Immirzi parameter model of \cite{Taveras:2008yf,TorresGomez:2008fj,Calcagni:2009xz}.
\item If if is only only ${\cal C}_{1}$ that depends on $\phi$, then we recover the scalar-Euler form gravity model of \cite{Toloza:2013wi}. 
\item If it is only ${\cal C}_{2}$ that depends on $\phi$ then, we recover the first-order Chern-Simons modified gravity model of \cite{Alexander:2008wi,Alexander:2009tp}.
\item If it is only ${\cal C}_{3}$ that depends on $\phi$, then we recover the Nieh-Yan gravity model of \cite{Mercuri:2009zi}.
\end{itemize}
%

\section{Phenomenology}
\label{phenomenology}
We are now in a position to consider the physical content of Cartan gravity and highlight the novel phenomenology that arises when we treat the gravitational Higgs field $V^A(x)$ as a genuine dynamical field.
\subsection{Peebles-Ratra quintessence and the dynamics of $V^{A}$}
\label{prq}
The first step here is to get a handle on precisely how the new field in the gravitational sector ($V^{A}$) behaves. The problem to come up with a natural action where $V^A$ is itself a dynamical field was labeled an open problem \cite{Randono:2010cq} and has inspired attempts at providing an action where $V^A$ can be regarded as a dynamical field. One approach to this is to simply include an action built from a metric $g_{\mu\nu}$ identified with $P_{AB}D_{\mu}V^{A}D_{\nu}V^{B}$ or $\eta_{AB}D_{\mu}V^{A}D_{\nu}V^{B}$ as follows \cite{Stelle:1979va}:

\begin{eqnarray}
S= \int \left(-\frac{1}{V^{2}}g^{\mu\nu}\partial_{\mu}V^{2}\partial_{\nu}V^{2}-U(V^{2})\right) \sqrt{-g} d^{4}x
\end{eqnarray}
Where $U$ is to be chosen so as to have a minimum at a non-zero, positive value of $V^{2}$. Note though that the kinetic term for $V^{2}$ is highly non-polynomial in $V^{A}$ (and $A_{\mu}^{\ph{\mu}AB}$) and so it is seems unsuitable for potentially describing a phase of gravity where $V^{A}=0$ and the originally $SO(1,4)$ symmetry is unbroken.

We now show that an action yielding dynamics for $V^{2}$ can instead be found among the general class of polynomial actions (\ref{genaction}): those for which $\{a_{2},b_{1},b_{2},b_{3},c_{1}\}$ are non-zero \cite{Westman:2013mf}: they are actions which for $V^{2}\neq 0$ and for invertible metric $g_{\mu\nu}= \eta_{IJ}e^{I}_{\mu}e^{J}_{\nu}$  possess a second-order formulation in which the degree of freedom $\phi$ picks up a kinetic term of the form $g^{\mu\nu}\partial_{\mu}\phi \partial_{\nu}\phi$.
The addition of $a_{1}$ and a $V^{2}$-dependent $a_{3}$ term present complications which we will discuss further on. For ease of illustration we will demonstrate the second-order formulation of the theory in detail for the case where only $b_{1}$ and $b_{2}$ are non-zero. Therefore we concentrate on the action:

\begin{eqnarray}\label{PRaction}
S_{b_{1}b_{2}}[A^{AB},V^{A}] &=&  \int \left(b_{1}\epsilon_{ABCDE}V^{E}+b_{2}V_{A}V_{C}\eta_{BD}\right)DV^{A}DV^{B}F^{CD}
\end{eqnarray}
When $V^{2}>0$ we may make the gauge choice $V^{A}=\phi(x^{\mu})\delta^{A}_{\ph{A}4}$ and the action becomes:

\begin{eqnarray}
\label{b1b2om}
S_{b_{1}b_{2}}[\omega^{IJ},e^{I},\phi] = \int b_{1}\phi \epsilon_{IJKL}e^{I}e^{J}\left(R^{KL}-\frac{1}{\phi^{2}}e^{K}e^{L}\right)-\frac{b_{2}}{2}d\phi^{2} T^{I}e_{I} 
\end{eqnarray}
To progress, we can use a familiar technique from Einstein-Cartan gravity. It is convenient to decompose $\omega^{IJ}$ as follows:

\begin{eqnarray}
\label{decomom}
\omega^{IJ} =  \bar{\omega}^{IJ}(e)+C^{IJ}
\end{eqnarray}
where $\bar{\omega}$ is defined to be the solution to the equation $de^{I}+\bar{\omega}^{I}_{\ph{I}J}e^{J}=0$. We then have that 

\begin{eqnarray}
R^{IJ}(\omega^{KL}) &=& \bar{R}^{IJ}(\bar{\omega})+C^{I}_{\ph{I}K}C^{KJ}+D^{(\bar{\omega})}C^{IJ}\\
T^{I} &=& C^{IJ}e_{J}
\end{eqnarray}
Thus we see that $C^{IJ}$ carries the information about the torsion two-form $T^{I}$ and hence the decomposition (\ref{decomom})
allows us to split $R^{IJ}$ into a torsion-free part and a torsional part; the torsion-free part of $R^{IJ}$ (i.e. $\bar{R}^{IJ}(\bar{\omega})$ becomes the Riemmanian curvature associated with the Christoffel symbols $\Gamma^{\mu}_{\alpha\beta}(g)$ when written entirely in space-time components (see Appendix \ref{palappendix}). We may now write (\ref{b1b2om}) as a functional of $C^{IJ}$, $\phi$, and $e^{I}$ , with a non-polynomial dependence on $e^{I}$ and we have that, up to boundary terms:

\begin{eqnarray}
S_{b_{1}b_{2}}[C^{IJ},e^{I},\phi] &=&  \int  b_{1}\phi\epsilon_{IJKL} e^{I}e^{J}\left(\bar{R}^{KL}-\frac{1}{\phi^{2}}e^{K}e^{L}+C^{K}_{\ph{K}M}C^{ML}\right)\nn \\
&&-\frac{1}{2}\left(\frac{1}{\phi}b_{1}\epsilon_{IJKL} - b_{2}\eta_{IK}\eta_{JL}\right)d\phi^{2} C^{IJ}e^{K}e^{L}
\end{eqnarray}
Then, varying with respect to $C^{IJ}$ we obtain an equation of motion for $C^{IJ}$ itself. If $a_{1}$ is zero and $a_{3}$ contains no dependence on $\phi$ then it can be shown that this equation contains no derivatives of $C^{IJ}$ and one can solve for $C^{IJ}$ algebraically yielding

\begin{eqnarray}
C_{IJ} &=&  \frac{1}{2\phi^{2}} e_{[I}\partial_{J]}\phi^{2}+ \frac{b_{2}}{8b_{1}\phi}\epsilon_{IJKL}\partial^{K}\phi^{2}e^{L}
\end{eqnarray}
where $\partial_{J} \equiv  e_{J}^{\mu}\partial_{\mu}$. As we have solved for $C_{IJ}$ algebraically, we may eliminate it from the variational principle by substituting into $S_{b_{1}b_{2}}[\phi,C_{IJ},e^{I}]$ to obtain the following action:

\begin{eqnarray}
S_{b_{1}b_{2}}[e^{I},\phi] &=&  \int  b_{1}\phi\epsilon_{IJKL}e^{I}e^{J}\left(\bar{R}^{KL}-e^{K}e^{L}\right)\nn \\
&&+\epsilon_{IJKL}\frac{b_{1}}{16\phi^{3}}\left(1-\frac{b_{2}^{2}\phi^{2}}{4b_{1}^{2}}\right)\left(6\partial_{M}\phi^{2}\partial^{J}\phi^{2}e^{I}e^{M}-\partial^{M}\phi^{2}\partial_{M}\phi^{2}e^{I}e^{J}\right)e^{K}e^{L}
\end{eqnarray}
Then, following the steps illustrated in the Appendices, we may write the subsequent action in metric variables:

\begin{eqnarray}
S_{b_{1}b_{2}}[g_{\mu\nu},\phi] &=&  \int b_{1}\left(2\bar{R}[g_{\mu\nu}]+\frac{3}{\phi^{2}}\left(1-\frac{b_{2}^{2}\phi^{2}}{4b_{1}^{2}}\right)g^{\mu\nu}\partial_{\mu}\phi\partial_{\nu}\phi-\frac{24\sigma}{\phi^{2}}\right)\phi \sqrt{-g} d^{4}x
\end{eqnarray}
where $\bar{R}[g_{\mu\nu}]$ is the Ricci scalar corresponding to $g_{\mu\nu}$. If we define a new metric tensor $h_{\mu\nu}=  (\phi /\phi_{0})g_{\mu\nu}$ (where $\phi_{0}$ is an arbitrary constant with the same dimensionality to $\phi$) then we have, up to a boundary term:

\begin{eqnarray}
S_{b_{1}b_{2}}[h_{\mu\nu},\phi] &=&  \int  \left(2b_{1}\phi_{0}\bar{R}[h_{\mu\nu}]-\frac{3\phi_{0}b_{2}^{2}}{4b_{1}}h^{\mu\nu}\partial_{\mu}\phi\partial_{\nu}\phi-\frac{24b_{1}\phi_{0}^{2}}{\phi^{3}}\right)\sqrt{-h}d^{4}x
\end{eqnarray}
Therefore the $b_{1}b_{2}$ model unifies General Relativity with a Peebles-Ratra (i.e. $V(\phi) \propto \phi^{-n}$) quintessence model \cite{Ratra:1987rm}. The case where all of $\{a_{2},b_{1},b_{2},b_{3},c_{1}\}$ are non-zero was considered in \cite{Westman:2013mf} and a similar result is recovered i.e. one can still solve for the field $C_{IJ}$ and eliminate it from the variational problem; the resulting second-order theory can be written as General Relativity plus a scalar field with canonical kinetic term and a potential term. We see then that `dynamics' for $\phi$-in the sense of the existence of a Klein-Gordon type kinetic term in the second-order formulation of the theory- is due to space-time gradients in the $\phi$ field sourcing torsion  \footnote{There exist claims \cite{Chamseddine:1977ih} that a particular sub-class of polynomial actions (specifically a combination of $a_{1}$, $b_{1}$, and $c_{1}$ terms) yield a second-order scalar-tensor theory description for  small perturbations of freely varied fields $A^{ab}$ and $V^{a}$ around a Minkowski space geometry for $\bar{g}_{\mu\nu}$ in a symmetry broken-phase where $V^{a}=\phi(x^{\mu})\delta^{a}_{4}$. However, for these actions the resulting scalar excitation's kinetic term is removed entirely by conformal transformation to the frame in which the tensor perturbation is described by the Einstein-Hilbert action i.e. the perturbed system is that of General Relativity alongside a scalar field $\varphi$ appearing algebraically and coupling only to the determinant of the metric. }.

\subsection{Inevitable dynamics of dark energy and the scalar field potential}
\label{darkdynamics}
We now note another interesting property of (\ref{4action}). The first term in the action can be made equal to the familiar Palatini action by a conformal rescaling of the co-tetrad $e^{I} = \sqrt{G(\phi)/G_{0}} \tilde{e}^{I}$  where $G_{0}$ is a constant with dimensionality the same as Newton's constant. The first two terms in the action are 

\begin{eqnarray}
S_{G,\Lambda}[\tilde{e}^{I},\omega^{IJ},\phi] = \int \frac{1}{32\pi G_{0}}\left(\tilde{e}^{I}\tilde{e}^{J}R^{KL}-\frac{\tilde{\Lambda}(\phi)}{6}\tilde{e}^{I}\tilde{e}^{J}\tilde{e}^{K}\tilde{e}^{L}\right) + \dots
\end{eqnarray}
where

\begin{eqnarray}
\tilde{\Lambda}(\phi)=  -\frac{6}{32\pi G_{0}}\frac{\left(a_{1}-b_{1}\phi^{2}+c_{1}\phi^{4}\right)}{\phi\left(2a_{1}-b_{1}\phi^{2}\right)^{2}} 
\label{lambdaeq}
\end{eqnarray}
where recall that we have allowed $a_{1},b_{1},c_{1}$ in principle to depend polynomially on $\phi^{2}$.
Thus we see that it is impossible to recover a constant cosmological term when any of the coefficients of (\ref{genaction}) have a polynomial dependence upon the norm $V^{2}$; consequently, any effective cosmological constant in Cartan gravity must arise from the dynamics of the field $V^{A}$. A `bare' cosmological term is forbidden by the requirement of a polynomial action for the gravitational variables. We shall see this remains the case if one adds in scalar and spinorial fields described by polynomial actions and considers constant contributions to what become their potential terms in the General Relativistic limit; the dynamics of $\phi$ will inevitably inform under what circumstances, if at all, such a limit is reached.

If we can assume that the curvature-squared ${\cal C}_{1}(\phi)$ and ${\cal C}_{3}(\phi)$ terms are negligible and other sources of torsion can be ignored then the action (\ref{4action}) can be written in a second order formalism as General Relativity- described by the Einstein-Hilbert action- coupled to matter and a scalar field $\phi$ described by the following action \cite{Westman:2013mf}: 

\begin{equation}
S_{\phi}=\int \left(- f(\phi)h^{\mu\nu}\partial_{\mu}\phi\partial_{\nu}\phi  -{\cal U}(\phi)-{\cal Y}(\chi^{M},\phi)\right)\sqrt{-h}d^{4}x
\end{equation}
where $f(\phi)$ is a complicated function dependent upon the theory's parameters, ${\cal Y}(\chi^{M},\phi)$ represents possible coupling between matter fields $\chi^{M}$ and $\phi$, and where 

\begin{eqnarray}
{\cal U}(\phi) \equiv k \frac{(a_{1}-b_{1}\phi^{2}+c_{1}\phi^{4})}{\phi(2a_{1}-b_{1}\phi^{2})^{2}}
\end{eqnarray}
We note then that if the effect of ${\cal C}_{1}$ and ${\cal C}_{3}$ are negligible then in some `background' of matter fields $\bar{\chi}^{M}$,  the scalar field can be attracted to settle at a value $\phi_{0}$ if it exists as a solution to the equation 

\begin{equation}
\frac{\partial \left({\cal U}(\phi)+{\cal Y}(\bar{\chi}^{M},\phi)\right)}{\partial (\phi)}|_{\phi=\phi_{0}} =  0
\end{equation}
and the inequality

\begin{equation}
\frac{\partial^{2} \left({\cal U}(\phi)+{\cal Y}(\bar{\chi}^{M},\phi)\right)}{\partial (\phi)^{2}} |_{\phi=\phi_{0}}>  0
\end{equation}
where $f(\phi)$ is assumed to be positive-definite, so giving the scalar field a right-sign kinetic term (see \cite{Westman:2013mf} for the explicit form of this function). Thus, in this regime the General Relativistic limit $V^{2}\rightarrow const.$ is seen to be approached dynamically within Cartan gravity.
As we shall see in Section \ref{cosmo}, the dynamics of the system can be considerably more exotic when the effect of ${\cal C}_{1}$ (via the $a_{1}$ term) is taken into account.

\subsection{Propagating Torsion}
\label{proptor}
In the polynomial action \eqref{genaction} the two terms corresponding to $a_1$ and $a_3$ are structurally different from the other terms. The complication introduced by the $a_{1}$ and $a_{3}$ terms (only if $a_{3}$ depends on $\phi^{2}$)  is that space-time gradients of $\phi$ can also couple to \emph{derivatives} of $C_{IJ}$ in the $C_{IJ}$ equation of motion, hence in this case $C^{IJ}$ cannot generally be solved for algebraically in terms of $e^{I}$ and $\phi$ (and derivatives thereof) and hence eliminated from the variational principle. Specifically, such terms are \cite{Westman:2013mf}:

\begin{eqnarray}
-2\int \left(\left(\frac{\partial a_{1}}{\partial \phi^{2}}+\frac{a_{1}}{2\phi^{2}}\right)\phi \epsilon_{IJKL}+\frac{\partial a_{3}}{\partial \phi^{2}} \eta_{IK}\eta_{JL}\right)d\phi^{2}C^{IJ}\left(\bar{R}^{KL}-\frac{1}{\phi^{2}}e^{K}e^{L}+\frac{1}{3}C^{K}_{\ph{K}L}C^{LJ}+\frac{1}{2}D^{(\bar{\omega})}C^{KL}\right)
\end{eqnarray}
Due to the total antisymmetry of components of differential forms, we see there can only be time derivatives of $C^{IJ}$  (via the term $D^{(\bar{\omega})}C^{KL}=dC^{KL}+\omega^{K}_{\ph{K}M}C^{ML}+\omega^{L}_{\ph{L}M}C^{KM}$) only when there are spatial gradients of $\phi$. This appearance of derivatives of $C^{IJ}$ in the action is rather different from the case of Poincar\'e gauge theory where such terms can arise from actions built from $e^{\mu}_{I}$ (which of course is non-polynomial in $e^{I}_{\mu}$) coupled to terms quadratic in the torsion two-form   \cite{Chen:2009at}. The behaviour of these potential new degrees of freedom in Cartan gravity remains an open question.
\subsection{Cosmological Solutions, the Hartle-Hawking no-boundary proposal and signature change}
\label{cosmo}
\begin{figure}[H]
\centering
\includegraphics[width=7cm]{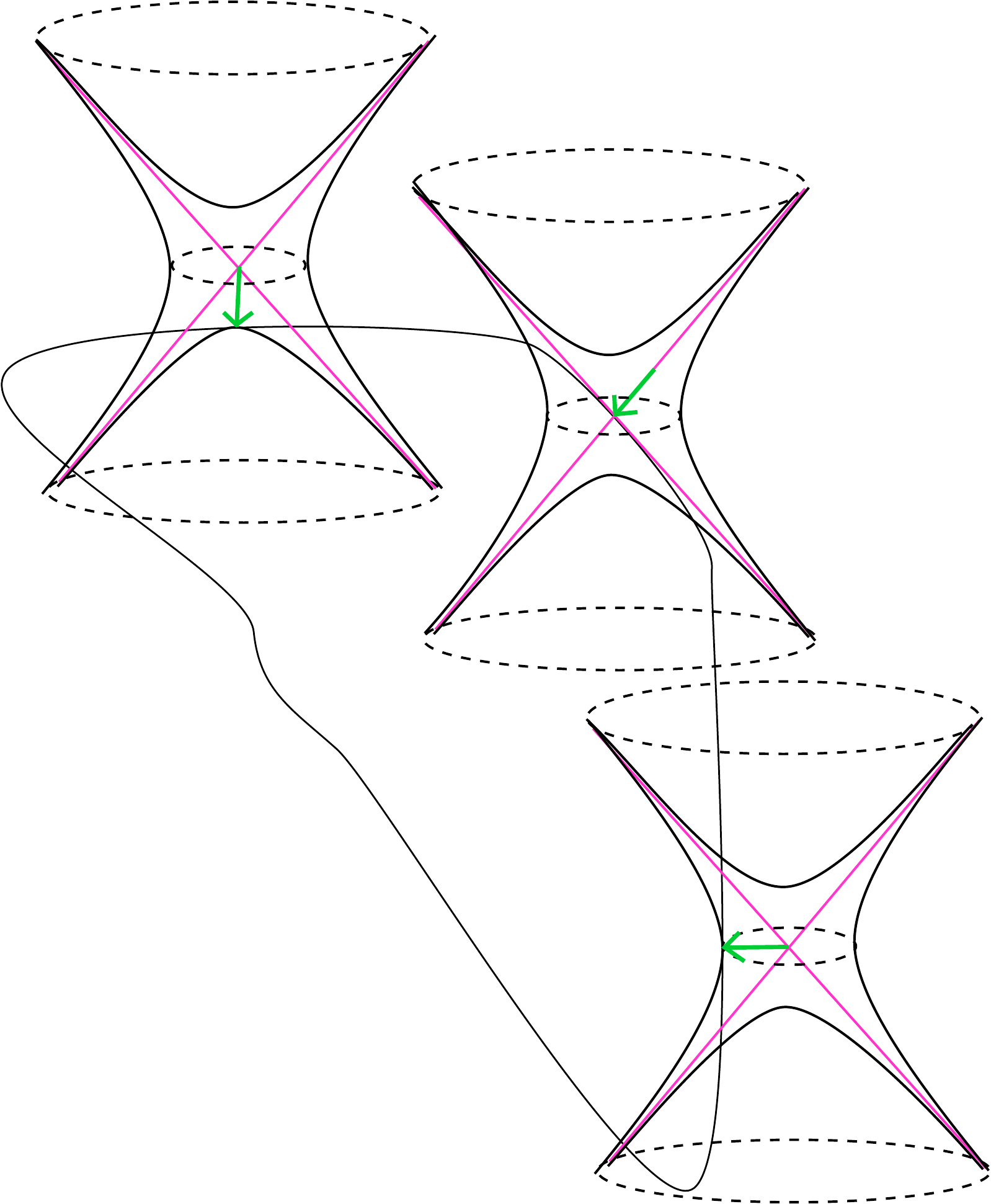} 
\caption{If we treat $V^A$ as a genuine dynamical field, no restrictions can be made on the sign of $V^2$ other than what the equations of motion dictate. In this figure we illustrate, using an embedding Lorentzian space $\mathbb{R}^{(1,2)}$, a region of a manifold ${\cal M}$ in which signature change happens. The green arrows represent (from left to right) the contact vector which is timelike, null, or spacelike. The `space being rolled' changes (from left to right) from the lower sheet of a hyperboloid of two sheets, to a null cone, to a hyperboloid of one sheet. In contrast to the metric formulation of gravity, signature change occurs naturally and smoothly and no {\em ad hoc} conditions need to be imposed. Instead, the equations of motion dictates the process completely. Note that signature change in the Einstein-Cartan theory is impossible as the metric always has the signature of the matrix $\eta_{IJ}$ }
 \label{siggy}
\end{figure}

If we truly regard $V^A$ as a genuine dynamical degree of freedom we are obliged to let the equations of motion dictate its behaviour. In principle this may  allow for solutions where $V^2$ changes sign or even  where the field vanishes altogether. In Fig. \ref{siggy} we illustrate the signature change process by making use of an embedding space.

Here we report on the cosmological solutions with FRW symmetry studied in \cite{Magueijo:2013yya}.  
As we have seen, for the group $SO(1,4)$ It is easy to show that the stabilizer group is $SO(1,3)$ whenever $V^2>0$ and $SO(4)$ whenever $V^2<0$. Thus, the signature of spacetime is not always Lorentzian $(-,+,+,+)$ but can also be Euclidean $(+,+,+,+)$. Indeed, once we have allowed the gravitational Higgs field $V^a$ to be a genuinely dynamical field, there is no guarantee that $V^2$ will be positive. It is then plausible that Cartan gravity permits signature change. 

\begin{figure}[H]
\centering
\includegraphics[width=8cm]{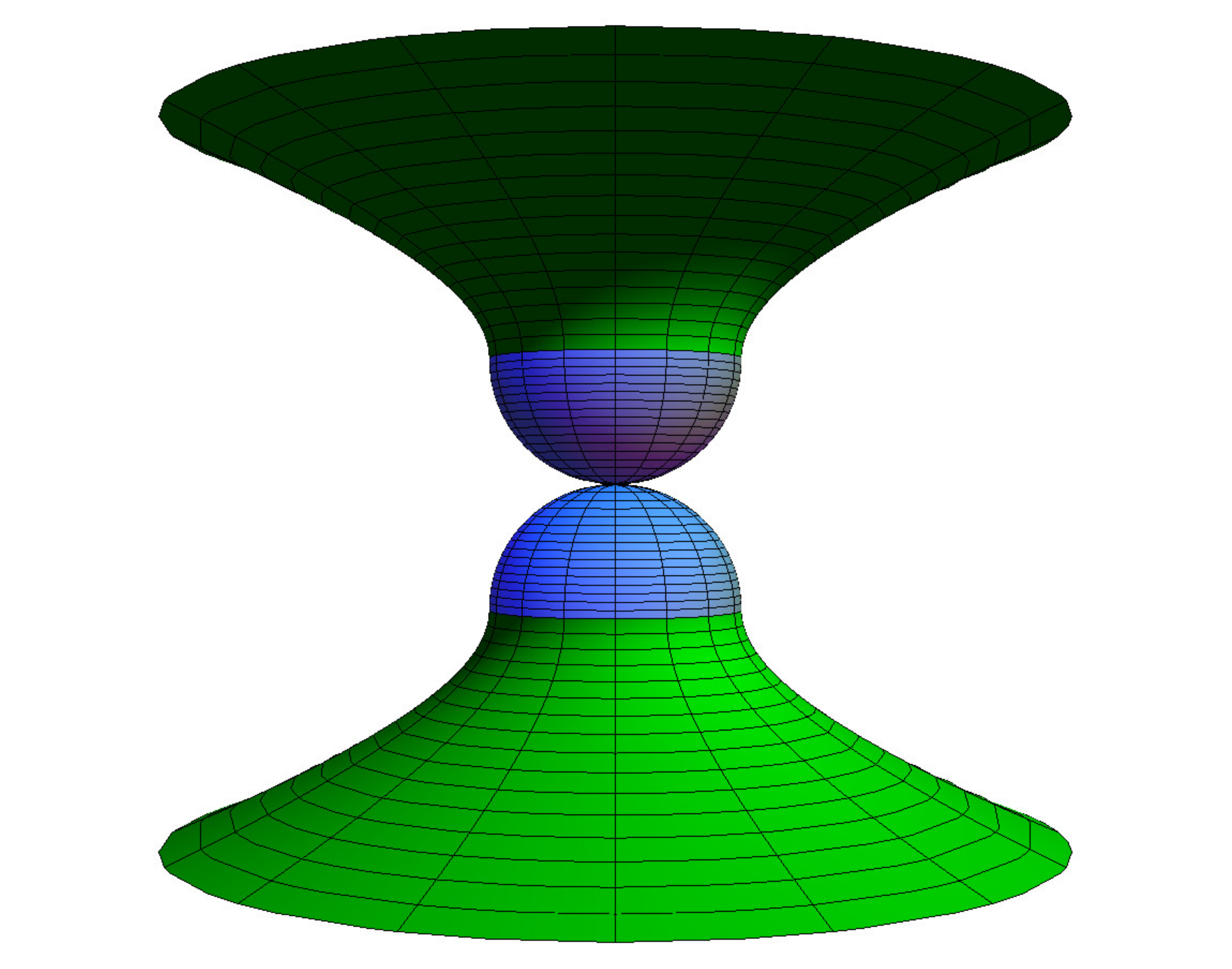} 
\caption{ Plotted is a solution to the equations of motion obtained from the MacDowell-Mansouri action with FRW symmetry imposed. The blue regions have Euclidean signature metric and the green Lorentzian. The whole manifold consists of Euclidean pole-less four-sphere hemispheres attached to halves of the de Sitter spacetime and joined via hyper-surface where the spatial triad is vanishing. Apart from the fact that the polynomial field equations allow for analytical extension through the event where the radius of the universe is zero (where the South and North poles join), this constitutes an exact classical realization of the Hartle-Hawking no-boundary proposal. Solutions with a deficit angle at the South Pole also exist and correspond to non-zero values for the contorsion form.
}
\label{fig:betag0}
\end{figure}

An intriguing and surprising consequence of the action \eqref{genaction} with only $a_{1}$ non-zero (henceforth called the Macdowell-Mansouri action due to it corresponding in the limit $V^{2}\rightarrow const.$ and in the standard gauge to the action proposed by Macdowell and Mansouri \cite{MacDowell:1977jt}) is that it generally provides solutions strongly reminiscent of the Hartle-Hawking no-boundary proposal. See Fig. \ref{fig:betag0}. This comes about by simply solving the equations of motion after imposing FRW symmetry.
This result makes for an amusing, but mathematically supported, answer to Hawking's question \cite{hawking1993brief}: what is South of the South Pole?  Answer: possibly another North Pole and a pre-big bang universe! The MacDowell-Mansouri action is polynomial in the basic variables and allows for an analytical continuation through the moment when the cosmic scale factor is zero. In fact, in Cartan gravity zero size of the universe need not correspond to a singularity in the sense that the equations of motion break down. Note that the above solution follows from the assumption that the spacetime manifold has the topology $\mathbb{R}\times \mathbb{S}^{3}$. If, instead, we had simply assumed that the topology of the manifold in the `upper' Euclidean regime was that of $\mathbb{S}^{4}$, we should identify the moment of `zero size' as the familiar coordinate singularity at poles in spherical coordinates yielding a solution genuinely consisting of a hemisphere of a four-sphere (the Euclidean regime) attached to half of de Sitter space (the Lorentzian regime). Clearly then, the choice of topology of the spacetime manifold can significantly affect the nature of solutions.

\begin{figure}[H]
\centering
\includegraphics[width=13cm]{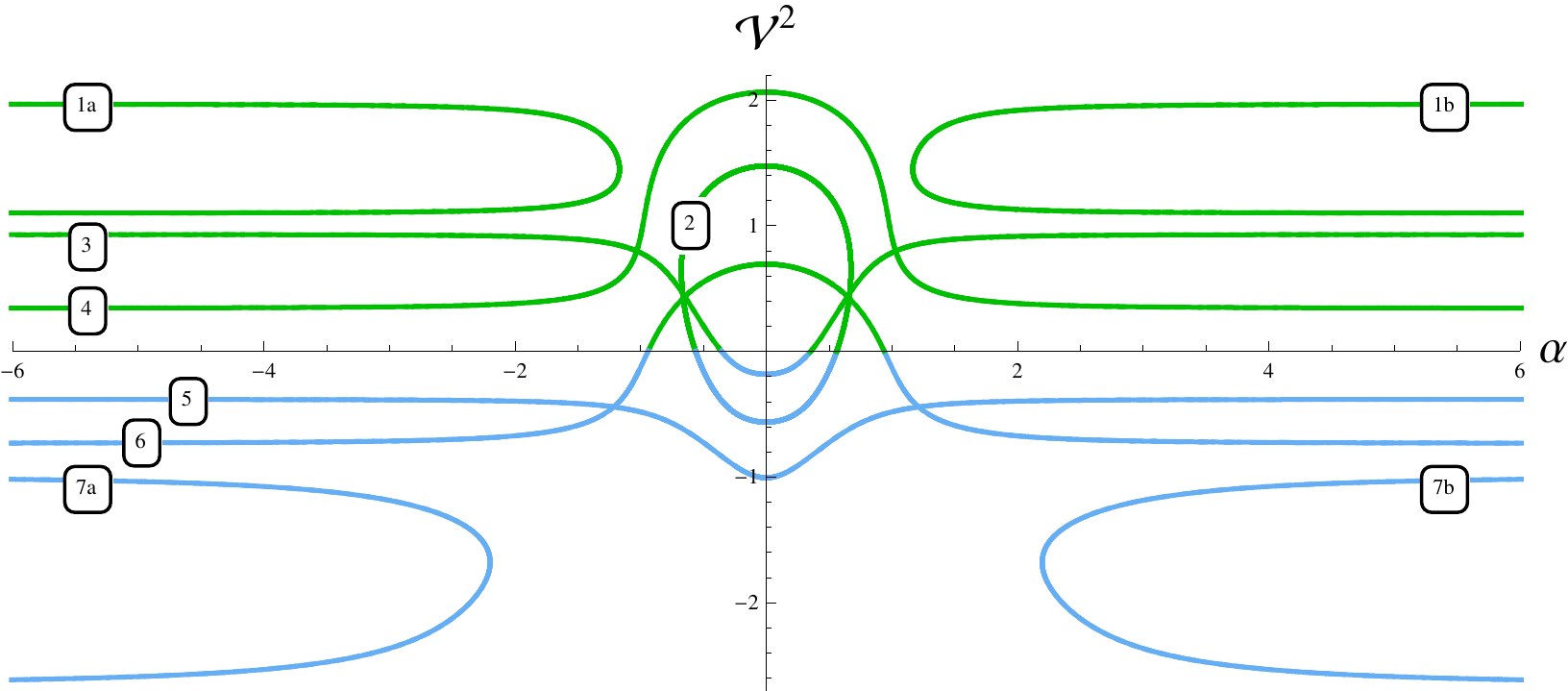} 
\caption{Parametric plot displaying a sample of the solutions to the $(a_1,b_2)$ system for spatially closed FRW geometries. $\alpha$ is a rescaled dimensionless radius of the universe $\alpha(t)$ and $\m V^2(t)$ the rescaled dimensionless norm of $V^2$. All but one solution exhibit the behaviour $\m V^2\rightarrow const$. Blue represents Euclidean regimes and green Lorentzian. No singular solutions were found.}
\label{fig:solspace}
\end{figure}

By way of comparison to other theories of gravity, we note that signature change has also been studied within General Relativity. However, in that context signature change does not occur naturally but must be imposed by hand. To quote Ellis et al. \cite{Ellis:1991st}: 
\begin{quote}
{\em ``The Einstein field equations by themselves do not determine the spacetime signature; that is imposed as an extra assumption.''}
\end{quote}
In contrast, as we have demonstrated, in Cartan gravity the signature change need not be imposed but happens naturally in the sense that the equations of motions predict it.
Does signature-change persist if we `switch-on' other constants in the polynomial action \eqref{genaction}?  A model that was explored in considerable detail was that 
when $a_{1}$ and $b_{2}$ only are non-zero, leading to the possible evolutions depicted in Fig. \ref{fig:solspace}. A detailed discussion of the various numbered cases is provided in \cite{Magueijo:2013yya}. Clearly signature change persists for some of the parameter space (i.e. for some choices of constants and initial data). As mentioned in Subsection \ref{darkdynamics}, a large number of solutions asymptote to $V^{2}\rightarrow const.$ (to the far future and far past of an intermediate cosmological bounce or signature change) indicating a General Relativistic limit being approached. This is intriguing behaviour because for the `$(a_{1},b_{2})$' action the `potential term' provided by (\ref{lambdaeq}) has a $1/\phi$ form and therefore would seem to combat the stabilization of $\phi(t)$ for finite values of $\phi$. Some other mechanism must be at work but it is not simple to cast this system in second-order form. Evident in Fig. \ref{fig:solspace} are a number of novel solutions:  bouncing universes which asymptote to
different values of $V^{2}$ and hence the effective cosmological constant (Type 1 solutions); universes which contract and non-singularly pass
through $\alpha(t_{0})=0$ before a period of infinite expansion (Type 4 solutions); signature change solutions reminiscent of Fig. \label{fig: betag0} (Type 3 solutions); and a solution which eternally oscillates between Euclidean and Lorentzian regimes (Type 2 solutions). There appears at this level to be a symmetry between solution types under exchange of metric signature (i.e . solution types 1,3,4 in Fig.  \ref{fig:solspace} are mirrored by solutions 5,6,7). 

\begin{figure}[H]
\centering
\includegraphics[width=16cm]{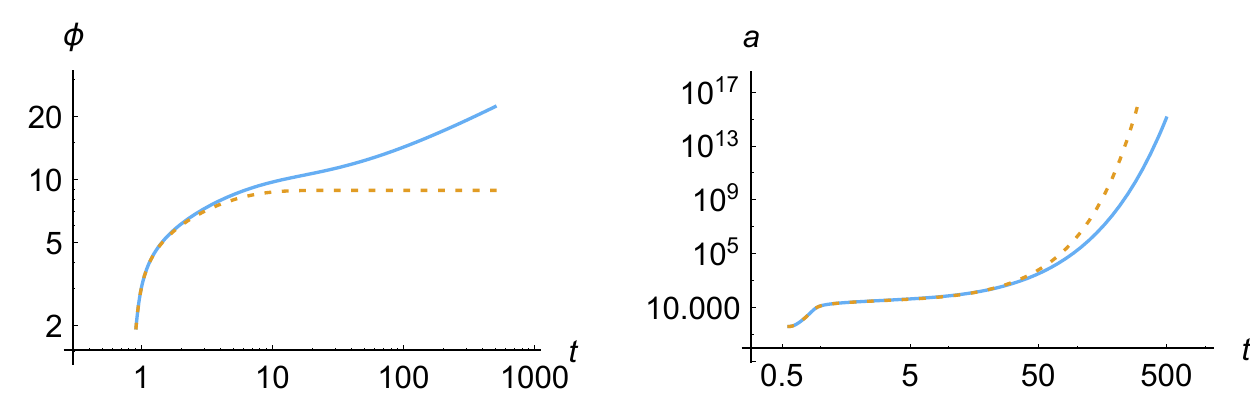} 
\caption{An illustration of the effect of adding on a $b_{1}$ term to the $(a_{1},b_{2})$ system.
The dotted lines represent evolution of $\phi(t)$ and $a(t)$ for the $(a_{1},b_{2})$ system,  corresponding to solution Type 3 from Figure \ref{fig:solspace}. The solid lines represent solutions with the same initial data and values of $(a_{1},b_{2})$ but with the addition of a non-zero $b_{2}$ term in the action. We see that this term can overcome the tendency of $\phi$ towards constancy at late times. Indeed at late times the behaviour is that of the Peebles-Ratra rolling quintessence.}
\label{fig:competition}
\end{figure}

Numerical evolution was explored for other parameters being non-zero beyond $a_{1}$ and $b_{2}$ and for various initial data \cite{Magueijo:2013yya}. For example, otherwise singular-behaviour in the Peebles-Ratra quintessence model of Subsection \ref{prq} appeared avoidable if the $a_{1}$ term was additionally present; rather, big-bang type singularities could be replaced by regions displaying signature change in the manner of Fig. \ref{fig:betag0}, indicating dominance of the $a_{1}$ term at high-curvature. An example of the effect of adding a $b_{1}$ term to the $(a_{1},b_{2})$ system is shown in Figure \ref{fig:competition} wherein the influence of the $a_{1}$ term would appear to dominate at early times (non-singular signature change behaviour is present) whereas at late times the tendency of $a_{1}$ and $b_{2}$ to lead to constancy of $V^{2}$ seems to be overridden by the $b_{1}$ term, with solutions instead asymptoting to the pure Peebles-Ratra quintessence of the $(b_{1},b_{2})$ model at late times. However, given the complexity of the $(a_{1},b_{2})$ system solution space, the $(a_{1},b_{1},b_{2})$ solution space will be yet more complicated. 

In summary, solutions have been explored in FRW symmetry for some of its parameter space. Clearly there are a wide range of possible solutions even in this restricted region. A next step would be to widen the exploration of the background behaviour and look at the evolution of perturbations to the gravitational fields on top of these backgrounds. Detailed comparison to the universe today would require inclusion of matter fields. In the next section we discuss the coupling of gravity to matter in Cartan gravity. 

\subsection{Coupling to matter fields}
\label{matter}
As shown in \cite{Westman:2012zk}, requiring manifest gauge invariance and {\em only} polynomial matter actions forces all field equations to be {\em first order} partial differential equations. This section highlights the main results. All Cartan-geometric matter actions exhibited below reduce to standard matter actions in the Lorentzian signature General Relativistic limit $V^2\rightarrow const.>0$. However, outside that limit the gravity-matter coupling is highly non-trivial.

Gravity in a Cartan-geometric formulation is nothing but a symmetry-broken Yang-Mills gauge theory. This strongly suggests that the coupling to matter fields should be similar to how a gauge field couples to scalar and spinor fields. According to the gauge prescription the matter field coupled to the gravitational field should always carry a $SO(1,4)$ gauge index and the partial derivatives should be replaced by a $SO(1,4)$ gauge covariant one. However, this is clearly not the case for the standard representation of matter fields. The scalar and Yang-Mills fields have no $SO(1,4)$ index and a Dirac spinor $\psi$ carries a (suppressed) $Spin(1,3$) but not an $Spin(1,4)$ index. In this section we shall show how one may couple matter fields to gravity in accordance with the gauge prescription. Interestingly, the Hodge dual pops up only when we reformulate the first order equations into second order ones.

 Note that this is not always the case in Einstein-Cartan theory. There, for example, spinor fields indeed carry a $Spin(1,3)$ ($Spin(1,3)$ being the double cover of $SO(1,3)$) index; hence the $Spin(1,3)$ covariant derivative contains the spin-connection $\omega_{\mu}^{\ph{\mu}IJ}$ and spinor fields can source torsion via this coupling. When it comes to scalar fields however, there is no $SO(1,3)$ index; rather the scalar field couples to gravity only via the volume-form and inverse-metric formed from $e^{I}_{\mu}$.
\subsubsection{Dirac fields}
We first consider the case of spinor fields coupled to Cartan-gravitational fields.  As the action must possess $SO(1,4)$ symmetry, the spinor fields must rather be representations of the group $Spin(1,4)$. As opposed to the case of $Spin(1,3)=SL(2,\mathbb{C})$, there are no two-dimensional representations of $Spin(1,4)$; rather one must consider a four-dimensional representation for which the generators are $\m J_{AB}= -\frac{i}{4}\left[\Gamma_{A},\Gamma_{B}\right]$ where $\Gamma_{A}=  (\gamma_{I},i\gamma_{5})$ and the $\Gamma_{A}$ satisfy the $Spin(1,4)$ Clifford algebra:

\begin{eqnarray}
\{\Gamma_{A},\Gamma_{B}\}= -2\eta_{AB}
\end{eqnarray}
We find that  the action \cite{Westman:2012zk} 
\begin{align}
S_{spinor}=\int\epsilon_{ABCDE}V^E DV^A  DV^B DV^C \left(\frac{i}{2}(\bar\psi\Gamma^DD\psi-D\bar\psi\Gamma^D\psi)-mDV^D\bar\psi\psi\right)
\label{diracact}
\end{align}
reduces to the standard  Dirac action in the General Relativistic limit. In accordance with the gauge prescription we have $D\psi = d\psi -\frac{i}{2}A^{AB}\m J_{AB}\psi$  and $\bar{\psi} \equiv \psi^\dagger\gamma_{0}$. Thus Dirac spinors can easily be accommodated in a harmonious way within a $Spin(1,4)$ Cartan-geometric description of gravity.
\subsubsection{Parity and the electroweak theory}
In the electroweak sector of the standard model of particle physics we have parity violation. The weak force discriminates between whether a spinor field is a left-handed representation of $SL(2,\mathbb C)$ or a right-handed one. The left-handed electron-neutrino $\ell^{\cal A}_{L}$ is a weak isospin doublet (${\cal A}$ is an $SU(2)$ index) and right-handed electron is a weak isospin singlet $e_{R}$. In a description based on $Spin(1,4)$ we are dealing with four-dimensional irreducible representations rather than the two-component Weyl spinors. Thus, it is clear that we need a way to isolate the left and right-handed components of the four-dimensional irreducible representations. To this end we may introduce the chiral projectors 
\begin{align}
P_L=\frac{1}{2}\left(1+\frac{i}{\sqrt{|V^2|}}V_A\Gamma^A\right)\qquad P_R=\frac{1}{2}\left(1-\frac{i}{\sqrt{|V^2|}}V_A\Gamma^A\right)
\end{align}
in terms of which we may now write down the electroweak action.

However, this does not seem harmonious with the underlying Cartan-geometric structure. First of all we note that the projectors contain inverses which are not well defined when $V^2=0$. Secondly, to combine $e_R$ with $\ell_L^{\m A}$ seems problematic due to the difference in index structure: $\ell_L^{\m A}$ has an $SU(2)$ index but $e_R$ does not. This observation may be taken to suggest a more speculative possibility: $e_{R}$ could be the right-handed part of a four-component spinor 
\begin{equation}
E=(\chi_{L},e_{R})
\end{equation}
where $\chi_{L}$ would be a left-handed spinor representing some new particle in nature. Similarly, $\ell^{\cal A}_{L}$ would become part of an object
\begin{equation}
{\cal L}^{\cal A}=(\ell^{\cal A}_{L},\chi^{{\cal A}}_{R})
\end{equation}
where the right-handed isospin doublet $\chi^{{\cal A}}_{R}$ is unknown to the standard model.

Regarding the nature of $\chi_{L}$ and $\chi^{{\cal A}}_{R}$, naively these fields and their particles may have thusfar eluded detection if too massive to be created in current experiments. Recall that in the standard model of particle physics, the mass of the electron comes from the Yukawa-type interaction term $({\phi^{\dagger}}_{\cal A}\ell^{\cal A}_{l}e^{\dagger}_{r}+C.C)$, where $\phi^{\cal A}$ is the electroweak Higgs boson. In the $Spin(1,4)$ case, we can also have Dirac mass terms for $E$ and ${\cal L}^{\cal A}$ and so collectively we may have the following mass terms:
\begin{equation}
\phi^{\dagger}_{{\cal A}}\bar{E}L^{{\cal A}}+C.C\quad ; \quad m_{1}\bar{E}E \quad;\quad  m_{2}\bar{L}_{{\cal A}}L^{\cal A}
\end{equation}
It has been argued that considerable fine-tuning would be needed for these terms to render the new fermionic degrees of freedom too massive to have been observed \cite{Chamseddine:2013hwa}.

\subsubsection{Scalar fields}
As mentioned above, scalar field $\Phi$ in the Einstein-Cartan or metric formalism has no index related to the gravitational gauge group $SO(1,3)$. In accordance with the gauge prescription in Cartan gravity we rather seek to attach a gravitational index to the scalar field, i.e. $\Phi\rightarrow\Phi^A$. This might seem puzzling since we have now have {\em five} scalar fields rather than one. However, gauge invariance and requiring the action to be polynomial forces the field equations to be {\em first order} partial differential equations. In fact, the  action
\begin{eqnarray}\label{HCaction}
S_{scalar}=\int \epsilon_{ABCDEF}V^E DV^A  DV^B  DV^C \left((\Phi^{\dagger D}D\Phi^F+D\Phi^{\dagger F}\Phi^D)V_F+\frac{1}{2}DV^D U(\Phi^{A}V_{A},\Phi^{A}\Phi_{A})\right)
\end{eqnarray}
for a complex $SO(1,4)$ vector is structurally similar to the Dirac action and reduces to the standard Klein-Gordon equation (with the term $U$ containing what eventually become `potential' terms) in the General Relativistic limit $V^2\rightarrow const.>0$. This comes about in the following way: at the level of the equations of motion in the standard gauge $V^{A}=\phi \delta^{A}_{\ph{A}4}$ , the four first components $\Phi^I$ ($I=0,\dots,3$) turn out to be nothing but the spatio-temporal derivatives of the fourth component $\Phi^4$. When we substitute in that solution in the action we arrive at the standard Klein-Gordon action. Thus, in the General Relativistic limit this action is equivalent on-shell to the standard Klein-Gordon action for the projection of $\Phi^{A}$ along $V^{A}$:  $\Phi^{A}V_{A}$.

We stress that, although very simple and natural within the Cartan-geometric formulation, the coupling between this scalar field and gravity looks in a second order Riemannian formulation rather contrived and unnatural outside the General Relativistic limit.
\subsubsection{Yang-Mills fields}
Aside from scalar fields and spinor fields, there are of course Yang-Mills fields related to non-gravitational symmetries.
The conventional, second-order formulation of Yang-Mills theory involves a Lagrangian term proportional to $g^{\alpha\beta}g^{\mu\nu}Tr \left(F_{\alpha\mu}F_{\beta\nu}\right)\sqrt{-g}$. One way to recover such a Lagrangian is to simply allow the presence of the inverse of the metric $g_{\mu\nu}=P_{AB}D_{\mu}V^{A}D_{\nu}V^{B}$ to exist in actions coupling gravity to other gauge fields. This of course would render the actions non-polynomial in both $V^{A}$ and $A^{AB}$. This approach was studied by Ha \cite{Ha:1995mj} and more recently Kerr \cite{Kerr:2014sma}. \footnote{In such a construction one may alternatively couple matter fields to a metric ${\cal G}_{\mu\nu}\equiv \eta_{AB}D_{\mu}V^{A}D_{\nu}V^{B} \overset{*}{=} g_{\mu\nu}+ \partial_{\mu}\phi\partial_{\nu} \phi$, thus obtaining a \emph{disformal} coupling between non-spinorial matter fields and the gravitational fields. The idea of disformal couplings has been an area of recent activity in cosmology  \cite{Magueijo:2008sx,Magueijo:2010zc,Zumalacarregui:2010wj,Koivisto:2008ak,Kaloper:2003yf,Bekenstein:2004ne,Skordis:2005xk}; it would be interesting to see whether variation of $V^{2}$ over spacetime may have a phenomenological role if such couplings are present.}

However, if we follow the same recipe as for the scalar field we should just add a gravity gauge index $B\rightarrow B^A=B_{\mu}^{\ph{\mu}A}dx^{\mu}$ (where internal indices have been suppressed for notational compactness) i.e. have the Yang-Mills gauge field transform as a vector under $SO(1,4)$ transformations. And similar to the case of the scalar field we find that the {\em first order} equations of motion for the action
\begin{eqnarray}
S_{Yang-Mills}=Tr\int\xi_{1}\epsilon_{ABCDE} V^E DV^A   DV^B B^C B^D+\xi_{2}V^{2}P_{AB}DV^{A} B^B G \label{ymc}
\end{eqnarray}
reduces to the standard Yang-Mills action in the General Relativistic limit. Here $\xi_{1}$ and $\xi_{2}$ are dimensionful constants and $G=dB+BB$ with $B=V_AB^A$. The Cartan-geometric Yang-Mills action is $SO(1,4)$ invariant under the skewed-looking gauge transformation

\begin{eqnarray}
B^{A} \rightarrow UB^{A}U^{-1}-\frac{i}{g}\frac{V^{A}}{V^{2}}dUU^{-1} 
\end{eqnarray}
where the matrix $U$ is an element of the Yang-Mills group in question. This indicates that $V_{A}B^{A}$ transforms precisely as a Yang-Mills field and should be identified with such a field. Meanwhile components of $B^{A}$ orthogonal to $V^{A}$ transform homogeneously under $U$;  indeed, inspection of the equations of motion following from varying (\ref{ymc}) with respect to $B^{A}$ shows that components of $B^{A}$ orthogonal to $V^{A}$ may be solved for in terms of $V_{C}B^{C}$ and its first derivatives. Substitution of these components into the action yields the familiar Yang-Mills action. The transformation property of $B^{A}$ under $U$ depending on the `direction' of $V^{A}$ is exotic. Is there a role to be played by more complicated objects such as one-forms $B^{AB}=B_{\mu}^{\ph{\mu}AB}dx^{\mu}$ with non-standard transformation properties under both gravitational and Yang-Mills transformations?

Alternatively, the formulation of gravity as a gauge theory may open the door to unification between gravity and some or all of the other forces of nature. For example one may consider theories built from an $SO(1,5)$ connection $A_{\mu}^{\ph{\mu}ab}$ (where we will briefly use $a,b,c,\dots$ to refer to $SO(1,5)$ indices) and `gravitational Higgs fields' that break the symmetry down to $SO(1,3)\times U(1)$; we could choose such fields to be $SO(1,5)$ vectors $V^{a}$ and $W^{a}$. If they are taken to satisfy $V^{a}V_{a} = const. >0$, $W^{a}W_{a} = const. >0$ and $W^{a}V_{a}=0$ (this may be achieved by Lagrange multipliers or conceivably by a dynamical mechanics as in the case of Cartan gravity). As in the case of Cartan gravity we may use these fields to define a projector ${\cal P}^{a}_{\ph{a}b} \equiv \delta^{a}_{\ph{a}b} - V^{a}V_{b}/V^{2} - W^{a}W_{b}/W^{2}$ which will covariantly project down to only the structure that transforms under residual Lorentz transformations after symmetry breaking. 
Then it may be shown \cite{Westman:2012zk} that the following polynomial action

\begin{eqnarray}
S[A^{ab},V^{a},W^{b}] &=&   \int \xi \epsilon_{abcdef}V^{e}W^{f}DV^{a}DV^{b}F^{cd}+ \chi V_{a}V_{c}W_{b}W_{d}F^{ab}F^{cd}
\end{eqnarray}
-where $\xi$ and $\chi$ are constants-reduces to the Palatini action of Einstein-Cartan theory plus a first order formulation of a Maxwell-type action i.e. in terms of one-forms ${\cal P}^{a}_{\ph{a}b}DW^{b}$ and $C \equiv A^{ab}W_{a}V_{b}/\sqrt{V^{2}W^{2}}$ which, respectively, play the role of $B_{\mu}^{\ph{\mu}I}$ and $B_{\mu}^{\ph{\mu}A}V_{A}$ in the above first-order formulation of gauge fields. However, it is unclear whether successful extension in this manner to the unification of gravity with non-Abelian gauge fields is possible. It seems likely that these approaches to unification will inevitably involve the soldering form $e^{I}_{\mu}$ being constructed from parts of the theory's gauge field and Higgs fields. An interesting alternative to this is to regard the existence of a generalised soldering-form to be basic in a putative unified theory \cite{Percacci:1984ai,Nesti:2009kk,Percacci:2009ij} i.e. a fundamental field $e_{\mu}^{{\cal A}}$ where ${\cal A}$ is an index in a vector representation of the `gravity+other gauge fields' unification group; this field is to play the role of $e_{\mu}^{I}$ after symmetry breaking.

Finally we note that though the matter actions here reduce to familiar ones in the limit of $V^{2}\rightarrow const. > 0$, we may equally look at the actions in the limit $V^{2}\rightarrow const. < 0$ i.e. in a regime of Euclidean metric signature. If there are certain instabilities in the Euclidean regime then this may suggest that Lorentzian signature (i.e. $V^{2}>0$) is dynamically preferred. For instance, it has been argued that electromagnetism in Euclidean-signature spacetime possesses instabilities not present in Minkowski spacetime \cite{Itin:2004qr}.
\subsection{How matter fields back-react on $A^{AB}$ and $V^A$}
Let us now discuss the back-reaction of matter fields on the descriptors of gravity, namely $V^A$ and $A^{AB}$. First we write
\begin{align}
S=S_{gravity}+S_{matter}
\end{align}
where $S_{gravity}$ is some specific action of the polynomial family \eqref{genaction} and $S_{matter}$ the sum of Cartan-geometric matter actions (i.e. scalar, spinor, and Yang-Mills type actions) discussed in the previous section. We can now define the objects $\m S_{AB}$ and $\m Q_A$
\begin{align}
\delta_{A}S_{matter}\equiv \m S_{AB}\delta A^{AB}\qquad \delta_{V}S_{matter}\equiv \m Q_A\delta V^A.
\end{align}
which serve as sources for $A^{AB}$ and $V^A$ together. The object $\m S_{AB}$ we call the spin-energy-momentum three-form and unifies the spin-density and energy-momentum density of the matter fields into a single Cartan-geometric object.

We now see that outside the General Relativistic limit $V^2=const.$ we have a non-trivial coupling between the matter fields to the gravitational variables $V^A$ and $A^{AB}$. Specifically, we see that dark energy is necessarily coupled to matter fields since $\m Q_A$ is non-zero for all matter fields be it scalar, spinor or gauge fields. The implications for cosmology for these non-trivial couplings of matter fields to dark energy should be investigated. Specifically, we must ask whether a natural Cartan geometric action for both gravity, dark energy, and matter fields exist that would be ruled out empirically. In addition, it is important to investigate whether {\em smooth} signature change is still possible with matter sources present. It would also be interesting to study perturbations around the Hartle-Hawking no-boundary solution to see whether the perturbations will propagate freely through the signature change hypersurface. We leave these here as open problems.
\section{Conclusions and outlook}\label{conclusions}
In its very essence Cartan geometry constitutes a mathematically distinct way of characterizing the geometry of a manifold which departs markedly from the more traditional Riemannian description. While the Riemannian approach puts strong emphasis on the metric tensor, Cartan geometry instead characterizes a geometry by how a symmetric space rotates when it is rolled without slipping along some path on the manifold. As such, the geometry of a manifold is no longer characterized by a metric tensor but instead by a pair of variables $\{V^i,A^{ij}\}$ which admits a crisp geometric interpretation in terms of idealized waywisers. The object $V^i$ represents the point of contact between the wheel of the waywiser, i.e. a symmetric space, and $A^{ij}$ dictates how much the wheel has rotated with rolled without slipping along some path on the manifold. Notably, the seemingly distinct geometric concepts of curvature and torsion are beautifully unified into a single object which is nothing but the curvature two form $F^{ij}=dA^{ij}+A^i_{\ph ik}A^{kj}$.

When generalized to the relativistic domain, accomplished by replacing the symmetric spaces with de Sitter spacetimes \footnote{The symmetric spaces can also be flat Minkowski spacetimes with associated gauge group $ISO(1,3)$ or the anti-de Sitter spacetime with the symmetry group $SO(2,3)$. Here we have focused on the de Sitter case.}, we find that Cartan geometry can be used to mathematically describe the gravitational field. In particular, we reproduce all of the predictions of General Relativity in the limit  where $V^2=const.$ One key aspect of Cartan gravity is the particular choice of representation for the Lie group $SO(1,4)$. In fact, when we write down action principles it is the fundamental representation of $SO(1,4)$ which is employed. This has the effect that a new scalar degree of freedom is naturally present the Cartan geometric description of gravity. This new degree of freedom is the invariant $V^2=\eta_{AB}V^AV^B$ which cannot be eliminated by a $SO(1,4)$ gauge transformation. Although it is tempting to regard this as an unwanted degree of freedom it is nonetheless there. It is commonplace to postulate this scalar quantity to be constant as a function on spacetime, i.e. $V^2(x)=const.$, thereby suppressing it as a possible new physical field in nature.

While mathematically unproblematic, the restriction $V^2(x)=const.$ is from a physicist's point of view rather {\em ad hoc} and unnatural. This feeling is greatly compounded when viewed alongside the electroweak theory whose dynamical variables mirrors that of Cartan gravity: the electroweak connection $W^{\alpha}_{\ph \alpha\beta}$ mirrors the rolling connection $A^{AB}$, and the symmetry breaking Higgs field $\Phi^\alpha$ mirrors the contact vector $V^A$. The quotient of the gauge group with the isometry group of the relevant Higgs field, i.e. $SU(2)_L\times U(1)_Y/U(1)_{EM}$ for the electro weak theory and $SO(1,4)/SO(1,3)$ for Cartan gravity, defines the Klein-geometries in both cases. In the case of $V^2>0$ the Klein geometry is the de Sitter spacetime and the three-sphere $S^3$ for the electroweak theory. From a geometric point of view we see that the electroweak theory is the theory of how to roll the three sphere on a four-dimensional manifold with $\Phi^\alpha$ representing the contact point and $W^\alpha_{\ph \alpha\beta}$ dictating how much the 3-sphere has rotated.\footnote{A spacetime metric can be defined by $h_{\mu\nu}=\delta_{\alpha\beta'}D^{(W)}_{\{\mu}\Phi^{\alpha} D^{(W)}_{\nu\}}\Phi^{*\beta'}$ (with $\delta_{\alpha\beta'}$ possibly replaced by the projector $P_{\alpha\beta'}=\left(\delta_{\alpha\beta'}-\frac{\Phi_{\alpha}\Phi^{*}_{\beta'}}{|\Phi|^2}\right)$ but since $h_{\mu\nu}(\Phi,W)\neq g_{\mu\nu}(V,A)$ in general the rolling of the three-sphere does not describe `rolling without slipping'. If the metric is defined using the projector $P_{\alpha\beta}$ it is easy to show that the metric is degenerate so that there exists a vector $v^\mu$ such that $g_{\mu\nu}v^\nu=0$. Thus, this Higgs metric does not admit an inverse. We can also define the anti-symmetric tensor $k_{\mu\nu}=i\delta_{\alpha\beta'}D^{(W)}_{[\mu}\Phi^{\alpha} D^{(W)}_{\nu]}\Phi^{*\beta'}$ whose geometric interpretation is perhaps not immediately clear.}

It now becomes rather strained to insist that while the quantum excitations of $|\Phi|^2$ play an integral part of the standard model of particle physics we must nonetheless postulate away the scalar field $V^2$ within Cartan gravity. 
restriction $V^2=const.$ It is therefore good news that the gravitational Higgs field can play the role of dark energy. Strong evidence for this comes from the fact that the simple polynomial action \eqref{PRaction} yields one of the most extensively studied models of dark energy, namely Peebles-Ratra slow rolling quintessence. This suggests that dark energy may be an integral component to Cartan gravity just as the Higgs field is an integral component of the standard model.

The Cartan-geometric description of gravity achieves something rather intriguing. Rather than driving a wedge between gravity and the electroweak and strong forces of the standard model, Cartan gravity highlights that gravity is `just another Yang-Mills field'. This opens up for study conceptually interesting questions: if gravity is nothing but another Yang-Mills field why does it end up playing the role of spacetime geometry? \footnote{The existence of a non-degenerate soldering map is necessary but not sufficient for why gravity should play the role of the geometry of spacetime. Specifically, this condition seems too weak to enforce the strong equivalence principle which is necessary for a geometric description of gravity.} What do we really mean when we say that it is possible to shield against the electromagnetic field but not gravity? Does gravity always act universally on all matter constituents and what is the ultimate fate of the equivalence principle \cite{PhysRevD.89.084053}? 

The geometry of space and time forms the very foundation of modern physics. It has been described as the dynamical stage on which all physics takes place. It is therefore not surprising that, if we change the mathematical representation of something so central as spacetime geometry, we will have ramifications on many aspects on modern physics and how it is mathematically formulated. In particular, we have seen that although a Riemannian metric formulation naturally operates with second order partial differential equations a Cartan geometric formulation is naturally a {\em first order} one. By insisting that gravity couples to matter fields in accordance with the gauge prescription for Yang-Mills field, we have seen that the mathematical description of all matter fields is a first order one. The move from second order to first order equations removes the need to restrict the metric tensor $g_{\mu\nu}(V,A)$ to be invertible. Instead all matter actions are polynomial ones which exhibit no problems for degenerate metrics.

We must also not fail to notice the non-trivial coupling between the gravitational Higgs field $V^A$ and matter fields. In the General Relativistic limit $V^2=const.$ all actions indeed reproduce the standard second order ones. But for $V^2\neq const.$ we have at our hands a very non-trivial, but from a Cartan geometric perspective elegant and simple, theory which is second order metric language would not easily have been guessed. Thus, the Cartan geometric machinery is here seen to function as a novel mathematical platform for modifications of General Relativity which should be explored further.

Another remarkable feature of Cartan gravity is that it predicts signature change for cosmological solutions with FRW symmetry imposed. Contrary to the somewhat contrived process of signature change in the Riemannian description, the Cartan geometric one is straightforward and is completely determined and described by the Cartan-geometric equations of motion. Quite surprisingly a classical realization of the Hartle-Hawking no-boundary proposal pops up as the simplest solution of the MacDowell-Mansouri equations of motion with $V^A$ fully dynamical. Importantly, the cosmological singularities were avoided and instead replaced by a change of spacetime signature, or a cosmic bounce. Given that no singular solutions were found it would be interesting to study Cartan-geometric black holes. Will the black hole singularity be replaced by a signature change, or perhaps by a bounce? Indeed, the interior solution of a black hole is nothing but a Kantowksi-Sachs cosmological spacetime.

Needless to say there are many outstanding open problems that should be pursued to further either strengthen or rule out Cartan gravity as an empirically viable description of gravity and its coupling to matter fields.  All that can be said at this moment is that a careful a systematic analysis is needed. We now list some outstanding open problems.
\begin{enumerate}
\item {\bf Linearized Cartan gravity:} Given a background solution $\{\bar V^A,\bar A^{AB}\}$ one can look to study the propagation of perturbations $\{\delta V^A,\delta A^{AB}\}$.
\begin{itemize}
\item Do perturbations propagate smoothly through signature change regions?
\item For some cosmological background, can perturbations in the Cartan-gravitational fields source primordial
fluctuations in the manner that the inflaton does?
\item Just as in the electroweak theory, will the gravitational Higgs field render perturbative gravity renormalizable?
\end{itemize}
\item {\bf Matter coupling:} The gauge prescription enforces a non-trivial coupling between matter fields and the gravitational sector. 
\begin{itemize}
 \item Does `test matter' propagate smoothly across signature change regions?
 \item Is signature change robust against the inclusion of back-reacting matter? 
 \item How does matter propagate outside the General Relativistic limit, i.e. $V^2\neq const.$?
\end{itemize}
\item {\bf Cosmological applications:}
\begin{itemize}
 \item We have seen that for much of the parameter space of possible Cartan gravity actions, the resulting theory takes the form of a scalar-tensor theory with potential coupling of the scalar field to matter. Can this new scalar degree of freedom in the gravitational sector act as a realistic inflation or dark energy candidate? 
 \item Furthermore, we may wonder whether Cartan gravity may contain a dark matter candidate. Firstly there is the field $V^{2}(x)$ which we have seen can have considerable influence on cosmology due to its own dynamics and likely due also to its coupling to other matter fields. Could excitations of this field also act as a dark matter candidate? Secondly, there is the contorsion $C_{IJ}$ which appears to have its own degrees of freedom for some choices of the theory's parameters; the effects of these degrees of freedom are currently unknown.
\end{itemize}
\item {\bf Cartan-geometric black holes:} It would be interesting to see how the gravitational Higgs field influences spherically symmetric and and static solutions. In view of the singularity free FRW solutions it is plausible that no singularity will be present for Cartan-geometric black holes. If signature change of the metric is involved here, would Euclidean regions necessarily be `hidden' by horizons? 
\item {\bf Parity and the gravitational Higgs field:} As noted in \cite{Westman:2012zk} the discrete transformation $V^a\ra -V^a$ changes the orientation of spacetime and is thus related to handedness. This suggests that the violation of parity in nature may be related to gravitation. To probe whether this is so or not one must write down a Cartan-geometric formulation of the electroweak theory. Whether this can be done ina natural way from Cartan-geometric perspective still remains to be seen. It particular, the chiral projectors $P_{L,R}=\frac{1}{2}(1\pm\frac{i}{\sqrt{|V^2|}} V_A\Gamma^A)\overset{*}{=}\frac{1}{2}(1\mp \gamma_{5})$ introduces non-polynomial dependence upon $V^{A}$. In addition, the fact that the irreducible representations of $Spin(1,4)$ are four-dimensional ones which may suggest the existence of new forms of fermions in nature; may they be detectable in future or would analysis reveal them necessarily to be incompatible with current data?
\item {\bf Cartan gravity as a limit of more general theory:} Just as Einstein-Cartan theory is a specific limit of Cartan-gravity, is Cartan-gravity a limit of a larger theory? A possible sign of this is the abundance of dimensional constants in the theory, such as $\{a_{i},b_{i},c_{1}\}$. Are these constants representative of expectation values of as-yet-unknown fields? In a future paper \cite{WestmanZlosnik2014} we will show in detail how Cartan gravity can be embedded into a larger theory based on conformal Cartan geometry which excludes the use of dimensionful constants. Progress towards understanding the expected size of the constants in the Cartan-gravity action (or indeed whether they are to be expected to have a functional dependence on $V^{2}$ and other scalars) would be a significant step towards making definitive predictions.
\end{enumerate}
We end this paper by stressing that Cartan gravity radically changes the geometric foundation upon which modern physics rests. 
As a consequence we have seen deviations from some of the basic structure of General Relativity such as the assumption of fixed metric signature or the ability to simply add a `cosmological constant term' to the gravitational action. Whether a Cartan-geometric description of nature will ultimately prove to be useful or not is something that only more research will reveal.

\vspace{0.5cm}
\noindent\textbf{Acknowledgements:} We would like to thank E. Anderson, F. Hehl, J. Magueijo, and A. Randono for useful discussions. We would also like to thank anonymous referees, whose suggestions
led to great improvements of the manuscript. This work is supported by Spanish MICINN Projects FIS2011-29287. H. Westman was also supported by the CSIC JAE-DOC 2011 program.
\appendix
%
\section{Notation}\label{unitsnconventions}
In this paper the following notation and conventions are in use. The symbol $\m M$ represents the manifold whose geometry we wish to characterize be it two-dimensional space or four-dimensional spacetime. Although pretty, we omit for notational compactness the wedge symbol $\w$ when multiplying forms. For example, if $y$ is a one-form and $z$ is a three-form then we have:
\begin{align}
yz = y\wedge z
\end{align}
In addition to the numerically invariant Kronecker delta tensor $\delta^\mu_\nu$ we also have the two numerically invariant Levi-Civita tensor {\em densities} $\varepsilon^{\mu\nu\rho\sigma}$ and $\epsilon_{\mu\nu\rho\sigma}$ related to each other as
\begin{align}
\varepsilon_{\mu\nu\rho\sigma}=g_{\mu\alpha}g_{\nu\beta}g_{\rho\gamma}g_{\sigma\delta}\varepsilon^{\alpha\beta\gamma\delta}=g\epsilon_{\mu\nu\rho\sigma}
\end{align}
with $g=det(g_{\mu\nu})$ and $\epsilon_{0123}=\varepsilon^{0123}=+1$.

\paragraph {\bf Indices:}
\begin{itemize}
\item $SO(3)$ rolling indices: $i,j,k,\dots=1,2,3$
\item Spatial 2D indices: $a,b,c,\dots= 1,2$
\item $SO(1,4)$ rolling indices: $A,B,C,\dots=0,\dots,4$
\item $SO(1,3)$ internal indices: $I,J,K,\dots = 0,\dots, 3$
\item Spacetime tensor 4D indices: $\mu,\nu,\rho,\dots=0,\dots,3$
\end{itemize}
\paragraph{\bf Geometric objects:}
\begin{itemize}
 \item Cartan (rolling) connection: $A^{AB}$ (4D spacetime) or $A^{ij}$ (2D space).
 \item Contact vector: $V^A$ (4D spacetime) or $V^i$ (2D space) 
 \item Curvature two-form: $F^{AB}$ (4D spacetime) or $F^{ij}$ (2D space)
 \item Torsion: ${\cal T}^A$ (4D spacetime) or ${\cal T}^i$ (2D space) 
\end{itemize}
\paragraph{\bf Natural units and dimensions:} $V^i$ and $V^A$ depict the size of the symmetric space we are rolling it is then appropriate to let these have dimensions of length. Since the connection $A^i_{\ph ij}$ yields infinitesimal rotations it should be dimensionless. Although different conventions exist in the physics literature we adopt the view that since, in generally covariant theories such as General Relativity, coordinates do not have an operational significance, it is natural to let $x^a$ and $x^\mu$ be dimensionless. 
\section{Exterior calculus}
Exterior calculus constitutes a powerful tool in differential geometry and this paper makes ample use of it. In order to  make this paper more accessible and self-contained we provide in the following appendices a crash-course in exterior calculus. The various operations, i.e. wedge product, exterior derivative, integration, are defined in such a way that they can be easily understood in terms of tensor operations seen in elementary textbooks in General Relativity.

\hiddensubsection{Definition of forms}
In a nutshell, forms are completely anti-symmetric covariant tensors. For example, a scalar $\Phi$ is a zero-form, a connection $A_\mu$ is a one-form, a curvature tensor $F_{\mu\nu}=-F_{\nu\mu}$ is a two-form. In general, we say that a completely antisymmetric covariant tensor of rank $(0,p)$ is a $p$-form. The number $p$ is called the degree of the form. If the manifold dimension is $N$ then no completely antisymmetric covariant tensor exists with more indices than $N$ and consequently no $p$-forms exists if $p>N$. In contradistinction to tensors we see that the number of types of forms is limited by the manifold dimension. Since the index structure of forms is simple and completely specified by its degree $p$ it is convenient to leave out the tensor indices. For example a $p$-form $\Omega_{\mu_1\mu_2\dots\mu_p}$ is written simply as $\Omega$.
\hiddensubsection{Exterior algebra}
Next we define a way of multiplying forms together that preserve the antisymmetry. This product is called the {\em wedge product} $\w$. Let $\Omega_1$ and $\Omega_2$ be two forms of degree $p$ and $q$ respectively. Then the wedge product $\Omega_1\w \Omega_2$ is a new form of degree $p+q$. For notational compactness we shall nevertheless omit the symbol $\w$ and simply write $\Omega_1\Omega_2$ since this will not cause any confusion. The basic idea of the wedge product is very simple and can be understood in terms of tensor methods as follows:
\begin{enumerate}
\item Write the forms as covariant tensors: $\Omega_{1\mu_1\mu_2\dots\mu_p}$ and $\Omega_{2\mu_1\mu_2\dots\mu_q}$
\item Multiply them as tensors: $\Omega_{1\nu_1\dots\nu_p}\Omega_{2\nu_{p+1}\dots\nu_{p+q}}$
\item Antisymmetrize: $\frac{(p+q)!}{p!q!}\Omega_{[1\nu_1\dots\nu_p}\Omega_{2\nu_{p+1}\dots\nu_{p+q}]}$.
\end{enumerate}
The last object defines the $p+q$-form $\Omega_1\Omega_2$ with tensor indices explicit. The following formal properties of the wedge product can easily be deduced. Let $\Omega_1$, $\Omega_2$, and $\Omega_3$ be a $p$-form, $q$-form, and $r$-form respectively, and $\alpha$ and $\beta$ real- or complex numbers.
\begin{itemize}
\item Linearity: $(\alpha\Omega_1+\beta\Omega_2) \Omega_3=\alpha\Omega_1\Omega_3+\beta\Omega_2\Omega_3$
\item Commutation law: $\Omega_1\Omega_2=(-1)^{pq}\Omega_2\Omega_1$ where $\Omega_1$ is a $p$-form and $\Omega_2$ is a $q$-form.
\item Associativity: $\Omega_1(\Omega_2\Omega_3)=(\Omega_1\Omega_2)\Omega_3$
\end{itemize}
The wedge-product of the two forms $\Omega_1$ and $\Omega_2$, of degree $p$ and $q$ say, produces a new form $\Omega_3=\Omega_1\Omega_2$ of degree $p+q$. Thus, if $p+q>N$ then $\Omega_1\Omega_2\equiv0$. The above rules defines the exterior algebra of forms.
\hiddensubsection{Coordinate basis}
A coordinate system is a collection of $N$ scalar fields $x^\mu=(x_1,\dots,x_N)$ on an $N$-dimensional manifold $\m M$. The gradients of these scalars $dx^\mu$ forms a set of $N$ one-forms which are normals to the equipotential surfaces $x^\mu=const$ for $\mu=1,2,\dots,N$. These normals are nothing but the gradients of the coordinate zero-forms $x^1,x^2,\dots$. As such they have one lowercase index and are therefore examples of one-forms. We write them $dx^1,dx^2,\dots,dx^N$ where the $d$ is here understood as a gradient. As such $dx^\mu$ are not infinitesimals. 

These one-forms collectively written as $dx^\mu$ are a set of co-vectors that span space of one-forms. Thus we can expand a one-form in terms of its coordinate coefficients $A_\mu$ as $A=A_\mu dx^\mu$. Similarly, the objects $dx^\mu\wedge dx^\nu\equiv dx^\mu dx^\nu$ are two-forms and they span the space of two forms. A two-form can then be expanded in terms of its coordinate coefficients $F_{\mu\nu}$ as $F=\frac{1}{2}F_{\mu\nu}dx^\mu dx^\nu$. More generally, any $p$-form $\Omega$ can be expanded in the coordinate one-form basis as follows
\begin{eqnarray}
\Omega=\frac{1}{p!}\Omega_{\mu_1\dots\mu_p}dx^{\mu_1} dx^{\mu_2}\dots dx^{\mu_p}.
\end{eqnarray}
Instead of forming the gradient of each scalar $x^\mu$ we can also consider the tangent vectors to the coordinate defined by varying one coordinate while holding all the others fixed. This yields $N$ tangent vectors which we here shall denote $\partial_\mu$ which then forms a set of basis vectors on the tangent space. Thus a vector may be written as
\begin{align}
V=V^\mu\partial_\mu.  
\end{align}
A general $(p,q)$ tensor $T$ is then expanded in the coordinate basis as 
\begin{align}
T=T^{\mu_1\dots\mu_p}_{\nu_1\dots\nu_q}dx^{\nu_1}\otimes\dots \otimes dx^{\nu_q}\otimes \partial_{\mu_1}\otimes \dots \otimes \partial_{\mu_p}
\end{align}
That we use the symbol $\partial_\mu$ which also denotes a partial derivative is no accident. The partial derivative is defined to take the derivative along the direction defined by changing the specific coordinate $x^\mu$ while holding the values $x^\nu$, $\nu\neq\mu$, of all other fixed. Thus, a vector has then a natural action on a scalar field $\phi$ by 
\begin{align}
V(\phi)=V^\mu\partial_\mu\phi\nn
\end{align}
Note, however, that a general contravariant tensor $T=T^{\mu_1\dots\mu_p}\partial_{\mu_1}\otimes\dots\otimes \partial_{\mu_p}$ does not have a natural coordinate independent action on a scalar. 

\hiddensubsection{Duality between forms and antisymmetric contravariant tensor densities}\label{duality} 
There is however another form of duality which always exists: The completely antisymmetric Levi-Civita tensor density $\varepsilon^{\mu_1\mu_2\dots\mu_N}$ establishes an isometry between the space of $p$-forms and the space of completely antisymmetric $(N-p,0)$-rank tensor densities of weight $+1$. We will use the symbol $\sim$ to denote the dual quantity. Specifically, let $\Omega$ be some $p$-form, then the dual contravariant antisymmetric $+1$ tensor density $\Omega^{\mu_{p+1}\dots\mu_N}$ is defined as

\begin{eqnarray}
\Omega=\frac{1}{p!}\Omega_{\mu_1\dots\mu_p}dx^{\mu_1}\dots dx^{\mu_p}\sim\frac{1}{p!}\Omega_{\mu_1\dots\mu_p}\varepsilon^{\mu_1\dots\mu_p\dots\mu_N}
\end{eqnarray}
where the two notationally distinct Levi-Civita symbols $\epsilon$ and $\varepsilon$ are defined so that 
\begin{eqnarray}
\epsilon_{\mu_1\mu_2\dots \mu_N}\varepsilon^{\mu_1\mu_2\dots \mu_N}=+N!.
\end{eqnarray}
As a simple concrete example we can see that, in the case of four spacetime dimensions, the object dual to the four-form $\m E=\frac{1}{4!}\epsilon_{IJKL}e^I\w e^J\w e^K\w e^L$, is nothing but the usual scalar density volume element $e\equiv det(e^I_\mu)$, i.e. we have
\begin{eqnarray}
\m E=\frac{1}{4!}\epsilon_{IJKL}e^Ie^Je^Ke^L&=&\frac{1}{4!}\epsilon_{IJKL}e^I_\mu e^J_\nu e^K_\rho e^L_\sigma dx^\mu dx^\nu dx^\rho dx^\sigma\nonumber\\
&\sim& \varepsilon^{\mu\nu\rho\sigma}\frac{1}{4!}\epsilon_{IJKL}e^I_\mu e^J_\nu e^K_\rho e^L_\sigma\equiv det(e^I_\mu)=e
\end{eqnarray}
This duality between differential forms and contravariant antisymmetric tensor densities is useful since it allows us to translate between expressions written in differential forms forms and the more common tensorial notation which is more common within the physics community.
\hiddensubsection{Forms as linear functionals}
A $p$-form written as $\Omega=\frac{1}{p!}\Omega_{\mu_1\dots\mu_p}dx^{\mu_1}\dots dx^{\mu_p}$ should not be interpreted as an infinitesimal quantity despite the appearance of the $dx^\mu$'s which might naively be interpreted as infinitesimal displacements which we in this paper instead denote as $\delta x^\mu$. Rather, forms are to be understood as completely antisymmetric multi-linear functionals $\Omega:T_p(\m M)\otimes\dots\otimes T_p(\m M)\ra \mb R$. For example, a one-form $A$ fed a vector $V$ yields the real number $A(V)$. If $A$ happens to be the exterior derivative of a scalar then we may note the following identity
\begin{align}
d\phi(V)=\partial_\mu\phi V^\mu=V^\mu\partial_\mu\phi=V(\phi)
\end{align}
We see that the coordinate basis one-form $dx^\mu$ fed the coordinate basis vector $\partial_\nu$ yields 
\begin{align}
dx^\mu(\partial_\nu)=\partial_\nu (x^\mu)=\frac{\partial x^\mu}{\partial x^\nu}=\delta^\mu_\nu
\end{align}
Thus we have $A(V)=A_\mu (dx^\mu)(V)=A_\mu V^\nu(dx^\mu)(\partial_\nu)=A_\mu V^\nu\delta^\mu_\nu=A_\mu V^\mu$.

The coordinate basis p-form $dx^{\mu_1}\dots dx^{\mu_p}$ fed $p$ coordinate basis vectors yields
\begin{align}
dx^{\mu_1}\dots dx^{\mu_p}(\partial_{\nu_1},\dots,\partial_{\nu_p})=\delta^{\mu_1\dots\mu_p}_{\nu_1\dots\nu_p}=\frac{1}{(N-p)!}\epsilon_{\nu_1\dots\nu_p\rho_{p+1}\dots \rho_N}\varepsilon^{\mu_1\dots\mu_p\rho_{p+1}\dots \rho_N}
\end{align}
so that we have
\begin{align}
\Omega(V_1,\dots,V_p)&=\frac{1}{p!}\Omega_{\mu_1\dots\mu_p}(dx^{\mu_1}\dots dx^{\mu_p})(V_1,\dots, V_p)=\frac{1}{p!}\Omega_{\mu_1\dots\mu_p}V_1^{\nu_1}\dots V_p^{\nu_p}(dx^{\mu_1}\dots dx^{\mu_p})(\partial_{\nu_1},\dots,\partial_{\nu_p})\nn\\
&=\frac{1}{p!(N-p)!}\epsilon_{\nu_1\dots\nu_p\rho_{p+1}\dots \rho_N}\varepsilon^{\mu_1\dots\mu_p\rho_{p+1}\dots \rho_N}\Omega_{\mu_1\dots\mu_p}V_1^{\nu_1}\dots V_p^{\nu_p}=\frac{1}{p!}\delta^{\mu_1\dots\mu_p}_{\nu_1\dots\nu_p}\Omega_{\mu_1\dots\mu_p}V_1^{\nu_1}\dots V_p^{\nu_p}\nn
\end{align}
The collection of vectors $(V_1,\dots, V_p)$ forms a $p$-dimensional parallelepiped in an $N$-dimensional tangent space.
\hiddensubsection{Exterior differentiation}

Next we define a coordinate independent derivative operator, called the {\em exterior derivative}, for forms that preserve the complete antisymmetry and generates from a $p$-form $\Omega$ a new form $d\Omega$ with degree $p+1$. The partial derivative $\partial_\mu$ will not do since: 1) it is coordinate dependent when acting on a $p$-form with $p>0$ and 2) it takes us out of the space of forms, i.e. completely antisymmetric tensors. The basic idea of the exterior derivative is simple and amounts to carrying out the following steps.
\begin{enumerate}
\item Write the form as a covariant tensor: $\Omega_{\mu_1\mu_2\dots\mu_p}$
\item Take the partial derivative: $\partial_{\mu_{p+1}}\Omega_{\mu_1\mu_2\dots\mu_p}$
\item Antisymmetrize: $(p+1)\partial_{[\mu_{p+1}}\Omega_{\mu_1\mu_2\dots\mu_p]}$.
\end{enumerate}
The last completely antisymmetric covariant vector defines the exterior derivative denoted $d\Omega$. This object is coordinate independent. The following formal properties of the exterior derivative can easily be checked:
\begin{itemize}
\item The components of the exterior derivative of a zero-form are its partial derivatives $(d\Phi)_\mu=\partial_\mu \Phi$ ($d\Phi=\partial_{\mu}\Phi dx^{\mu}$).
\item Linearity: $d(\alpha\Omega_1+\beta\Omega_2)=\alpha d\Omega_1+\beta d\Omega_2$
\item Leibniz rule: $d(\Omega_1\Omega_2)=d\Omega_1\Omega_2+(-1)^{p}\Omega_1 d\Omega_2$ where $\Omega_1$ is a $p$-form.
\item $d^2\Omega=d(d\Omega)\equiv0$ for all $p$-forms $\Omega$ and all $p$.
\end{itemize}
The factor of $(-1)^p$ in the Leibniz rule is there to compensate for the commutation rule for forms. The last property is nothing but a restatement of the commutativity of partial derivatives. The exterior derivative of an $N$-form is automatically zero since there are no forms with degree $N+1$.

\hiddensubsection{Integration of forms}

Consider an integral of some quantity on some $p$-dimensional surface in an $N$-dimensional space. If the surface is parametrized by $\xi^\alpha$ so that $x^\mu=x^\mu(\xi)$ consists of points in that surface such integral is written as
\begin{align}
\int \Phi(x(\xi))\delta^p\xi.
\end{align}
We write $\delta^{p}\xi$ instead of the standard $d^{p}\xi$ so as to avoid confusion with the `$d$' appearing in the formalism of differential forms. The appearance of $\delta^p\xi$ and $x(\xi)$ clearly shows that this integral is not written in a manifestly coordinate and parametrization independent manner. The language of differential forms allows for a neat coordinate and parametrization-free notation. In fact, forms are precisely those elementary mathematical objects which appear under integral signs. A one-form $A$ can be integrated along a one-dimensional curve on the manifold, a two form $F$ over a two-dimensional surface, and a $p$-form $\Omega$ over a $p$-dimensional sub-manifold $\Sigma$.

To see this clearly and establish a concrete connection with standard notation let us consider the integral of some $p$-form $\Omega$ on some $p$-dimensional surface. Although we write $\int \Omega$ one should not fall prey to the temptation of thinking of $\Omega$ as an infinitesimal quantity. Rather we should think of evaluating the integration of the $p$-form $\Omega$ on a $p$-dimensional surface $\Sigma$ in the following way. Using $x^\mu$ as coordinates on $\m M$ and some parametrization $\xi^i$ for the surface $\Sigma$ we can span the tangent space $T_\xi(\Sigma)$ by the vectors
\begin{align}
\overrightarrow{\xi^1}=\frac{\partial x^\mu}{\partial \xi^1}\partial_\mu\qquad \overrightarrow{\xi^2}=\frac{\partial x^\mu}{\partial \xi^2}\partial_\mu\qquad \dots\qquad \overrightarrow{\xi^p}=\frac{\partial x^\mu}{\partial \xi^p}\partial_\mu
\end{align}
From these we can now define $p$ infinitesimal displacement vectors 
\begin{align}
\overrightarrow{\delta\xi^1}=\delta\xi^1\frac{\partial x^\mu}{\partial \xi^1}\partial_\mu\qquad \overrightarrow{\delta\xi^2}=\delta\xi^2\frac{\partial x^\mu}{\partial \xi^2}\partial_\mu\qquad \dots\qquad \overrightarrow{\delta\xi^p}=\delta\xi^p\frac{\partial x^\mu}{\partial \xi^p}\partial_\mu
\end{align}
where $\delta \xi^i$, $i=1,\dots,p$ are infinitesimals. This collection of infinitesimal vectors forms a $p$-dimensional parallelepiped. We can now form at each point on $\Sigma$ an infinitesimal real number by feeding the form $\Omega$ the infinitesimal parallelepiped $(\overrightarrow{\delta\xi^1},\dots,\overrightarrow{\delta\xi^p})$, i.e. $\Omega(\overrightarrow{\delta\xi^1},\dots,\overrightarrow{\delta\xi^p})$. The evaluation of the integral $\int \Omega$ then simply consists of summing all these infinitesimal real numbers together. Specifically, the evaluation goes as follows:
\begin{align}
\int \Omega&=\int \Omega(\overrightarrow{\delta\xi^1},\dots,\overrightarrow{\delta\xi^p})=\int\frac{1}{p!}\Omega_{\mu_1\dots\mu_p}(dx^{\mu_1}\dots dx^{\mu_p})(\overrightarrow{\delta\xi^1},\dots,\overrightarrow{\delta\xi^p})\nn\\
&=\int \frac{1}{p!}\delta\xi^1\frac{\partial x^{\nu_1}}{\partial \xi^1}\dots \delta\xi^p\frac{\partial x^{\nu_p}}{\partial \xi^p}\Omega_{\mu_1\dots\mu_p}(dx^{\mu_1}\dots dx^{\mu_p})(\partial_{\nu_1},\dots,\partial_{\nu_p})\nn\\
&=\int \frac{1}{p!}\frac{\partial x^{\nu_1}}{\partial \xi^1}\dots\frac{\partial x^{\nu_p}}{\partial \xi^p}\Omega_{\mu_1\dots\mu_p}\delta^{\mu_1\dots\mu_p}_{\nu_1\dots\nu_p}\delta^p\xi\nn\\
&=\int \frac{1}{p!(N-p)!}\frac{\partial x^{\nu_1}}{\partial \xi^1}\dots\frac{\partial x^{\nu_p}}{\partial \xi^p}\Omega_{\mu_1\dots\mu_p}\epsilon_{\nu_1\dots\nu_p\rho_{p+1}\dots\rho_N}\varepsilon^{\mu_1\dots\mu_p\rho_{p+1}\dots\rho_N}\delta^p\xi.
\end{align}
We note again that the coordinate volume element $\delta^p\xi$ is usually written as $d^p\xi$ but here we have used the symbol $\delta$ rather than $d$ so as to not confuse it with the exterior derivative symbol which appears in $dx^\mu$ for example. 

For concreteness let us consider a standard flux integral over a two-dimensional surface in a three-dimensional flat Euclidean space. A typical notation for this is
\begin{align}
\Phi=\int B\cdot n\delta A
\end{align}
where $B$ is some vector field, $n$ the field of normals on the surface, and $\delta A$ the area element. We write $\delta A$ rather than the standard $dA$ to avoid confusion with the exterior derivative symbol $d$. To compute the normal $n$ and area element $\delta A$ we first parametrize the surface $X(u,v)=(x(u,v),y(u,v),z(u,v))$ and then compute the tangent vectors 
\begin{align}
X_u=\frac{\partial X^{i}}{\partial u}\partial_{i}\qquad X_v=\frac{\partial X^{i}}{\partial v}\partial_{i}
\end{align}
and defining the infinitesimal vectors
\begin{align}
\overrightarrow{\delta u}=\delta u\frac{\partial X^i}{\partial u}\partial_i\qquad \overrightarrow{\delta v}=\delta v\frac{\partial X^i}{\partial v}\partial_i
\end{align}
so that the normal and area element become
\begin{align}
n=\frac{X_u\times X_v}{|X_u\times X_v|}\qquad \delta A=|X_u\times X_v|\delta u\delta v
\end{align}
where it may be checked explicitly that indeed $\delta A$ is the area of a parallelepiped spanned by vectors $\overrightarrow{\delta u}$ and $\overrightarrow{\delta u}$. The flux integral now reads
\begin{align}
\Phi=\int B\cdot (X_u\times X_v)\delta u\delta v \label{flux1}
\end{align}
where the dot denotes the metric inner-product $\delta_{ij}B^{i}(X_u\times X_v)^{j}$. Indeed, in components (\ref{flux1}) reads
\begin{align}
\int B^i(X_u\times X_v)_i\delta u\delta v=\int \frac{1}{2}\varepsilon^{ijk}F_{kl}\epsilon_{imn}\frac{\partial X^m}{\partial u}\frac{\partial X^n}{\partial v}\delta u\delta v
\end{align}
where we have introduce the dual antisymmetric object $F_{ij}$ defined via $B^i=\frac{1}{2}\varepsilon^{ijk}F_{jk}$. Using the identity
\begin{align}
(dx^jdx^k)(\partial_m,\partial_n)=\delta^{jk}_{mn}=\delta^j_m\delta^k_n-\delta^j_n\delta^k_m=\varepsilon^{ijk}\epsilon_{imn}
\end{align}
we get
\begin{align}
\Phi=\int \frac{1}{2}F_{kl}(dx^jdx^k)(\partial_m,\partial_n)\frac{\partial X^m}{\partial u}\frac{\partial X^n}{\partial v}\delta u\delta v=\int F(\partial_m,\partial_n)\frac{\partial X^m}{\partial u}\frac{\partial X^n}{\partial v}\delta u\delta v
\end{align}
hence we have 
\begin{align}
\Phi=\int F(\overrightarrow {\delta u},\overrightarrow {\delta v})=\int F
\end{align}
We note that the orientation is specified by the normal $n$ and that this orientation is automatically accounted for in the forms language. We also note that a flux integral in coordinate and parameterization independent language naturally involves the two-form $F=\frac{1}{2}F_{ij}dx^idx^j$ rather then the vector $F^i$. It is in this precise sense we may say that forms `{\em are the things which occur under integral signs.}' \cite{flanders2012differential}. 
\section{Gauge connections, curvature, and Bianchi identities}
We provide here a brief exposition of the basic techniques and ideas of gauge connections in the language of forms. Although the formulas of this section is valid for any gauge group we will mostly use the waywiser variables to illustrate the ideas. 

The contact vector $V^A$ appears with a gauge index $A$ and transforms under a spacetime-dependent gauge transformation as $V^A\rightarrow \theta(x)^A_{\ph AB}V^B$. Objects with gauge index downstairs, e.g. $U_A$, transforms as $U_A\rightarrow U_B(\theta^{-1})^B_{\ph BA}$ so that $U_AV^A$ is invariant under arbitrary gauge transformations. This fixes the transformation law of mixed objects $W^A_{\ph AB}$ as $W^A_{\ph AB}\rightarrow \theta^A_{\ph AC}W^C_{\ph CD}(\theta^{-1})^D_{\ph DB}$.

The exterior derivative of $dV^A$ transforms inhomogeneously $d(\theta^A_{\ph AB}V^B)\neq \theta^A_{\ph AB}dV^B$ and $dV^A$ under a spacetime-dependent gauge transformation $V^A\rightarrow \theta(x)^A_{\ph AB}V^B$. It is therefore not a gauge-covariant object. In order to restore gauge-covariance the exterior derivative is replaced by the gauge covariant exterior derivative $d\rightarrow D^{(A)}$:
\begin{eqnarray}
D^{(A)}V^A\equiv dV^A+A^A_{\ph AB}V^B \qquad D^{(A)}U_A\equiv dU_A-A^B_{\ph BA}U_B 
\end{eqnarray} 
with the minus sign on the right equation guaranteeing that $D(U_AV^A)=d(U_AV^A)$. The requirement of gauge-covariance, i.e. $D^{(A')}(\theta^A_{\ph AB}V^B)=\theta^A_{\ph AB}D^{(A)}V^B$, implies immediately that the connection $A^A_{\ph AB}$ transforms inhomogeneously under local gauge transformation:
\begin{eqnarray}
A^A_{\ph AB}\rightarrow A^{\prime A}_{\ph{\prime A}B}=-d\theta^A_{\ph AC}(\theta^{-1})^C_{\ph CB}+\theta^A_{\ph AC}A^C_{\ph CD}(\theta^{-1})^D_{\ph DB}.
\end{eqnarray}
We will often write $D$ for the gauge-covariant instead of the more cumbersome notation $D^{(A)}$ wherever no confusion can arise. The gauge covariant exterior derivative of some $p$-form, $\Omega^A_{\ph AB}$ say, is given by
\begin{eqnarray}
D\Omega^A_{\ph AB}=d\Omega^A_{\ph AB}+A^A_{\ph AC} \Omega^C_{\ph CB}-A^C_{\ph CB} \Omega^A_{\ph AC}
\end{eqnarray}
The curvature two-form $F^A_{\ph AB}$ defined by
\begin{eqnarray}\label{standef}
F^A_{\ph AB}\equiv dA^A_{\ph AB}+A^A_{\ph CC} A^C_{\ph CB}
\end{eqnarray}
can straightforwardly be shown to transform as $F^A_{\ph AB}\rightarrow \theta^A_{\ph AC}F^C_{\ph CD}(\theta^{-1})^D_{\ph DB}$ and is therefore gauge covariant. Note however that the gauge covariant derivative applied to the (gauge non-covariant) connection 
\begin{eqnarray}
DA^{AB}=dA^{AB}+A^A_{\ph{A}C} A^{CB}+A^B_{\ph{B}C} A^{AC}=F^{AB}+A^A_{\ph{A}C} A^{CB}
\end{eqnarray}
is not gauge covariant. 

The identity $DF^A_{\ph AB}\equiv 0$ is extremely useful and is called the {\em first Bianchi identity}. It follows immediately from the definition of the gauge-covariant exterior derivative and the rules of exterior calculus:
\begin{multline}
DF^A_{\ph AB}\equiv D^2A^A_{\ph AB}\equiv dF^A_{\ph AB}+A^A_{\ph AC} F^C_{\ph CB}-A^C_{\ph CB} F^A_{\ph AC}=d(dA^A_{\ph AB}+A^A_{\ph AC} A^C_{\ph CB})\nonumber\\
+A^A_{\ph AC} (dA^C_{\ph CB}+A^C_{\ph CD} A^D_{\ph DB})-A^C_{\ph CB} (dA^A_{\ph AC}+A^A_{\ph AD} A^D_{\ph DC})\\
=dA^A_{\ph AC} A^C_{\ph CB}-A^A_{\ph AC} dA^C_{\ph CB}+A^A_{\ph AC} dA^C_{\ph CB}
+A^A_{\ph AC} A^C_{\ph CD} A^D_{\ph DB}-A^C_{\ph CB} dA^A_{\ph AC}-A^C_{\ph CB} A^A_{\ph AD} A^D_{\ph DC}\nonumber\\
=dA^A_{\ph AC} A^C_{\ph CB}+A^A_{\ph AC} A^C_{\ph CD} A^D_{\ph DB}-dA^A_{\ph AC} A^C_{\ph CB} -A^A_{\ph AD} A^D_{\ph DC} A^C_{\ph CB}\equiv0\nonumber
\end{multline}
By taking the gauge-covariant derivative of the torsion tensor defined by $T^A\equiv F^A_{\ph AB}V^B$ and making use of the Leibniz rule and the first Bianchi identity $DF^A_{\ph AB}\equiv0$ we obtain the {\em second Bianchi identity}
\begin{eqnarray}
DT^A\equiv D(F^A_{\ph AB}V^B)=F^A_{\ph AB} DV^B
\end{eqnarray}
\section{The Palatini action in the language of forms}\label{tensorformtrans}
\label{palappendix}
To help make contact with standard notation we illustrate how the Palatini action of the Einstein-Cartan theory (written as a four-form) corresponds to the more familiar Einstein-Hilbert action (written in terms of the density $\sqrt{-g}$ and coordinate displacement product $d^{4}x\equiv \delta^{4}x$). The Palatini action is as follows: 
\begin{eqnarray}
{\cal S}_P=\int \epsilon_{IJKL}e^I e^J R^{KL}.
\end{eqnarray}
The one-form $e^I$ is the co-tetrad (the inverse tetrad) and $R^{IJ}$ is the Riemann curvature two-form defined by  $R^{IJ}=d\omega^{IJ}+\omega^I_{\ph IK} \omega^{KJ}$ with $\omega^{IJ}$ a  one-form valued in the Lie algebra of $SO(1,3)$. 

This action is written in a manifestly coordinate independent way. In order to relate this action to the more well-known Einstein-Hilbert action $\int \sqrt{-g}Rd^4x$ which is not written in a manifestly coordinate independent way we must introduce a coordinate system, $x^\mu$ say. We can now expand the forms $e^I$ and $R^{IJ}$ in the basis $dx^\mu$: $e^I=e^I_\mu dx^\mu$ and $R^{KL}=\frac{1}{2}R_{\mu\nu}^{\ph {\mu\nu}KL}dx^\mu dx^\nu$. Thus we have,
\begin{eqnarray}
{\cal S}_P&=&\int \epsilon_{IJKL}e^I e^J R^{KL}=\int \frac{1}{2}\epsilon_{IJKL}e_\mu^I e_\nu^J R_{\rho\sigma}^{\ph{\rho\sigma}KL}dx^\mu dx^\nu dx^\rho dx^\sigma\nn
\end{eqnarray}
Next we define the infinitesimal four-dimensional parallelepiped\footnote{Since we are integrating over all of the four-dimensional manifold $\m M$ rather than some subsurface we have without loss of generality let the parametrization $\xi$ coincide with the coordinates $x$.}
\begin{align}
\overrightarrow{\delta x^0}=\delta x^0\frac{\partial x^\nu}{\partial x^0}\partial_\nu=\delta x^0\partial_0\quad \overrightarrow{\delta x^1}=\delta x^1\frac{\partial x^\nu}{\partial x^1}\partial_\nu=\delta x^1\partial_1\quad\overrightarrow{\delta x^2}=\delta x^2\frac{\partial x^\nu}{\partial x^2}=\delta x^2\partial_2\quad\overrightarrow{\delta x^3}=\delta x^3\frac{\partial x^\nu}{\partial x^3}\partial_\nu=\delta x^3\partial_3
\end{align}
which when fed to the four-form $\epsilon_{IJKL}e^Ie^JR^{KL}$ yields
\begin{align}
(\epsilon_{IJKL}e^Ie^JR^{KL})(\overrightarrow{\delta x^0},\overrightarrow{\delta x^1},\overrightarrow{\delta x^2},\overrightarrow{\delta x^3})&=\frac{1}{2}\epsilon_{IJKL}e_\mu^I e_\nu^J R_{\rho\sigma}^{\ph{\rho\sigma}KL}(dx^\mu dx^\nu dx^\rho dx^\sigma)(\overrightarrow{\delta x^0},\overrightarrow{\delta x^1},\overrightarrow{\delta x^2},\overrightarrow{\delta x^3})\nn\\
&=\frac{1}{2}\epsilon_{IJKL}e_\mu^I e_\nu^J R_{\rho\sigma}^{\ph{\rho\sigma}KL}\delta^{\mu\nu\rho\sigma}_{0123}\delta^4x=\int \frac{1}{2}\epsilon_{IJKL}e_\mu^I e_\nu^J R_{\rho\sigma}^{\ph{\rho\sigma}KL}\varepsilon^{\mu\nu\rho\sigma}\underbrace{\epsilon_{0123}}_{=+1}\delta^4x\nn\\
&=\frac{1}{2}\epsilon_{IJKL}e_\mu^I e_\nu^J R_{\rho\sigma}^{\ph{\rho\sigma}KL}\varepsilon^{\mu\nu\rho\sigma}\delta^4x\nn
\end{align}
In order to see that the action ${\cal S}_P$ is nothing but (twice) the Einstein-Hilbert action $\m S_{EH}$ written in the variables $e^I$ and $\omega^{IJ}$ we do the following rewriting
\begin{eqnarray}
\m S_{P}&=&\int\epsilon_{IJKL}e^Ie^JR^{KL}=\int \frac{1}{2}\epsilon_{IJMN}e_\mu^I e_\nu^J e^M_\kappa e^N_\tau e^\kappa_Ke^\tau_L R_{\rho\sigma}^{\ph{\rho\sigma}KL}\varepsilon^{\mu\nu\rho\sigma}\delta^4x\nonumber\\
&=&\int\frac{1}{2}\epsilon_{IJKL}e_\mu^I e_\nu^J R_{\rho\sigma}^{\ph{\rho\sigma}KL}\varepsilon^{\mu\nu\rho\sigma}\delta^4x=\int \frac{1}{2}e\epsilon_{\mu\nu\kappa\tau}\varepsilon^{\mu\nu\rho\sigma}e^\kappa_Ke^\tau_L R_{\rho\sigma}^{\ph{\rho\sigma}KL}\delta^4x\nonumber\\
&=&\int \frac{1}{2}e2(\delta^\rho_\tau\delta^\sigma_\kappa-\delta^\rho_\kappa\delta^\sigma_\tau)e^\kappa_Ke^\tau_L R_{\rho\sigma}^{\ph{\rho\sigma}KL}\delta^4x=\int 2e e^\mu_Ie^\nu_J R_{\mu\nu}^{\ph{\mu\nu}IJ}\delta^4x\nn\\
&=&\int 2\sqrt{-g} R\delta^4x=2\m S_{EH}\nonumber
\end{eqnarray}
where we made use of the identities
\begin{eqnarray}
\sqrt{-g}=e\quad R=e^\mu_Ie^\nu_J R_{\mu\nu}^{\ph{\mu\nu}IJ}\quad\epsilon_{\mu\nu\kappa\tau}\varepsilon^{\mu\nu\rho\sigma}=
2(\delta^\rho_\kappa\delta^\sigma_\tau-\delta^\rho_\tau\delta^\sigma_\kappa)\quad e^\mu_I e_\mu^J=\delta^J_I\quad e\epsilon_{\mu\nu\rho\sigma}=\epsilon_{IJKL}e^I_\mu e^J_\nu e^K_\rho e^L_\sigma\nonumber
\end{eqnarray}
with $e$ the co-tetrad determinant and $e^\mu_I$ its inverse. As before we have written $\delta^4x$ rather than $d^4x$ as to not confuse it with the symbol $d$ for the exterior derivative. 
\section{The variational calculus of differential forms}\label{variationalforms}
A spacetime action ${\cal S}$ is per definition an integral ${\cal S}=\int\mathcal{L}$ of some four-form ${\cal L}$ over some spacetime region $V$. Since all the basic variables in Cartan waywiser geometry are themselves differential forms, and the equations of motions are obtained by requiring the action to be extremized, we provide, for completeness and accessibility, an exposition of the variational calculus of differential forms and related helpful tricks which simplify calculations immensely. For the sake of simplicity, our Lagrangian four-forms $\mathcal{L}$ will be assumed to be polynomial in the basic forms.

The variation of a p-form $\Omega$ is as usual defined as $\Omega\rightarrow \Omega+\delta\Omega$. The variation symbol $\delta$ commutes with the exterior derivative $\delta d\Omega=d\delta\Omega$ which follows immediately from the linear property of the exterior derivative: $\delta d\Omega\equiv d(\Omega+\delta\Omega)-d\Omega =d\Omega+d\delta\Omega-d\Omega=d\delta\Omega$.

Let us now consider some action ${\cal S}=\int_V \mathcal{L}$ where $\mathcal{L}$ is a four-form that for concreteness depends on some form $\Omega$ and it's first exterior derivative $d\Omega$, i.e. $\mathcal{L}=\mathcal{L}(\Omega,d\Omega)$. In order to obtain the equations of motion for $\Omega$ we wish to vary the action with respect to the differential form $\Omega$. The variation $\delta_\Omega {\cal S}$ is defined by
\begin{eqnarray}
\delta_\Omega {\cal S}=\int_V \delta_\Omega \mathcal{L}(\Omega,d\Omega)\equiv \int_V \mathcal{L}(\Omega+\delta\Omega,d\Omega+d\delta\Omega)-\mathcal{L}(\Omega,d\Omega)=\int_V\mathcal{L}(\delta\Omega,d\Omega)+\mathcal{L}(\Omega,d\delta\Omega)
\end{eqnarray}
In order to extract equations of motion we as usual integrate by parts which we now turn to. 
\hiddensubsection{Integration by parts}
After a variation of a Lagrangian four-form $\mathcal{L}$ with respect to a form $\Omega$ we might end up with terms like $d(\delta_\Omega\omega)$ where $\omega$ is some three-form. If we now assume that the variation of $\Omega$ is zero at the boundary $\partial V$, i.e. $\delta\Omega|_{\partial V}=0$, we also have that $\delta_\Omega\omega|_{\partial V}=0$. Gauss theorem then yields
\begin{eqnarray}
\int_V \delta_\Omega d(\omega)=\int_V d(\delta_\Omega\omega)=\int_{\partial V} \delta_\Omega\omega=0
\end{eqnarray}
and we conclude that terms like in a Lagrangian which are a exterior derivatives of a three-forms, e.g. $d\omega$ above, do not alter the equations of motion. These are also called topological terms. 

Suppose now that we have obtained
\begin{eqnarray}
\int_V \delta\Omega \Psi + d\delta\Omega\Phi
\end{eqnarray}
after a variation with respect to $\Omega$. By making use of the Leibniz rule for exterior derivatives 
\begin{eqnarray}
d(\delta\Omega\Phi)=d\delta\Omega\Phi+(-1)^p\delta\Omega d\Phi
\end{eqnarray}
we see that we can simplify the above variation using Gauss theorem and the fact that the variation $\delta\Omega$ vanishes at the boundary
\begin{align}
\int_V \delta\Omega \Psi + d\delta\Omega\Phi&=\int_V \delta\Omega \Psi + d(\delta\Omega\Phi)-(-1)^p\delta\Omega d\Phi\nonumber\\
&=\int_V \delta\Omega \Psi-(-1)^p\delta\Omega d\Phi + \underbrace{\int_{\partial V} \delta\Omega\Phi}_{=0}\nonumber\\&=\int_V \delta\Omega(\Psi-(-1)^p d\Phi)\nonumber
\end{align}
If the action is supposed to extremized its variation must be zero for all choices of $\delta\Omega$. This means that 
\begin{eqnarray}
\Psi-(-1)^p d\Phi=0
\end{eqnarray}
which then constitute the equations of motion.
\hiddensubsection{Methods using the gauge covariant exterior derivative}
We can now extend the above discussion to include gauge covariant exterior derivatives $D$. Strictly speaking there is no need to do this but it simplifies calculations immensely and keeps the expressions manifestly gauge covariant throughout the calculation. 

For concreteness we use the waywiser forms and their gauge-covariant derivatives to illustrate the computational techniques involved. As in the case of the exterior derivative, we infer from linearity that the variation symbol $\delta$ commutes with the gauge covariant exterior derivative $D$. In the case of the curvature two-form we have the important relation
\begin{eqnarray}
\delta_A F^{AB}=\delta_A (dA^{AB}+A^A_{\ph AC} A^{CB})=d\delta A^{AB}+\delta A^A_{\ph AC} A^{CB}+A^A_{\ph AC} \delta A^{CB}=D\delta A^{AB}
\end{eqnarray}
Because the gauge covariant exterior derivative satisfies the Leibniz rule, e.g. 
\begin{eqnarray}
D(\Phi^{ABC\dots}\Psi^{DEF\dots})=D\Phi^{ABC\dots}\Psi^{DEF\dots}+(-1)^p \Phi^{ABC\dots} D\Psi^{DEF\dots}
\end{eqnarray}
where $\Phi^{ABC\dots}$ is some Lie-algebra-valued p-form, and the gauge covariant exterior derivative reduces to the ordinary exterior derivative for a form with no free gauge indices, e.g.
\begin{eqnarray}
D\Phi^{A}_{\ph AA}=d\Phi^{A}_{\ph AA}
\end{eqnarray}
we can make use of the same tricks as above to vary a Lagrangian four-form which per definition contains no free gauge indices. See Appendix \ref{MMaction} for a concrete example.
\hiddensubsection{Topological terms}\label{topologicalterms}
When writing down actions is it important to quickly be able to recognize topological terms since they do not alter the classical equations of motion. These all have the form $d\Omega$ where $\Omega$ is some three-form. Let $A^{AB}$ and $\omega^{IJ}$ be two connections with $F^{AB}$ and $R^{IJ}$ the corresponding curvature two forms. Two examples of topological terms (i.e. exterior derivatives of three-forms) are then
\begin{align}
F^{AB} F_{AB}&=d\left(A^{AB} F_{AB}+\frac{1}{3}A^{AC} A_{A}^{\phantom{A}D} A_{CD}\right) \label{top1}\\
\epsilon_{IJKL} R^{IJ} R^{KL}&=d\left(\epsilon_{IJKL}\omega^{IJ}( R^{KL}-\frac{1}{3}\omega^{K}_{\ph{K}M} \omega^{ML})\right) \label{top2}.
\end{align}
Another topological can be formed when both $F^{AB}$ and $V^{A}$ are used. Consider the following three-form:
\begin{eqnarray}
F^{AB} DV_AV_B.
\end{eqnarray}
by taking its exterior derivative (which is amounts to taking the divergence of its dual)
\begin{eqnarray}\label{nyny}
d(F^{AB} DV_AV_B)&=&D(F^{AB} DV_AV_B)=F^{AB} F_{AC}V^CV_B-F^{AB} DV_A DV_B
\end{eqnarray}
where we have used the identities $DF^{AB}\equiv0$ and $D^2V^A=F^A_{\ph AB}V^B$. In the General Relativistic limit where $V^{2}\rightarrow const.$ it can be seen that $F^{AB}DV_{A}V_{B}$ is proportional to the Nieh-Yan three-form $T^{I}e_{I}$.
\hiddensubsection{Example: MacDowell-Mansouri action}\label{MMaction}
As a concrete example of the calculus of variations for forms we consider the variation of the Mansouri-MacDowell action (i.e. the action (\ref{genaction}) with $a_{1}=1$ and all other coefficients set to zero) with respect to the one-form $A^{AB}=A_{\mu}^{\ph{\mu}AB}dx^{\mu}$  with all the essential calculational steps included:  %
\begin{align}
\delta_A {\cal S}_P&=\int_V \delta_A(\epsilon_{ABCDE}V^E F^{AB} F^{CD})=\int_V \epsilon_{ABCDE}V^E (D\delta A^{AB} F^{CD}+F^{AB} D\delta A^{CD})\nonumber \\
&=2\int_V \epsilon_{ABCDE}V^E D\delta A^{AB} F^{CD}\nonumber\\
&=2\int_V D(\epsilon_{ABCDE}V^E \delta A^{AB} F^{CD})+\delta A^{AB} D(\epsilon_{ABCDE}V^E F^{CD})\nonumber\\
&=2\int_V d(\epsilon_{ABCDE}V^E \delta A^{AB} F^{CD})+\delta A^{AB} \epsilon_{ABCDE} (DV^E F^{CD}+V^E \underbrace{DF^{CD}}_{\equiv0})\nonumber\\
&=2\underbrace{\int_{\partial V} \epsilon_{ABCDE}V^E \delta A^{AB} F^{CD}}_{=0}+\int_V\delta A^{AB} \epsilon_{ABCDE} DV^E F^{CD}\nonumber\\
&=2\int_V\delta A^{AB} (\epsilon_{ABCDE} DV^E F^{CD})\nonumber\\
\end{align}
from which the equations of motions, which naturally appear as a set of three-forms, are readily identified as
\begin{eqnarray}\label{MMwaywiserfieldeq}
\epsilon_{ABCDE} DV^E F^{CD}=0.
\end{eqnarray}
\section{Bibliography}
\bibliographystyle{hunsrt}
\bibliography{references}

\begin{thebibliography}{10}

\bibitem{Weinberg:1996kr}
Steven Weinberg.
\newblock {The quantum theory of fields. Vol. 2: Modern applications}.
\newblock 1996.

\bibitem{Zee:2003mt}
A.~Zee.
\newblock {Quantum field theory in a nutshell}.
\newblock 2003.
\newblock Book, Princeton University Press.

\bibitem{Randono:2010cq}
Andrew Randono.
\newblock {Gauge Gravity: a forward-looking introduction}.
\newblock 2010, 1010.5822.

\bibitem{Westman:2012zk}
Hans~F. Westman and Tom~G. Zlosnik.
\newblock {Cartan gravity, matter fields, and the gauge principle}.
\newblock {\em Annals Phys.}, 334:157--197, 2013, 1209.5358.

\bibitem{Weinberg:100595}
Steven Weinberg.
\newblock {\em Gravitation and Cosmology: Principles and Applications of the
  General Theory of Relativity}.
\newblock Wiley, New York, NY, 1972.

\bibitem{Trautman:2006fp}
Andrzej Trautman.
\newblock {Einstein-Cartan theory}.
\newblock 2006, gr-qc/0606062.

\bibitem{Leclerc:2005qc}
M.~Leclerc.
\newblock {The Higgs sector of gravitational gauge theories}.
\newblock {\em Annals Phys.}, 321:708--743, 2006, gr-qc/0502005.

\bibitem{Wise:2006sm}
Derek~K. Wise.
\newblock {MacDowell-Mansouri gravity and Cartan geometry}.
\newblock {\em Class.Quant.Grav.}, 27:155010, 2010, gr-qc/0611154.

\bibitem{SharpeCartan}
R.W Sharpe.
\newblock {Cartan's Generalization of Klein's Erlangen Program}.
\newblock 1997.
\newblock Book, Springer.

\bibitem{Shirafuji:1988ia}
Takeshi Shirafuji and Masahumi Suzuki.
\newblock {Gauge Theory of Gravitation: A Unified Formulation of Poincare and
  Anti-de Sitter Gauge Theories}.
\newblock {\em Prog.Theor.Phys.}, 80:711, 1988.

\bibitem{Westman:2013mf}
H.F. Westman and T.G. Zlosnik.
\newblock {Exploring Cartan gravity with dynamical symmetry breaking}.
\newblock {\em Class.Quant.Grav.}, 31:095004, 2014, 1302.1103.

\bibitem{Jennen:2014mba}
Hendrik Jennen.
\newblock {Cartan geometry of spacetimes with a nonconstant cosmological
  function $\Lambda$}.
\newblock 2014, 1406.2621.

\bibitem{Vaid:2013woa}
Deepak Vaid.
\newblock {Superconducting and Anti-Ferromagnetic Phases of Spacetime}.
\newblock 2013, 1312.7119.

\bibitem{Gronwald:1995em}
Frank Gronwald and Friedrich~W. Hehl.
\newblock {On the gauge aspects of gravity}.
\newblock 1995, gr-qc/9602013.

\bibitem{Hehl:1994ue}
Friedrich~W. Hehl, J.~Dermott McCrea, Eckehard~W. Mielke, and Yuval Ne'eman.
\newblock {Metric affine gauge theory of gravity: Field equations, Noether
  identities, world spinors, and breaking of dilation invariance}.
\newblock {\em Phys. Rept.}, 258:1--171, 1995, gr-qc/9402012.

\bibitem{Blagojevic:2002du}
M.~Blagojevic.
\newblock {Gravitation and gauge symmetries}.
\newblock Bristol, UK: IOP (2002) 522 p.

\bibitem{westman2009first}
Hans~F Westman.
\newblock A first-principles implementation of scale invariance using best
  matching.
\newblock {\em arXiv preprint arXiv:0910.1631}, 2009.

\bibitem{WestmanZlosnik2014}
Westman. H.F and Zlosnik. T.G.
\newblock {Work In Preparation}.
\newblock 2014.

\bibitem{Gryb:2012qt}
Sean Gryb and Flavio Mercati.
\newblock {2+1 gravity on the conformal sphere}.
\newblock 2012, 1209.4858.

\bibitem{Wald:1984rg}
Robert~M. Wald.
\newblock {General Relativity}.
\newblock 1984.
\newblock Book, The University of Chicago Press.

\bibitem{Krasnov:2011pp}
Kirill Krasnov.
\newblock {New Action Principle for General Relativity}.
\newblock {\em Phys. Rev. Lett.}, 106:251103, 2011, 1103.4498.

\bibitem{Krasnov:2012pd}
Kirill Krasnov.
\newblock {A Gauge Theoretic Approach to Gravity}.
\newblock 2012, 1202.6183.

\bibitem{Lego}
{Lego is a popular Danish toy brand.}
\newblock http://www.lego.dk.

\bibitem{AndersonRelativity}
J.L. Anderson.
\newblock {Principles of Relativity Physics}.
\newblock 1967.
\newblock Book, Academic Press Inc.

\bibitem{WestmanSonego2007b}
Hans Westman and Sebastiano Sonego.
\newblock {Coordinates, observables and symmetry in relativity}.
\newblock 2007, 0711.2651.

\bibitem{Stelle:1979va}
K.~S. Stelle and Peter~C. West.
\newblock {De Sitter gauge invariance and the geometry of the Einstein-Cartan
  theory}.
\newblock {\em J. Phys.}, A12:L205--L210, 1979.

\bibitem{Pagels:1983pq}
Heinz~R. Pagels.
\newblock {Gravitational gauge fields and the cosmological constant}.
\newblock {\em Phys. Rev.}, D29:1690, 1984.

\bibitem{Randono:2010ym}
Andrew Randono.
\newblock {Gravity from a fermionic condensate of a gauge theory}.
\newblock {\em Class. Quant. Grav.}, 27:215019, 2010, 1005.1294.

\bibitem{Mercuri:2009zi}
Simone Mercuri.
\newblock {Peccei-Quinn mechanism in gravity and the nature of the
  Barbero-Immirzi parameter}.
\newblock {\em Phys.Rev.Lett.}, 103:081302, 2009, 0902.2764.

\bibitem{Taveras:2008yf}
Victor Taveras and Nicolas Yunes.
\newblock {The Barbero-Immirzi Parameter as a Scalar Field: K-Inflation from
  Loop Quantum Gravity?}
\newblock {\em Phys.Rev.}, D78:064070, 2008, 0807.2652.

\bibitem{TorresGomez:2008fj}
Alexander Torres-Gomez and Kirill Krasnov.
\newblock {Remarks on Barbero-Immirzi parameter as a field}.
\newblock {\em Phys.Rev.}, D79:104014, 2009, 0811.1998.

\bibitem{Calcagni:2009xz}
Gianluca Calcagni and Simone Mercuri.
\newblock {The Barbero-Immirzi field in canonical formalism of pure gravity}.
\newblock {\em Phys.Rev.}, D79:084004, 2009, 0902.0957.

\bibitem{Toloza:2013wi}
Adolfo Toloza and Jorge Zanelli.
\newblock {Cosmology with scalar-Euler form coupling}.
\newblock {\em Class.Quant.Grav.}, 30:135003, 2013, 1301.0821.

\bibitem{Alexander:2008wi}
Stephon Alexander and Nicolas Yunes.
\newblock {Chern-Simons Modified Gravity as a Torsion Theory and its
  Interaction with Fermions}.
\newblock {\em Phys.Rev.}, D77:124040, 2008, 0804.1797.

\bibitem{Alexander:2009tp}
Stephon Alexander and Nicolas Yunes.
\newblock {Chern-Simons Modified General Relativity}.
\newblock {\em Phys.Rept.}, 480:1--55, 2009, 0907.2562.

\bibitem{Ratra:1987rm}
Bharat Ratra and P.J.E. Peebles.
\newblock {Cosmological Consequences of a Rolling Homogeneous Scalar Field}.
\newblock {\em Phys.Rev.}, D37:3406, 1988.

\bibitem{Chamseddine:1977ih}
Ali~H. Chamseddine.
\newblock {Massive Supergravity from Spontaneously Breaking Orthosymplectic
  Gauge Symmetry}.
\newblock {\em Annals Phys.}, 113:219, 1978.

\bibitem{Chen:2009at}
Hsin Chen, Fei-Hung Ho, James~M. Nester, Chih-Hung Wang, and Hwei-Jang Yo.
\newblock {Cosmological dynamics with propagating Lorentz connection modes of
  spin zero}.
\newblock {\em JCAP}, 0910:027, 2009, 0908.3323.

\bibitem{Magueijo:2013yya}
Joao Magueijo, Matias Rodriguez-Vazquez, Hans Westman, and T.G. Zlosnik.
\newblock {Cosmological signature change in Cartan Gravity with dynamical
  symmetry breaking}.
\newblock {\em Phys.Rev.}, D89:063542, 2014, 1311.4481.

\bibitem{MacDowell:1977jt}
S.~W. MacDowell and F.~Mansouri.
\newblock {Unified Geometric Theory of Gravity and Supergravity}.
\newblock {\em Phys. Rev. Lett.}, 38:739, 1977.
\newblock [Erratum-ibid.38:1376,1977].

\bibitem{hawking1993brief}
Stephen Hawking.
\newblock {\em A brief history of time}.
\newblock Bantam, 1993.

\bibitem{Ellis:1991st}
G.~Ellis, A.~Sumeruk, D.~Coule, and Charles Hellaby.
\newblock {Change of signature in classical relativity}.
\newblock {\em Class.Quant.Grav.}, 9:1535--1554, 1992.

\bibitem{Chamseddine:2013hwa}
Ali~H. Chamseddine and Viatcheslav Mukhanov.
\newblock {Who Ordered the Anti-de Sitter Tangent Group?}
\newblock {\em JHEP}, 1311:095, 2013, 1308.3199.

\bibitem{Ha:1995mj}
Yuan~K. Ha.
\newblock {Coupling of gravity to matter via SO(3,2) gauge fields}.
\newblock {\em Gen.Rel.Grav.}, 27:713--719, 1995, gr-qc/0409058.

\bibitem{Kerr:2014sma}
Steven Kerr.
\newblock {Gauge theory of gravity and matter}.
\newblock 2014, 1408.1994.

\bibitem{Magueijo:2008sx}
Joao Magueijo.
\newblock {Bimetric varying speed of light theories and primordial
  fluctuations}.
\newblock {\em Phys. Rev.}, D79:043525, 2009, 0807.1689.

\bibitem{Magueijo:2010zc}
Joao Magueijo, Johannes Noller, and Federico Piazza.
\newblock {Bimetric structure formation: non-Gaussian predictions}.
\newblock {\em Phys. Rev.}, D82:043521, 2010, 1006.3216.

\bibitem{Zumalacarregui:2010wj}
M.~Zumalacarregui, T.S. Koivisto, D.F. Mota, and P.~Ruiz-Lapuente.
\newblock {Disformal Scalar Fields and the Dark Sector of the Universe}.
\newblock {\em JCAP}, 1005:038, 2010, 1004.2684.

\bibitem{Koivisto:2008ak}
Tomi~S. Koivisto.
\newblock {Disformal quintessence}.
\newblock 2008, 0811.1957.

\bibitem{Kaloper:2003yf}
Nemanja Kaloper.
\newblock {Disformal inflation}.
\newblock {\em Phys. Lett.}, B583:1--13, 2004, hep-ph/0312002.

\bibitem{Bekenstein:2004ne}
Jacob~D. Bekenstein.
\newblock {Relativistic gravitation theory for the MOND paradigm}.
\newblock {\em Phys.Rev.}, D70:083509, 2004, astro-ph/0403694.

\bibitem{Skordis:2005xk}
Constantinos Skordis, D.~F. Mota, P.~G. Ferreira, and C.~Boehm.
\newblock {Large Scale Structure in Bekenstein's theory of relativistic
  Modified Newtonian Dynamics}.
\newblock {\em Phys. Rev. Lett.}, 96:011301, 2006, astro-ph/0505519.

\bibitem{Percacci:1984ai}
R.~Percacci.
\newblock {Spontaneous Soldering}.
\newblock {\em Phys.Lett.}, B144:37--40, 1984.

\bibitem{Nesti:2009kk}
F.~Nesti and R.~Percacci.
\newblock {Chirality in unified theories of gravity}.
\newblock {\em Phys.Rev.}, D81:025010, 2010, 0909.4537.

\bibitem{Percacci:2009ij}
R.~Percacci.
\newblock {Gravity from a Particle Physicists' perspective}.
\newblock {\em PoS}, ISFTG2009:011, 2009, 0910.5167.

\bibitem{Itin:2004qr}
Yakov Itin and Friedrich~W. Hehl.
\newblock {Is the Lorentz signature of the metric of space-time electromagnetic
  in origin?}
\newblock {\em Annals Phys.}, 312:60--83, 2004, gr-qc/0401016.

\bibitem{PhysRevD.89.084053}
Eolo Di~Casola, Stefano Liberati, and Sebastiano Sonego.
\newblock Weak equivalence principle for self-gravitating bodies: A sieve for
  purely metric theories of gravity.
\newblock {\em Phys. Rev. D}, 89:084053, Apr 2014.

\bibitem{flanders2012differential}
Harley Flanders.
\newblock {\em Differential forms with applications to the physical sciences}.
\newblock Courier Dover Publications, 2012.

\end{thebibliography}
\end{document}